\documentclass[12pt]{article}
\usepackage{amsmath,amssymb,epsfig,ulem}
\usepackage[bookmarks=false]{hyperref}  

\allowdisplaybreaks     
 
\newcommand{\ord}{{\cal O}}
\newcommand{\ie}{{\it i.e.}}  
   
\newcommand{\etc}{{\it etc.}}  
\newcommand{\beq}{\begin{equation}}
\newcommand{\eeq}{\end{equation}}
\newcommand{\bea}{\begin{eqnarray}} 
\newcommand{\eea}{\end{eqnarray}}
\newcommand{\Fig}[1]{Figure~\ref{#1}}

\newcommand{\Tab}[1]{Table~\ref{#1}}
\newcommand{\Sec}[1]{Section~\ref{#1}}
\newcommand{\App}[1]{Appendix~\ref{#1}}
\newcommand{\eq}[1]{(\ref{#1})}

\newcommand{\GeV}{\hspace{1mm} {\rm GeV}}
\newcommand{\TeV}{\hspace{1mm} {\rm TeV}}
\newcommand{\ff}{f\hspace{-0.4em}f}

\begin{document}

\begin{titlepage}

\begin{flushright}
MZ-TH/11-14 \\
OUTP-11-41P
\end{flushright}

\vspace{15pt}
\begin{center}
  \Large\bf 
  Massive Color-Octet Bosons: \\
  Bounds on Effects in Top-Quark Pair Production
\end{center}

\vspace{5pt}
\begin{center}
{\sc Ulrich~Haisch$^{a,c}$ and Susanne Westhoff$^{a,b}$}\\
\vspace{10pt} {\sl
$^a$Institut f\"ur Physik (THEP),  
Johannes Gutenberg-Universit\"at \\
D-55099 Mainz, Germany} \\
\vspace{10pt} {\sl
$^b$Helmholtz-Institut Mainz,  
Johannes Gutenberg-Universit\"at \\
D-55099 Mainz, Germany} \\
\vspace{10pt} {\sl
$^c$Rudolf Peierls Centre for Theoretical Physics, 
University of Oxford\\
OX1 3PN Oxford, United Kingdom}
\end{center}

\vspace{10pt}
\begin{abstract}
\vspace{2pt} 
\noindent
A critical survey of the existing direct and indirect constraints on
massive spin-one color octets is presented. Since such new degrees of
freedom appear in any extension of the color gauge group to a product
of at least two $SU(3)$ factors, we keep our discussion as independent
as possible from the underlying theory. In the framework of scenarios
that involve flavor non-universal couplings, we show that excessive
flavor-changing neutral currents can be avoided by a suitable
alignment in flavor space. Constraints from electroweak precision
observables and direct production at hadron colliders still leave
space for sizable new-physics effects in top-quark pair production, in
particular a large forward-backward asymmetry.  In this context, we
derive a model-independent upper bound on the asymmetry that applies
whenever top-antitop production receives the dominant corrections from
$s$-channel exchange of a new single color-octet resonance.
\end{abstract}

\vfill
\end{titlepage}

\tableofcontents

\section{Introduction}

The CERN Large Hadron Collider (LHC) has launched a new era of
particle physics research. With the first $45 \, {\rm pb}^{-1}$ of
luminosity recorded by both ATLAS and CMS during stable $pp$ beams at
$7 \TeV$ center-of-mass (CM) energy, it has started to supersede the
Fermilab Tevatron at the energy frontier.  While much of the attention
concerning new-physics searches has centered around theories that
offer new insights into the mechanism of electroweak symmetry breaking
(EWSB), all initial states at hadron colliders consist of particles
that are charged under the color gauge group. Any new strongly-coupled
colored resonance with TeV-scale mass will therefore be copiously
produced at the LHC. Experimental studies of the properties of such
resonances will provide us with valuable information about the
underlying theory and may also give us some clues about the dynamics
of EWSB and/or other deep questions left unanswered by the standard
model (SM) of particle physics.

Of general phenomenological interest are massive color-octet vector
bosons, whose production channels interfere with the dominant parton
decay channels of quantum chromodynamics (QCD).  Resonances of this
kind arise in a wide spectrum of new-physics models, for instance in
extra dimension scenarios of Randall-Sundrum type
\cite{Randall:1999ee}, which explain the gauge hierarchy by
gravitational red-shifting, or in technicolor theories, where EWSB is
triggered by new strong dynamics (for a review see
\cite{Hill:2002ap}).  Many of these models have in common that they
involve an extension of the SM color gauge group $SU(3)_c$ to a chiral
product group of two (or more) $SU(3)$ factors, which is spontaneously
broken to its diagonal subgroup at energies above the electroweak
scale. The most important model-independent prediction of the original
proposal of chiral color \cite{Frampton:1987dn}, as well as its
numerous variations, is the existence of a massive color-octet vector
boson. This new resonance, dubbed axigluon \cite{Frampton:1987dn,
  Frampton:1987ut, Bagger:1987fz}, interacts strongly and possesses
axial-vector couplings to fermions, whose chiral states are charged
under different $SU(3)$ factors. While the earliest chiral color
models were treating the different quark flavors universally, it is
also possible to construct flavor non-universal scenarios by choosing
anomaly-free fermion representations with appropriate color quantum
numbers. In fact, in concepts like topcolor \cite{Hill:1991at,
  Hill:1994hp} and certain coloron models \cite{Hill:1993hs,
  Chivukula:1996yr}, the massiveness of the top quark and/or the
mechanism of EWSB are explained by the exchange of a TeV-scale
gluon-like object with enhanced couplings to heavy quarks. Due to the
compositeness of the top quark, the very same feature is shared by Kaluza-Klein (KK)
gluons in warped extra dimensions and their dual conformal field
theory interpretations \cite{Agashe:2003zs, Agashe:2006hk,
  Lillie:2007yh}. In flavor non-universal set-ups, the existence
of a massive colored octet might hence serve a higher purpose, rather
than being an accident.

Another more pragmatic motivation to study the physics of flavor
non-universal color-octet bosons is provided by the puzzling picture
of results in top-quark pair production obtained at the Tevatron: the
accurately measured total cross section \cite{CDFnotetot, D0notetot}
and the spectrum in bins of the invariant mass of the top-antitop pair
\cite{Aaltonen:2009iz} are both in good agreement with their SM
predictions, but the inclusive $t \bar t$ forward-backward asymmetry
and its distribution at high invariant masses $M_{t \bar t} > 0.45
\TeV$ \cite{Aaltonen:2011kc} are not.  The inclusive asymmetry has
been measured several times in the lepton plus jets channel
\cite{Schwarz:2006ud, Abazov:2007qb, Aaltonen:2008hc, D0brandnew} and
very recently also in the dilepton channel \cite{CDFdileptonnote}.
The central values of all measurements turned out to be larger than
expected from the sole presence of SM physics. The experimental
situation is tantalizing because the anomaly at high $M_{t\bar t}$ has
a statistical significance of $3.4 \sigma$, which translates into a
discrepancy of about $2\sigma$ in the inclusive measurements. The
sharp growth of the excess with the invariant mass of the $t \bar t$
system suggests the presence of a new heavy particle in top-quark pair
production. By its very nature a heavy color-octet resonance with
appropriate axial-vector couplings to the top quark and the light
quarks seems to be able to generate a sizable $t \bar t$
forward-backward asymmetry. But how big can the effect of such a
flavor non-universal colored resonance in the asymmetry be, given the
strong constraints imposed by the symmetric top-quark observables?
How restrictive are other constraints that follow from flavor physics,
precision measurements at the $Z$ pole, and direct production?  In
particular, do they leave enough space to explain the anomaly? 

The purpose of this article is to provide quantitative answers to
these questions by critically assessing the existing direct and
indirect constraints on axigluons and their doppelg\"angers. We find
that the dominant indirect constraints arise from the electroweak
precision observables (EWPOs) \cite{LEPEWWG:2005ema}. At the one-loop
level, the presence of heavy gluon partners alters the $Z$-boson
couplings to quarks, which are tightly constrained by the precise
measurements of the bottom-quark pseudo observables (POs), as well as
the total and hadronic decay widths of the $Z$ boson. Less
significant, but nevertheless non-trivial, are the constraints that
stem from two-loop contributions to the Peskin-Takeuchi parameters
which encode oblique corrections to the weak gauge boson propagators
\cite{Peskin:1991sw}. While axigluon models with flavor non-universal
couplings are rife with new and dangerous tree-level flavor-changing
neutral currents (FCNCs), the constraints resulting from neutral meson
mixing turn out to be highly model-dependent. Nevertheless, there
exists a minimal, though quite weak, constraint from $D$-meson mixing
that has to be respected by any scenario with flavor non-universal
interactions. Direct constraints derive from resonance searches and
analyses of angular distributions in dijet production at hadron
colliders. The recent LHC bounds \cite{Khachatryan:2010jd,
  Khachatryan:2011as, Collaboration:2011aj} significantly restrict the
properties of any strongly-coupled $s$-channel resonance produced by
the annihilation of light quarks and antiquarks. Taking direct and
indirect constraints into account, we show that the effects of a
flavor non-universal axigluon in top-quark pair production yield a
satisfactory fit to the data, but that the size of effects is limited
by direct searches. Throughout the article, we try to keep our
discussion as general as possible by presenting analytic formulas that
are applicable to a wide class of scenarios involving massive
color-octet bosons. Other recent studies that discuss the physics of
heavy gluon partners, which partially overlap with our work, include
\cite{Frampton:2009rk, Cao:2010zb, Chivukula:2010fk, Han:2010rf,
  Bai:2011ed, AguilarSaavedra:2011vw, Gresham:2011pa, Hewett:2011wz,
  Barcelo:2011fw, AguilarSaavedra:2011hz}.

This article is organized as follows. After reviewing in
\Sec{sec:model} the basic ingredients of the flavor non-universal
axigluon model, we study in \Sec{sec:flavor} the additional sources of
flavor breaking present in the scenario and calculate their impact on
$K$--$\bar K$, $B_{d,s}$--$\bar B_{d,s}$, and $D$--$\bar D$
mixing. \Sec{sec:ewpo} is devoted to a discussion of the indirect
constraints coming from the EWPOs. In \Sec{sec:collider}, we elaborate
on effects of heavy colored resonances at hadron colliders. After
setting the scene for new physics in top-quark observables, we derive
the constraints from dijet production at the LHC. The actual bounds on
the parameter space of models with massive color-octet vectors are
presented in \Sec{sec:global}, which also contains a detailed study of
axigluon effects in top-quark pair production. We conclude in
\Sec{sec:concl}. A series of appendices contains useful details
concerning some technical aspects of our calculations.

\section{Flavor Non-Universal Axigluon Model}
\label{sec:model}

We consider the extended gauge group $SU(3)_A\times SU(3)_B$ with
coupling constants $g_A$ and $g_B$, which is spontaneously
broken\footnote{We ignore possible effects of the scalar sector
  responsible for this breaking in our work.} at a scale of $f =
\ord{(1\,\rm TeV)}$ down to the QCD gauge group $SU(3)_c$.  This
breaking pattern yields two mass eigenstates of color-octet gauge
bosons. A massless particle $g$, which can be identified with the
regular gluon, and an axigluon $G$, that acquires a mass $M_G = g f$
with $g = \sqrt{g_A^2 +g_B^2}$. The corresponding fields are related
to the gauge eigenstates $A_{\mu}$ and $B_{\mu}$ via
\beq \label{eq:GgAB}
\left(\begin{array}{c}
    G_{\mu}\\
    g_{\mu}
      \end{array}
    \right) = \left(\begin{array}{cc}
        \sin\theta & -\cos\theta\\
        \cos\theta & \phantom{+} \sin\theta
                \end{array}
              \right)\left(\begin{array}{c}
                  A_{\mu}\\
                  B_{\mu}
             \end{array}
\right)\,,\qquad \tan\theta = \frac{g_A}{g_B}\,.
\eeq
Notice that physical predictions are symmetric under the exchange of
$g_A$ with $g_B$, so that only half of the parameter space in the
mixing angle $\theta$ is physical. This means that instead of
considering values of $\theta$ in the range $[0^\circ, 90^\circ]$, one
can restrict $\theta$ to lie within $[0^\circ,45^\circ]$.  The gauge
sector of the model is therefore fully characterized by two parameters
only, the axigluon mass $M_G$ and the mixing angle $\theta$.

If left- and right-chiral quarks carry different charges under the
$SU(3)_A$ and $SU(3)_B$ gauge factors, the axigluon exhibits
axial-vector couplings $g_A^q$ to quarks. This is an interesting
feature because it leads to a non-zero forward-backward asymmetry
$A_{\rm FB}^t$ in $t \bar t$ production at tree
level~\cite{Antunano:2007da, Ferrario:2008wm}. In order to achieve a
positive shift in $A_{\rm FB}^t$ the axigluon has to couple to the
first and the third generation of quarks with opposite axial-vector
couplings \cite{Ferrario:2009bz}, implying $g_A^q \hspace{0.25mm}
g_A^t < 0$.\footnote{Strictly speaking, this is true only under the
  assumption that the interference between the SM and the new-physics
  $q \bar q \to t \bar t$ amplitudes gives the dominant contribution
  to the asymmetric cross section.}  This observation motivates us to
choose the charge assignments for the quark fields as in
\Tab{tab:charges}, following the proposal in \cite{Frampton:2009rk}.

\begin{table}[!t]
\begin{center}
\begin{tabular}{|c|c c c|c c c|}
\hline
 & $Q_{1,2}$ & $u_{1,2}^c$ & $d_{1,2}^c$ & $Q_{3,4}$ & $u_{3,4}^c$ & $d_{3,4}^c$\\
\hline
$SU(3)_A$ & $3$ & $1$ & $1$ & $1$ & $\bar 3$ & $\bar 3$\\
$SU(3)_B$ & $1$ & $\bar 3$ & $\bar 3$ & $3$ & $1$ & $1$\\
\hline
\end{tabular}
\end{center}
\begin{center}
\parbox{15.5cm}{\caption{\label{tab:charges} 
  Charge assignments of the quark fields under the $SU(3)_A$ 
  and $SU(3)_B$ gauge factors. The symbols $Q_p$ and $u_p^c$, 
  $d_p^c$ with $p = 1,2,3,4$ denote electroweak doublets and 
  conjugate singlets, respectively. 
}}
\end{center}
\end{table}

We emphasize that in order to cancel the gauge anomaly our flavor
non-universal axigluon model contains a sequential fourth generation
of heavy quarks and leptons (the charge assignments of the lepton
fields have not been reported in \Tab{tab:charges} for brevity). While
the absolute mass scale of the new fermions is bounded from below by
direct searches,\footnote{The 95\% confidence limits (CLs) on the
  masses of fourth-generation leptons obtained by LEP II are
  $m_{\nu_4} > 101.5 \, {\rm GeV}$ and $m_{\ell_4} > 101.9 \, {\rm
    GeV}$ \cite{Achard:2001qw}, while the latest published Tevatron
  bounds on the fourth-generation up- and down-type quark masses read
  $m_{u_4} > 311 \, {\rm GeV}$ \cite{Lister:2008is} and $m_{d_4} > 372
  \, {\rm GeV}$ \cite{Aaltonen:2011vr} at 95\% CL. The first LHC
  bounds \cite{Collaboration:2011em} on the masses $m_{u_4}$ and
  $m_{d_4}$ are still weaker than those from the Tevatron.} both the
mass splitting and inter-generational mixing of extra chiral matter is
strongly constrained by EWPOs and flavor physics. The latter
constraints as well as those following from the bottom-quark POs, are
readily evaded by setting the mixing between the first three and the
fourth generation to zero. In this case, one is left with the
constraints that arise from the oblique parameters $S$ and $T$, which
are mostly sensitive to the mass splitting of the fourth-generation
up- and down-type quarks $u_4$ and $d_4$. We will be more precise
below (see the discussion in \Sec{sec:ewpo}), but suffice to say that
the wide spread of possible fourth-generation contributions to $S$ and
$T$ does not allow to derive any model-independent bound on the
parameter space of the flavor non-universal axigluon model from the
oblique corrections.

In the further discussion, we will restrict ourselves for most of the
time, to the first three generations of quarks. Their couplings to the
axigluon are given by ($i,j = 1, 2, 3$)
\beq  \label{eq:quarkaxi}
\mathcal{L} \, \supset \, g_s \hspace{0.5mm} \Big [ \hspace{0.5mm}
(\Gamma_L^u)_{ij} \, \bar u_{L}^i G \!\!\!\! /\,\hspace{0.25mm}
u_{L}^j + (\Gamma_L^d)_{ij} \, \bar d_{L}^i G \!\!\!\!
/\,\hspace{0.25mm} d_{L}^j + (\Gamma_R^u)_{ij} \, \bar u_{R}^i G
\!\!\!\! /\,\hspace{0.25mm} u_{R}^j + (\Gamma_R^d)_{ij} \, \bar
d_{R}^i G \!\!\!\! /\,\hspace{0.25mm} d_{R}^j \Big ] \,,
\eeq
where $g_s = g\sin\theta\cos\theta$ is the strong coupling constant,
$u_{L,R}^i$ ($d_{L,R}^i$) are left- and right-handed chiral up-type
(down-type) quark fields of generation $i$. We use the abbreviation $G
\!\!\!\!  /=\gamma^{\mu} \hspace{0.25mm} G_{\mu}^a
\hspace{0.25mm}T^a$, where $T^a$ is a generator of $SU(3)_c$ in the
adjoint representation. In the weak interaction basis indicated by a
superscript ``$I\hspace{0.25mm}$'', the axigluon couplings are
diagonal $3 \times 3$ matrices in flavor space,
\beq \label{eq:gAI}
\Gamma_L^I = (\Gamma_L^{u,d})^I = \text{diag} \left
  (g_L^{\ell},g_L^{\ell},g_L^h \right )\,, \qquad \Gamma_R^I =
(\Gamma_R^{u,d})^I = \text{diag} \left (g_R^{\ell},g_R^{\ell},g_R^h
\right )\,.
\eeq
If the quark charge assignments are chosen as in \Tab{tab:charges},
the coupling strengths for light ($\ell$) and heavy ($h$) quarks are
given by
\beq \label{eq:glgh}
g_L^{\ell} = g_R^h = \tan\theta \,,\qquad g_L^h = g_R^{\ell} =
-\cot\theta \,.
\eeq
Notice that for $\theta \in [0^\circ, 45^\circ]$ these couplings
fulfill
\begin{gather} \label{eq:gAgV}
  g_A^\ell \hspace{0.5mm} g_A^h = \left (g_L^\ell - g_R^\ell \right )
  \left (g_L^h - g_R^h \right ) = -\frac{4}{\sin^2 \left (2 \theta
    \right )} \, < \, 0 \,,
  \nonumber \\[-3mm] \\[-3mm]
  \frac{| g_V^{\ell, h} |}{ | g_A^{\ell, h} |} = \frac{|g_L^{\ell, h}
    + g_R^{\ell, h} |}{ |g_L^{\ell, h} - g_R^{\ell, h} |} = \cos \left
    (2 \theta \right ) \, < \, 1 \,. \nonumber
\end{gather}
In particular, one has $g_A^\ell \hspace{0.5mm} g_A^h \to -4$ and
$|g_V^{\ell, h}|/|g_A^{\ell, h} | \to 0$ in the limit $\theta \to
45^\circ$, which is crucial to obtain a large forward-backward
asymmetry in $t \bar t$ production, while leaving the symmetric cross
section largely unaffected.

\section{Flavor Physics}
\label{sec:flavor}

Naively one might think that the number of additional flavor
parameters in our axigluon model is given by $4 \cdot 6 = 24$ mixing
angles and $4 \cdot 3 = 12$ CP-violating phases, \ie, the number of
elements of the hermitian matrices $\Gamma_L^{u,d}$ and
$\Gamma_R^{u,d}$. Yet, most of the parameters which appear in the
axigluon couplings are unphysical. In order determine the number of
physical degrees of freedom in (\ref{eq:quarkaxi}), we first recall
that in the absence of the Yukawa couplings,
\beq \label{eq:SMY}
{\cal L} \, \supset \, - \hspace{0.25mm} (Y_u)_{ij} \, \bar Q_{L}^i
\hspace{0.25mm} \tilde \phi \hspace{0.75mm} u_{R}^j - (Y_d)_{ij}
\hspace{0.5mm} \bar Q_{L}^i \hspace{0.25mm} \phi \hspace{0.75mm}
d_{R}^j + {\rm h.c.} \,,
\eeq
the SM would possess a large global non-abelian $SU(3)_{Q_L} \times
SU(3)_{u_R} \times SU(3)_{d_R}$ flavor symmetry.  Here $\phi$ denotes
the SM Higgs doublet and $\tilde \phi = i \tau_2 \hspace{0.25mm}
\phi^\ast$.

Before spontaneous EWSB, the terms that lead to (\ref{eq:quarkaxi})
can be written in the following way
\beq  \label{eq:quarkaxiSU2}
\mathcal{L} \, \supset \, g_s \hspace{0.5mm} \Big [ \hspace{0.25mm}
(\Gamma_L)_{ij} \, \bar Q_{L}^i \hspace{0.25mm} G \!\!\!\!
/\,\hspace{0.5mm} Q_{L}^j + (\Gamma_R)_{ij} \, \bar Q_{R}^i
\hspace{0.25mm} G \!\!\!\! /\,\hspace{0.5mm} Q_{R}^j \Big ] \,,
\eeq
where $Q_{L,R}^i = (u_{L,R}^i\hspace{0.25mm}, d_{L,R}^i)^T$ are
$SU(2)_{L,R}$ doublets. Using now the $SU(3)_{Q_L} \times SU(3)_{u_R}
\times SU(3)_{d_R}$ symmetry transformations, one can choose (without
loss of generality) to work in the basis with
\beq \label{eq:dbase}
Y_u = V^\dagger \hspace{0.25mm} \lambda_u \,, \qquad Y_d = \lambda_d
\,, \qquad \Gamma_L = U_d^\dagger \hspace{0.75mm} \Gamma_L^I
\hspace{0.5mm} U_d \,, \qquad \Gamma_R = \Gamma_R^I \,,
\eeq
where $V$ denotes the Cabibbo-Kobayashi-Maskawa (CKM) matrix and
$\lambda_{u,d}$ are of the form
\beq \label{eq:lambdaud} 
\lambda_u = \frac{\sqrt{2}}{v} \, {\rm diag} \left ( m_u, m_c, m_t
\right ) \,, \qquad \lambda_d = \frac{\sqrt{2}}{v} \, {\rm diag} \left
  ( m_d, m_s, m_b \right ) \,, 
\eeq 
with $v \approx 246 \hspace{0.5mm} \GeV$ being the Higgs vacuum
expectation value (VEV) and $m_u$, $m_d$, \etc \ denoting the quark
masses. The unitary matrix $U_d$ in (\ref{eq:dbase}) parametrizes the
misalignment of the left-handed operator appearing in
(\ref{eq:quarkaxiSU2}) relative to the down-type quark mass basis.

Alternatively, one can choose to work in a basis where
\beq \label{eq:ubase}
Y_u = \lambda_u \,, \qquad Y_d = V \hspace{0.25mm} \lambda_d \,,
\qquad \Gamma_L = U_u^\dagger \hspace{0.75mm} \Gamma_L^I
\hspace{0.5mm} U_u \,, \qquad \Gamma_R = \Gamma_R^I \,,
\eeq
and $U_u$ is a unitary matrix which parametrizes the misalignment of
the left-handed operator in (\ref{eq:quarkaxiSU2}) with the up-type
quark mass basis. Importantly, the unitary matrices $U_{u,d}$ are not
independent from each other, but related via the CKM
matrix. Explicitly, one has
\beq \label{eq:UuUdV}
U_u = U_d \hspace{0.5mm} V^\dagger \,.
\eeq

By appropriate rotations of the chiral quark fields, we can choose a
basis where the mass matrices are diagonal and the axigluon couplings
take the form
\beq \label{eq:Gammafinal}
\Gamma_L^{u,d} =  U_{u,d}^\dagger \hspace{0.75mm} \Gamma_L^I
\hspace{0.5mm} U_{u,d} \,, \qquad 
\Gamma_R^{u,d} =  \Gamma_R^I \,,
\eeq
with $U_u$ and $U_d$ satisfying (\ref{eq:UuUdV}). We conclude that the
total number of additional real and imaginary flavor parameters
amounts to $6$ mixing angles and $3$ phases. These appear all in the
sector of left-handed quarks, namely in $\Gamma_L^{u,d}$. In contrast,
the right-handed quark sector does not involve new sources of flavor
breaking, since the corresponding chiral rotations are not observable
in the SM. This allows one to diagonalize the axigluon couplings
$\Gamma_R^{u,d}$ simultaneously.

\begin{figure}[!t]
\begin{center}
\includegraphics[height=3cm]{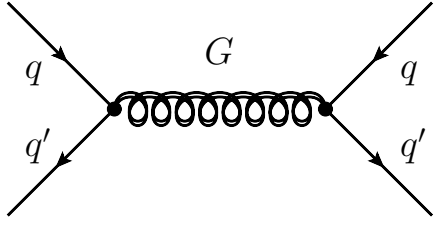}
\begin{picture}(0,0)(0,0)
\put(-92,60){\large $G$}
\put(-160,25){\large $q^\prime$}
\put(-160,54.5){\large $q$}
\put(-18,25){\large $q^\prime$}
\put(-18,54.5){\large $q$}
\end{picture}
\end{center}
\begin{center} 
  \parbox{15.5cm}{\caption{\label{fig:flavor} Tree-level axigluon
      contribution to the mixing of neutral mesons.}}
\end{center}
\end{figure}

The structure of (\ref{eq:Gammafinal}) implies that the
non-universality of the axigluon couplings induces tree-level FCNCs
among both left-handed up- and down-type quarks.  However, the
flavor-breaking terms in the two sectors are connected via 
\beq \label{eq:GuVGdV}
\Gamma_L^u = V \hspace{0.5mm} \Gamma_L^d \hspace{0.5mm} V^\dagger \,.
\eeq
This equation implies that the axigluon couplings $\Gamma_L^u$ and
$\Gamma_L^d$ are to first approximation identical, since the CKM
matrix is close to a unit matrix. Explicitly, we find that the
left-handed axigluon coupling to down-type quarks takes the form
\beq\label{eq:gld}
\Gamma_L^d = g_L^{\ell} \; 1 +
(g_L^h-g_L^{\ell})\left(\begin{array}{ccc} |(U_d)_{31}|^2 &
    (U_d^{\ast})_{31}(U_d)_{32} &
    (U_d^{\ast})_{31}(U_d)_{33}\\[1mm]
    (U_d)_{31}(U_d^{\ast})_{32} & |(U_d)_{32}|^2 &
    (U_d^{\ast})_{32}(U_d)_{33}\\[1mm]
    (U_d)_{31}(U_d^{\ast})_{33} & (U_d)_{32}(U_d^{\ast})_{33} &
    |(U_d)_{33}|^2\\ \end{array} \right), 
\eeq
where $1$ denotes the $3 \times 3$ identity matrix.  Notice that the
splitting of heavy- and light-quark couplings, $(g_L^h-g_L^{\ell})$,
determines the overall strength of flavor breaking and that all of
these terms arise from the mixing with the third generation. For
example, the $s_L \to d_L$, $b_L \to d_L$, and $b_L \to s_L$ processes
involve the combinations $(U_d^{\ast})_{31}
(U_d)_{32}\hspace{0.25mm}$, $(U_d^{\ast})_{31}
(U_d)_{33}\hspace{0.25mm}$, and $(U_d^{\ast})_{32} (U_d)_{33}$ of
mixing-matrix elements.

The presence of the new unitary matrix $U_d$, which (together with
$V$) controls the amount of flavor mixing in the left-handed quark
sector, can have a considerable impact on particle-antiparticle mixing
of neutral mesons. The relevant Feynman diagram is shown in
\Fig{fig:flavor}. In order to determine the structure of the mixing
matrix $U_d$, we could use the present knowledge on the $\Delta F = 2$
amplitudes and obtain upper bounds on most of the elements of $U_d$
for given values of $M_G$ and $\theta$. In particular, we could
investigate which pattern of $U_d$ would cure potential SM
inconsistency concerning the size of CP violation in
$K\hspace{0.25mm}$--$\bar K$ and $B_{d,s}\hspace{0.25mm}$--$\bar
B_{d,s}$ mixing.\footnote{Recent measurements of the CDF and D{\O}
  collaborations of both the mixing-induced CP asymmetry in $B_s \to
  J/\psi \phi$ and the like-sign di-muon charge asymmetry in
  $B_{d,s}$-meson samples show deviations from their SM predictions. A
  tension also exists in $|\epsilon_K|$ if the exclusive determination
  of $|V_{cb}|$ is used. Yet, the disagreements are not large enough
  to exclude the possibility of statistical and/or systematic origins
  of the deviations.} In the following, we will however take a
different path in order to work out the weakest possible bounds on
$M_G$ and $\theta$ by minimizing the amount of flavor violation in the
$\Delta F = 2$ sector. By virtue of \eq{eq:Gammafinal}, the axigluon
contributions to $K\hspace{0.25mm}$--$\bar K$ and
$B_{d,s}\hspace{0.25mm}$--$\bar B_{d,s}$ mixing can be set to zero by
alignment in the down-type quark sector, \ie, $U_d = 1$. Corrections
to $D\hspace{0.25mm}$--$\bar D$ mixing can be removed by alignment in
the up-type quark sector, \ie, $U_u = 1$ or equivalently $U_d =
V$. However, it is not possible to set the contributions to both down-
and up-type quark $\Delta F = 2$ amplitudes simultaneously to zero by
any choice of $U_d\hspace{0.25mm}$. This feature has also been
observed independently in \cite{Bai:2011ed}. Notice that if the
down-type (up-type) quark sector is aligned one has $\Gamma_L^u = V
\hspace{0.5mm} \Gamma_L^I \hspace{0.5mm} V^\dagger$ ($\Gamma_L^d =
V^\dagger \hspace{0.5mm} \Gamma_L^I \hspace{0.5mm} V$). This implies
that the $c_L \to u_L G$, $t_L \to u_L G$, and $t_L \to c_L G$ ($s_L
\to d_L G$, $b_L \to d_L G$, and $b_L \to s_L G$) amplitudes are to
leading order proportional to the combinations $V_{cb}^{\ast}
V_{ub}\hspace{0.25mm}$, $V_{tb}^\ast V_{ub}\hspace{0.25mm}$, and
$V_{tb}^\ast V_{cb}$ ($V_{td}^\ast V_{ts}\hspace{0.25mm}$,
$V_{td}^\ast V_{tb}\hspace{0.25mm}$, and $V_{ts}^\ast V_{tb}$) of CKM
elements. In this case the flavor-changing axigluon contributions
follow the pattern of minimal flavor violation (MFV). In particular,
there will be no new sources of CP violation beyond the CKM phase.
Flavor alignment is crucial to respect the severe constraints arising
from the quark flavor sector, especially from CP-violating $\Delta
F=2$ observables. This requirement constrains the structure of viable
non-universal axigluon models that are accessible at the existing
hadron colliders.

Before deriving and comparing the constraints on $M_G$ and $\theta$
from the relevant $\Delta F = 2$ mixing amplitudes in the two
scenarios of flavor alignment, let us briefly comment on partial
misalignment in the down-type quark sector.  Generically partial
flavor alignment implies axigluon effects in both up- and down-type
quark sectors. Given the freedom in the choice of $U_d$, it is of
course possible to set the axigluon contributions to $K$--$\bar K$,
$B_d$--$\bar B_d$, and/or $B_s$--$\bar B_s$ mixing to zero. For
example, choosing\footnote{Textures of this type arise in the context
  of topcolor models from the requirement to produce the observed
  non-degenerate quark masses \cite{Buchalla:1995dp, Burdman:2000in,
    Martin:2004ec}. The possible textures are controlled by the
  breaking patterns of horizontal global flavor symmetries, which we
  leave unspecified.}
\beq \label{eq:UdBdmix}
U_d = \begin{pmatrix}1 + {\cal O} (\lambda^2) & 0 & V_{td}^\ast
  \\[1mm] \phantom{i} 0 \phantom{i} & \phantom{i} 1 \phantom{i} &
  \phantom{i} 0 \phantom{i} \\[1mm] V_{td} & 0 & V_{tb} + {\cal O}
  (\lambda^2) \end{pmatrix} ,
\eeq
one arrives at $(\Gamma_L^d)_{13} = \left (g_L^h-g_L^\ell \right )
V_{td}^\ast V_{tb} + \ord (\lambda^5)$ and $(\Gamma_L^d)_{12} =
(\Gamma_L^d)_{23} = 0$, \ie, new-physics effects of MFV pattern in
$B_d$--$\bar B_d$ mixing and no corrections in the $K$--$\bar K$ and
$B_s$--$\bar B_s$ observables. Since $\Gamma_L^d$ and $\Gamma_L^u$
necessarily have to obey \eq{eq:GuVGdV}, the choice \eq{eq:UdBdmix}
will fix the pattern of flavor violation in the $D$-meson
sector. Explicitly, we find
\beq \label{eq:gammaLu12}
\begin{split}
  (\Gamma_L^u)_{12} & = \left (g_L^h-g_L^\ell \right ) \hspace{0.25mm}
  \Big [ V_{cb}^\ast V_{ub} \left |V_{tb} \right |^2 + V_{cb}^\ast
  V_{ud} V_{td}^\ast V_{tb} + V_{cd}^\ast V_{ud} \left |V_{td} \right
  |^2 + V_{cd}^\ast V_{ub} V_{tb}^\ast V_{td} \Big ]
  + \ord(\lambda^5) \\[1mm]
  & = \left (g_L^h-g_L^\ell \right) A^2 \lambda^5 + \ord (\lambda^7)
  \,,
\end{split}
\eeq 
where in the second line we have parametrized the CKM elements in
terms of the Wolfenstein parameters $A$, $\lambda$, $\bar{\rho}$,
$\bar{\eta}$, and expanded in $\lambda\approx 0.23$.  This result has
to be compared with the expression $(\Gamma_L^u)_{12} =
(g_L^h-g_L^\ell) \hspace{0.5mm} V_{cb}^\ast V_{ub} = A^2 \lambda^5
\hspace{0.5mm} (\bar \rho - i \bar \eta) + \ord (\lambda^7)$ obtained
in the case of flavor alignment, $U_d = 1$.  We see that MFV effects
in $B_d\hspace{0.25mm}$--$\bar B_d$ mixing of $\mathcal{O}(\lambda^3)$
imply flavor violation in $D\hspace{0.25mm}$--$\bar D$ mixing of
$\mathcal{O}(\lambda^5)$, but can lead to new CP violation with
respect to the case of flavor alignment. Yet, unless the element
$(U_d)_{13}$ in \eq{eq:UdBdmix} is smaller than
$\mathcal{O}(\lambda^5)$, the constraints on the axigluon parameters
$M_G$ and $\theta$ from $B_d\hspace{0.25mm}$--$\bar B_d$ mixing turn
out to be more stringent than those arising from the
$D\hspace{0.25mm}$--$\bar D$ observables.  Notice that a scenario of
partial flavor alignment in the down-type quark sector similar to the
one in \eq{eq:UdBdmix} has been considered in
\cite{Chivukula:2010fk}. The bound on $M_G$ and $\theta$ derived in
the latter article from $B_d$--$\bar B_d$ mixing is thus the relevant
constraint for the specific pattern of flavor violation, but not the
weakest constraint in general.

We now collect the formulas necessary to analyze the $\Delta F = 2$
observables. The effects of axigluons in neutral meson mixing can be
described by an effective theory. Assuming alignment in the up-type
quark sector, we find in terms of the four-quark operator
\beq \label{eq:Ods}
{\cal O}_K = (\bar s_L \hspace{0.25mm} \gamma^\mu \hspace{0.25mm} d_L)
(\bar s_L \hspace{0.25mm} \gamma_\mu \hspace{0.25mm} d_L) \,,
\eeq
the following axigluon contribution to the effective $\Delta S = 2$
Hamiltonian
\beq \label{eq:HeffDS}
{\cal H}_K^G = \left ( 1 - \frac{1}{N_c} \right ) \frac{\pi
  \alpha_s}{M_G^2} \, (g_L^h - g_L^\ell)^2 \hspace{0.5mm} (
V_{ts}^\ast V_{td} )^2\, {\cal O}_K = \frac{8 \pi \alpha_s}{3 M_G^2}\,
\frac{(V_{ts}^\ast V_{td} )^2}{ \sin^2\left (2 \theta \right)} \,
{\cal O}_K \,,
\eeq
which induces $K\hspace{0.25mm}$--$\bar K$ mixing. Here we have set
$N_c = 3$ and used \eq{eq:glgh} to obtain the final result. The strong
coupling constant $\alpha_s = g_s^2/(4\pi)$ entering the above formula
is understood to be normalized at the scale $M_G$. The expressions for
the $\Delta B = 2$ Hamiltonians, describing
$B_d\hspace{0.25mm}$--$\bar B_d$ or $B_s\hspace{0.25mm}$--$\bar B_s$
mixing, are obtained from \eq{eq:HeffDS} by simply replacing $
V_{ts}^\ast V_{td}$ with $V_{tb}^\ast V_{td} $ or $V_{tb}^\ast V_{ts}$
and employing the relevant four-quark operator. The $\Delta C = 2$
observables will receive no correction in this case.  Assuming instead
alignment in the down-type quark sector, there will be no axigluon
contributions to the $\Delta S =2$ and $\Delta B =2$ transitions, but
the prediction for $D\hspace{0.25mm}$--$\bar D$ mixing will be altered
with respect to its SM expectation. The relevant effective $\Delta C =
2$ Hamiltonian is obtained from \eq{eq:HeffDS} by changing the factor
$V_{ts}^\ast V_{td}$ into $V_{cb}^{\ast} V_{ub}$ and using again the
appropriate four-quark operator.

The tree-level expressions for the Wilson coefficient given in
\eq{eq:HeffDS} must be evolved, using the renormalization group (RG),
down to the low-energy scale $\mu_K = 2 \hspace{0.25mm} \GeV$ in the
case of $K$--$\bar K$ mixing. For the $B_{d,s}$--$\bar B_{d,s}$ and
$D$--$\bar D$ observables the appropriate low-energy scales are $\mu_B
= 4.6 \hspace{0.25mm} \GeV$ and $\mu_D = 2.8 \hspace{0.25mm} \GeV$,
respectively. The running is accomplished by means of the well-known
RG formulas \cite{Buras:2001ra}. From $M_G$ to the top-quark threshold
$m_t$ we use the leading order (LO) approximation, while from $m_t$
down to the appropriate low-energy scale we work at next-to-leading
order (NLO) precision. At the low-energy scale the hadronic matrix
element of the operator \eq{eq:Ods} is customarily expressed in terms
of the parameter $B_K$. For the operator relevant to our analysis, we
write
\beq \label{eq:BK}
\big \langle K | {\cal O}_K | \bar K \big \rangle = \frac{1}{3} \, m_K
f_K^2 B_K \,.
\eeq
Analogous definitions are used for the other mesons. In our numerical
analysis we will employ $m_K = 497.6 \hspace{0.5mm} {\rm MeV}$, $f_K =
(155.8 \pm 1.7) \hspace{0.5mm} {\rm MeV}$, $B_K = 0.527 \pm 0.022$,
$m_{B_d} = 5.2796 \hspace{0.5mm} {\rm GeV}$, $f_{B_d} = (192.8 \pm
9.9) \hspace{0.5mm} {\rm MeV}$, $B_{B_d} = 0.82 \pm 0.07$, $m_{B_s} =
5.3664 \hspace{0.5mm} {\rm GeV}$, $f_{B_s} = (238.8 \pm 9.5)
\hspace{0.5mm} {\rm MeV}$, and $B_{B_s} = 0.86 \pm 0.04$
\cite{Laiho:2009eu} to calculate the $\Delta S =2$ and $\Delta B =2$
observables. In the case of the $\Delta C = 2$ transition we use
instead $m_D = 1.8645 \hspace{0.5mm} {\rm GeV}$, $f_D = (212 \pm 14)
\hspace{0.5mm} {\rm MeV}$, and $B_D = 0.85 \pm 0.09$
\cite{Lubicz:2008am}.

In order to constrain the axigluon mass $M_G$ and the mixing angle
$\theta$ in the case of alignment in the up-type quark sector, we will
consider the three observables
\beq \label{eq:epsKDMds}
\begin{split}
  \epsilon_K & = \frac{\kappa_\epsilon \hspace{0.5mm} e^{i
      \varphi_\epsilon}}{\sqrt{2} \left (\Delta m_K \right )_{\rm
      exp}} \; {\rm Im} \hspace{0.25mm} \big \langle K | {\cal
    H}_K^{\rm SM} + {\cal H}_K^G | \bar K \big \rangle , \\[2mm] &
  \hspace{-2mm} \Delta m_{B_{d,s}} = 2 \hspace{0.5mm} \big | \big
  \langle B_{d,s} | {\cal H}_{B_{d,s}}^{\rm SM} + {\cal H}_{B_{d,s}}^G
  | \bar B_{d,s} \big \rangle \big |\,,
\end{split}
\eeq
where $\varphi_\epsilon = (43.51 \pm 0.05)^\circ$, $\kappa_\epsilon =
0.94 \pm 0.02$ \cite{Buras:2010pza}, and $(\Delta m_K)_{\rm exp} =
3.483 \cdot 10^{-12} \, {\rm MeV}$, while ${\cal H}_{K}^{\rm SM}$
and ${\cal H}_{B_{d,s}}^{\rm SM}$ denote the SM contributions to the
effective $\Delta S =2$ and $\Delta B = 2$ Hamiltonians. We do not
attempt a prediction for $\Delta m_K$, which is plagued by very large
hadronic uncertainties, making it less restrictive than the
CP-violating parameter $\epsilon_K$.

Since we are dealing with the effect of a single operator only, the
resulting expressions for the axigluon contributions to $|\epsilon_K|$
and $\Delta m_{B_{d,s}}$ turn out to be very compact. We find the
following analytic expressions 
\beq \label{eq:epsKDMG}
\begin{split}
  |\epsilon_K|_G & = \frac{\kappa_\epsilon}{\sqrt{2} \left (\Delta m_K
    \right )_{\rm exp}} \, \frac{16\pi \alpha_s}{9} \frac{m_K f_K^2
    P_K}{M_G^2} \frac{A^4 \lambda^{10}\hspace{0.5mm} \bar \eta \left
      (1- \bar \rho \right
    )}{\sin^2\left(2\theta\right )} \, + \, \ord (\lambda^{12}) \,, \\
  & \hspace{+4mm} (\Delta m_{B_d})_G = \frac{16 \pi \alpha_s}{9} \,
  \frac{m_{B_d} f_{B_d}^2 P_{B_d}}{M_G^2} \, \frac{A^2 \lambda^6 \left
      (\bar \eta^2 + (1-\bar \rho)^2 \right ) }{\sin^2 \left (2 \theta
    \right )} \, + \, \ord (\lambda^8) \,.
\end{split} 
\eeq 
The latter formula also applies to $B_s$--$\bar B_s$ mixing after
obvious replacements. The correct CKM factor reads in this case $A^2
\lambda^4$. In our numerical analysis we will use $\lambda = 0.22543
\pm 0.00077$, $A = 0.812 \pm 0.015$, $\bar \rho = 0.148 \pm 0.022$,
and $\bar \eta = 0.344 \pm 0.014$ \cite{Charles:2004jd} and will
include the full CKM dependence. The factors $P_M$ with $M = K, B_d,
B_s$ entail the RG effects below $M_G$ as well as the hadronic
parameters $B_M$ calculated at low energies. In the case of the
$K$--$\bar K$ transition, we obtain
\beq \label{eq:PK} 
P_K = 0.416 \hspace{0.75mm} \eta_6^{6/21} \approx 0.392 \left [ 1-
  0.024 \ln \left ( \frac{M_G}{1 \TeV} \right ) \right ] \,,
\eeq 
where $\eta_6 =\alpha_s(M_G)/\alpha_s(m_t)$. In order to get the
result for the case of $B_{d}$--$\bar B_{d}$ ($B_s$--$\bar B_s$)
mixing one simply replaces $0.416$ by $0.68$ ($0.416$ by $0.72$) in
the above expression. The rescaling factor $0.944 \left [ 1- 0.024 \ln
  \left ( M_G/(1 \TeV) \right ) \right ]$ arising from inserting the
leading-logarithmic expression for $\eta_6$ is universal, \ie,
independent from the considered meson.

In order to determine the allowed parameter space in the
$M_G\hspace{0.5mm}$--$\hspace{0.5mm}\theta$ plane, we will
compare the ratios $C_K = |\epsilon_K|_{\rm exp}/|\epsilon_K|_{\rm
  SM}$ and $C_{B_{d,s}} = (\Delta m_{B_{d,s}})_{\rm exp}/(\Delta
m_{B_{d,s}})_{\rm SM}$ to the corresponding predictions in our
axigluon model. Combining the state-of-the-art SM calculations of
$|\epsilon_K|$ \cite{Brod:2010mj} and $\Delta m_{B_{d,s}}$
\cite{Buras:2001ra} with the current experimental results
\cite{Nakamura:2010zzi}, we find
\beq \label{eq:CM}
C_K = 1.17 \pm 0.16 \,, \qquad C_{B_d} = 0.96 \pm 0.14 \,, \qquad
C_{B_s} = 0.91 \pm 0.09 \,,
\eeq
where the quoted errors have been obtained by adding individual
uncertainties of both theoretical and experimental nature in
quadrature.

In the case of alignment in the down-type quark sector, we will
consider 
\beq \label{eq:xD}
x_D = \frac{2 \left | {\rm Re} \left \langle D | {\cal H}_D^{\rm SM} +
      {\cal H}_D^{\rm G} | \bar D \right \rangle \right |}{\Gamma_D}
\,,
\eeq
to determine the allowed parameter space. Here $\Gamma_D = 1/ \tau_D$
with $\tau_D$ being the $D$-meson lifetime. The axigluon contribution
to $x_D$ reads   
\beq \label{eq:xDG} 
(x_D)_G =  \frac{16 \pi \alpha_s}{9} \, (\tau_D)_{\rm exp} \, \frac{m_D
  f_D^2 P_D}{M_G^2} \frac{A^4 \lambda^{10} \left (\bar \eta^2 - \bar
    \rho^2 \right )}{\sin^2(2\theta)} \, + \, \ord (\lambda^{12}) \,,
\eeq 
where $(\tau_D)_{\rm exp} = 0.4101 \, {\rm ps}$
\cite{Nakamura:2010zzi}. The expression for $P_D$ is obtained from
\eq{eq:PK} by simply replacing $0.416$ by $0.69$.  Like $\Delta m_K$
also $x_D$ is plagued by large theoretical uncertainties. Assuming
that there are no accidental cancellations between the SM and the
axigluon contributions to $x_D\hspace{0.25mm}$, we bound the axigluon
parameters by requiring that the new-physics effects alone do not
exceed the measured value of $x_D$
\cite{TheHeavyFlavorAveragingGroup:2010qj},
\beq \label{eq:xDconstraint}
(x_D)_G \leq (x_D)_{\rm exp} = \left (0.419 \pm 0.211 \right ) \% \,.
\eeq

In principle further constraints on the off-diagonal elements of the
axigluon couplings $\Gamma_{L,R}^{u,d}$ also follow from $\Delta F =1$
transitions (\ie, radiative and rare weak decays). From the above
discussion it should have however become clear that all relevant
constraints in the down-type quark sector, \ie, $B \to X_s
\hspace{0.5mm} (K^\ast) \hspace{0.5mm} \gamma$, $B \to X_s
\hspace{0.5mm} (K^\ast) \hspace{0.5mm} \ell^+ \ell^-$, $B_s \to \mu^+
\mu^-$, and $K^+ \to \pi^+ \nu \bar \nu$, are readily satisfied by
aligning the down-type quark sector. In the case of alignment in the
down-type quark sector the axigluon contributions to the
short-distance amplitudes of the radiative $D$-meson decays $D \to
\bar K^\ast \hspace{0.5mm} (\phi) \hspace{0.5mm} \gamma$ are of the
order of $\alpha_s/(4\pi) \hspace{0.75mm} m_c/M_G^2 \hspace{0.5mm}
(V_{cb}^\ast V_{ub})/\sin^2 (2\theta)$. Since radiative charm decays are
fully dominated by non-perturbative physics, the strongly suppressed
axigluon corrections cannot be separated from the much larger SM
long-distance contributions. Similar statements apply to other rare
$D$-meson decays like $D \to \pi \hspace{0.5mm} (\rho) \hspace{0.5mm}
\ell^+ \ell^-$ and $D \to \mu^+ \mu^-$. The poor experimental bounds
on rare top-quark decays do not lead to any sensible restriction
neither.  The bottom line is that at present there are no additional
constraints on $M_G$ and $\theta$ from $\Delta F =1$ processes in the
up-quark sector beyond those already imposed by the $\Delta F = 2$
transitions.

\section{Precision Measurements}
\label{sec:ewpo}

The axigluon interactions with the SM quarks give rise to corrections
to measurements at $e^+ e^-$ machines as well as hadron colliders. In
this and the next section we provide generic calculations of these
effects. As already mentioned in \Sec{sec:model}, the presence of a
fourth generation of sequential fermions would lead to large one-loop
corrections unless the parameters of the model are tuned.  For most of
the following discussion, we hence consider only the effects of
massive color octets and discard the model-dependent contributions
from massive fermions beyond those in the SM. We will furthermore
neglect possible flavor-violating effects. Assuming an appropriate
flavor alignment, such effects are strongly Cabibbo-suppressed.

\begin{figure}[!t]
\begin{center}
\vspace{-1cm}
\includegraphics[height=4cm]{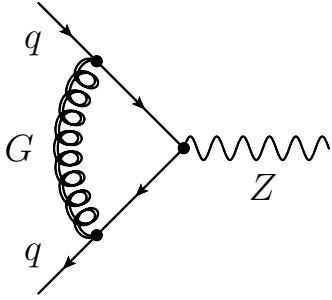}
\begin{picture}(0,0)(0,0)
\put(-127.5,52.5){\large $G$}
\put(-120,95){\large $q$}
\put(-120,15){\large $q$}
\put(-35,37.5){\large $Z$}
\end{picture}
\end{center}
\begin{center} 
  \parbox{15.5cm}{\caption{\label{fig:zqq} One-loop axigluon
      contribution to the $Zq\bar q$ vertex.}}
\end{center}
\end{figure}

We begin our survey by studying the axigluon correction to the
$Z$-boson vertices.  The corresponding Feynman diagram involving the
virtual exchange of an axigluon between the quark lines is shown in
\Fig{fig:zqq}. For light quarks ($q = u,d,s,c,b$) the main corrections
to the $Zq_P\bar q_P$ couplings ($P = L,R$) occur at second order in
the expansion with respect to the external momenta. In terms of
effective operators, this means that these corrections are encoded in
the Wilson coefficients of the following operators
\beq \label{eq:dipoleop}
{\cal O}_{Q_PW} = (\bar Q_{P} \hspace{0.25mm} \gamma^\mu \tau^i
\hspace{0.25mm} Q_{P}) \hspace{0.25mm} D^\nu W_{\mu \nu}^i \,, \qquad
{\cal O}_{\hspace{0.25mm} Q_PB} = (\bar Q_{P} \hspace{0.25mm}
\gamma^\mu \hspace{0.25mm} Q_{P}) \hspace{0.5mm} \partial^\nu B_{\mu
  \nu} \,.
\eeq
Here $\tau^i$ are $SU(2)_P$ generators, $D_\mu = 1/2 \hspace{0.75mm}
\big ( \overset{\rightarrow}{D}_\mu - \overset{\leftarrow}{D}_\mu \big
)$ with $\overset{\rightarrow}{D}_\mu = \partial_\mu + ig
\hspace{0.25mm} \tau^i \hspace{0.5mm} W_\mu^i + i \hspace{0.25mm}
g^\prime \hspace{0.5mm} Y/2 \hspace{0.25mm} B_\mu$ is the covariant
derivative with $g$ ($g^\prime$) denoting the $SU(2)_L$ ($U(1)_Y$)
gauge coupling associated with the field $W_\mu^i$ ($B_\mu$), and
$W_{\mu\nu}^i$ ($B_{\mu \nu}$) is the corresponding field-strength
tensor.

Using the equations of motion in the broken phase of the theory, \ie,
including the mass terms of the $W$ and $Z$ bosons, the off-shell
operators in \eq{eq:dipoleop} induce corrections to the current
couplings of the light quarks to the electroweak gauge bosons.  In the
case of the left- and right-handed bottom-quark couplings to the $Z$
boson, we find that the tree-level $Zb_P\bar b_P$ couplings $g_P^b$
are modified, yielding 
\beq \label{eq:calGbapprox} 
{\cal G}_{P}^{b} \approx g_{P}^{b} \left [1 - \frac{2}{3} \,
  \frac{\alpha_s}{4\pi} \, C_F \left ( g_P^h \right )^2
  \frac{M_Z^2}{M_G^2} \, \ln \left ( \frac{M_Z^2}{M_G^2} \right )
\right ] \,,
\eeq
where $C_F = 4/3$ is the Casimir invariant in the fundamental
representation. The expressions for the quarks of the first two
generations are obtained from the above formula by simply replacing
$g_P^h$ with $g_P^\ell$. The result in \eq{eq:calGbapprox} implies
that the axigluon correction to the current coupling of the $Z$ boson
interferes constructively with the SM for $M_G > M_Z$, resulting in
$|{\cal G}_P^q| > |g_P^q|$ for all light quarks.  Yet, the corrections
are flavor non-universal. The new-physics corrections to the $Z b_L
\bar b_L$ coupling are enhanced by a factor of $\big (g_L^h/g_R^h \big
)^2 = \cot^4 \theta$ relative to those affecting the $Z b_R \bar b_R$
coupling, while for the quarks of the first and second family the
pattern of deviations is reversed, given that $\big (g_L^\ell/g_R^\ell
\big )^2 = \tan^4 \theta$. The corrections to the $Z$-boson vertex
involving bottom quarks are hence dominantly left-handed, while those
involving the remaining light quarks are mostly right-handed. We
emphasize also that the latter formula, originally derived in
\cite{Hill:1994di}, contains only the leading-logarithmic corrections
in the limit of an infinitely heavy axigluon, $M_Z^2/M_G^2 \to 0$. In
consequence, this result can be found by integrating out the axigluon
and calculating the anomalous dimension of the ``penguin diagram''
obtained from the graph in \Fig{fig:zqq} by pinching the axigluon
propagator. The full one-loop result for the complex effective
couplings ${\cal G}_P^q$ is given in \App{app:Zqq}. From the limiting
behavior \eq{eq:ReKImKlim}, we see that non-logarithmic corrections to
the formula \eq{eq:calGbapprox} are not fully negligible for $M_G =
{\cal O} (1 \, {\rm TeV})$. We will therefore use the full real parts
of ${\cal G}_P^q$ in our numerical analysis.

In the case of the top quark, on the other hand, the axigluon
corrections to the $Z$-boson vertex are contained in the Wilson
coefficients of the effective operators
\beq \label{eq:Zop}
{\cal O}_{Q_P\phi}^{\,1} = i \hspace{0.5mm} (\bar Q_{P} \hspace{0.25mm}
\gamma^\mu \hspace{0.25mm} \tau^i \hspace{0.25mm} Q_{P})
\hspace{0.5mm} \phi^\dagger \hspace{0.25mm} \tau^i \hspace{0.25mm}
D_\mu \phi \,, \qquad {\cal O}_{Q_P\phi}^{\,2} = i \hspace{0.5mm} (\bar
Q_{P} \hspace{0.25mm} \gamma^\mu Q_{P}) \hspace{0.5mm} \phi^\dagger
D_\mu \phi \,.
\eeq
Evaluating the Feynman diagram in \Fig{fig:zqq} with an internal top
quark at zero external momenta, we find that after EWSB the matching
corrections to the operators in \eq{eq:Zop} shift the tree-level $Zt_P
\bar t_P$ couplings $g_P^t$ by
\beq \label{eq:calGtapprox} 
{\cal G}_{P}^{t} \approx g_{P}^{t} \pm \frac{g}{c_w} \,
\frac{\alpha_s}{4\pi} \, C_F \left ( g_P^h \right )^2
\frac{m_t^2}{M_G^2} \, \ln \left ( \frac{m_t^2}{M_G^2} \right ) \,,
\eeq 
where the plus (minus) sign applies in the case $P = L$ ($P=R$).  Our
result agrees with \cite{Hill:1994di} and again contains only the
leading-logarithmic corrections in the limit $m_t^2/M_G^2 \to 0$. The
complete expression for the $Z \to t \bar t$ form factor with
vanishing external momenta can be found in \App{app:Zqq}. Like in the
case of the on-shell $Z \to q \bar q$ form factor, non-logarithmic
corrections to ${\cal G}_P^t$ are non-negligible and should be
included if one aims for precision. Glancing at \eq{eq:calGtapprox},
we see that the one-loop axigluon corrections decrease (increase) the
value of the left-handed (right-handed) coupling of the $Z$ boson to
$t \bar t$ pairs, \ie, ${\cal G}_{L}^{t} < g_{L}^t$ (${\cal G}_{R}^{t}
> g_{R}^t$). Since $\big (g_L^h/g_R^h \big )^2 = \cot^4 \theta$, the
corrections to the left-handed coupling are always more pronounced than
those to the right-handed coupling.

Constraints on the $Z$-boson couplings to quarks of the first
generation are imposed by the measurement of the forward-backward
asymmetry of $e^+ e^-$ in $q \bar q \to Z/\gamma \to e^+ e^-$ and a
combined fit to the $Z$-boson lineshape, the lepton forward-backward
asymmetries, and asymmetry parameters. In the former case one arrives
at ${\cal G}_L^u = 0.355 \pm 0.025$, ${\cal G}_R^u = -0.147 \pm
0.017$, ${\cal G}_L^d = -0.436 \pm 0.008$, and ${\cal G}_R^d = 0.058
\pm 0.031$ \cite{Abazov:2011ws}, while in the latter case one has
${\cal G}_L^u = 0.356 \pm 0.035$ and ${\cal G}_R^u =
-0.11^{+0.30}_{-0.07}\hspace{0.25mm}$, ${\cal G}_L^d = -0.423 \pm
0.012$, and ${\cal G}_R^d = 0.10 ^{+0.04}_{-0.06}$
\cite{LEPEWWG:2005ema}. The limited precision of these extractions
makes it impossible to derive any sensible constraint on the
$M_G\hspace{0.5mm}$--$\hspace{0.5mm}\theta$ plane for axigluon masses
at the order of a TeV. The situation is improved for what concerns the
second family. In the case of the charm quark, precision measurements
of the relevant couplings ${\cal G}_L^c = 0.3453 \pm 0.0036$ and
${\cal G}_R^c = -0.1580 \pm 0.0051$ \cite{LEPEWWG:2005ema} follow from
a combination of the measurements of $R_c$, $A_c$, and $A_{\rm
  FB}^c$. Yet, the obtained constraints on $M_G$ and $\theta$ turn out
to be less strict than those that derive from the same set of
observables in the bottom-quark sector. These will be discussed in a
moment. Obvious limitations also precluded measurements of the
$Z$-boson couplings to top quarks so far. There was insufficient CM
energy at LEP to produce top-quark pairs in $e^+ e^- \to \gamma/Z \to
t \bar t$. At hadron colliders, $t \bar t$ production is fully
dominated by the QCD processes $gg \to t \bar t$ and $q \bar q \to g
\to t \bar t$, so that it is impossible to extract a signal from the
sought couplings entering via $q \bar q \to \gamma/Z \to t \bar
t$. However, it is feasible to determine the $Z$-boson couplings to
top quarks via the process of $gg \to Z t \bar t$ at the LHC with
integrated luminosity of at least $300 \, {\rm fb}^{-1}$. The expected
relative precision amounts to around 10\% (45\% to 85\%) for the
axial-vector (vector) coupling \cite{Baur:2004uw, Baur:2005wi,
  Berger:2009hi}. These uncertainties are too large to allow to derive
stringent bounds on the parameter space of the axigluon model.

Important constraints on the
$M_G\hspace{0.5mm}$--$\hspace{0.5mm}\theta$ plane follow from the
measurement of the bottom-quark POs performed at LEP and SLC. In the
following, we will consider the impact of the one-loop axigluon
corrections on the ratio of the $Z$-boson decay width into bottom
quarks and the total hadronic width, $R_b$, the bottom-quark
left-right asymmetry, $A_b$, and the forward-backward asymmetry for
bottom quarks, $A_{\rm FB}^b$. The dependences of these quantities on
the left- and right-handed bottom-quark couplings are
given by \cite{Field:1997gz}
\begin{equation}\label{eq:bPOtheory}
\begin{split}
    R_b &= \left [ 1 + \frac{4 \; {\displaystyle \sum}_{q=u,d} \left[
          ({\cal G}_L^q)^2 + ({\cal G}_R^q)^2\right]}%
      {\eta_{\rm QCD}\,\eta_{\rm QED} \left[ (1-6z_b) ({\cal
            G}_L^b-{\cal G}_R^b)^2 + ({\cal G}_L^b+{\cal G}_R^b)^2
        \right]} \right]^{-1}\! , \\
    A_b &= \frac{2\sqrt{1-4z_b}\,\, {\displaystyle \frac{{\cal
            G}_L^b+{\cal G}_R^b}{{\cal G}_L^b-{\cal G}_R^b}}}%
    {1-4z_b+(1+2z_b) {\displaystyle \left( \frac{{\cal G}_L^b+{\cal
              G}_R^b}{{\cal G}_L^b-{\cal G}_R^b} \right)^2}} \,,
    \qquad A_{\rm FB}^b = \frac34\,A_e\hspace{0.25mm}A_b \,.
\end{split}
\end{equation}
Radiative QCD and QED corrections are encoded in $\eta_{\rm
  QCD}=0.9954$ and $\eta_{\rm QED}=0.9997$, while the parameter $z_b=
m_b^2(M_Z)/M_Z^2=0.997\cdot 10^{-3}$ describes the effects of the
non-zero bottom-quark mass.

Since to the considered order axigluon corrections do not affect the
asymmetry parameter of the electron, $A_e$, we will fix this quantity
to its SM value $(A_e)_{\rm SM} =0.1464$. For the SM couplings, we use
$({\cal G}_L^u)_{\rm SM}=0.34665$, $({\cal G}_R^u)_{\rm SM}
=-0.15477$, $({\cal G}_L^d)_{\rm SM}=-0.42429$, $({\cal G}_R^d)_{\rm
  SM}=0.077379$, $\big ( {\cal G}_L^b \big )_{\rm SM}=-0.42106$, and
$\big ( {\cal G}_R^b \big)_{\rm SM}=0.077451$
\cite{Arbuzov:2005ma}.\footnote{The default flags of {\tt ZFITTER}
  version 6.43 are used, except for setting ${\tt ALEM}=2$ to take
  into account the externally supplied value of $\Delta \alpha_{\rm
    had}^{(5)} =0.02758 \pm 0.00035$. The other relevant input
  parameters read $M_Z = (91.1875 \pm 0.0021) \GeV$, $m_t = (173.3 \pm
  1.1) \GeV$ \cite{Group:1900yx}, $m_h=150 \GeV$, and $\alpha_s (M_Z)
  = 0.118 \pm 0.001$. } Evaluating the relations (\ref{eq:bPOtheory})
using this input, we obtain for the bottom-quark POs
\beq\label{eq:bPOSM}
\begin{split}
 & \big (R_b \big )_{\rm SM} = 0.21578 \pm 0.00004 \,, \\ 
 & \big ( A_b \big )_{\rm SM} = 0.9347 \pm 0.0001 \,, \\ 
 & \big ( A_{\rm FB}^b \big )_{\rm SM} = 0.1026 \pm 0.0007 \,. 
\end{split}
\eeq
One should compare these numbers with the experimental results
\cite{LEPEWWG:2005ema}
\begin{equation}\label{eq:bPOsexp}
  \begin{array}{l}
    \big ( R_b \big)_{\rm exp} = 0.21629\pm 0.00066 \,, \\[0.25mm] 
    \big ( A_b \big)_{\rm exp} = 0.923\pm 0.020 \,, \\[1mm]
    \big ( A_{\rm FB}^b \big )_{\rm exp} = 0.0992\pm 0.0016 \,, 
  \end{array}
  \qquad 
  \rho = \begin{pmatrix}
    1.00 \, & \, -0.08 & \, -0.10 \\ 
    \, -0.08 & \, 1.00 & \, 0.06 \\ 
    -0.10 \, & \, 0.06 & \, 1.00 
  \end{pmatrix} ,
\end{equation}
where $\rho$ is the correlation matrix. While the $R_b$ and $A_b$
measurements agree with their SM predictions within $+0.8\sigma$ and
$-0.6\sigma$ for $m_h = 150 \, {\rm GeV}\hspace{0.25mm}$, the $A_{\rm
  FB}^b$ measurement is almost $-2.0\sigma$ away from its SM
expectation.\footnote{For $m_h=115$\,GeV the discrepancy in $A_{\rm
    FB}^b$ would amount to around $-2.4\sigma$.} Shifts of order
$+20\%$ and $-0.5\%$ in the right- and left-handed bottom-quark
couplings relative to the SM could explain the observed
discrepancy. Such a pronounced correction in ${\cal G}_R^b$ would
affect $A_b$ and $A_{\rm FB}^b$, which both depend linearly on the
ratio ${\cal G}_R^b/{\cal G}_L^b$ in a significant way, while it
would not spoil the good agreement in $R_b \propto ({\cal G}_L^b)^2 +
({\cal G}_R^b)^2$.

\begin{figure}[!t]
\begin{center}
\vspace{-1cm}
\includegraphics[height=7.5cm]{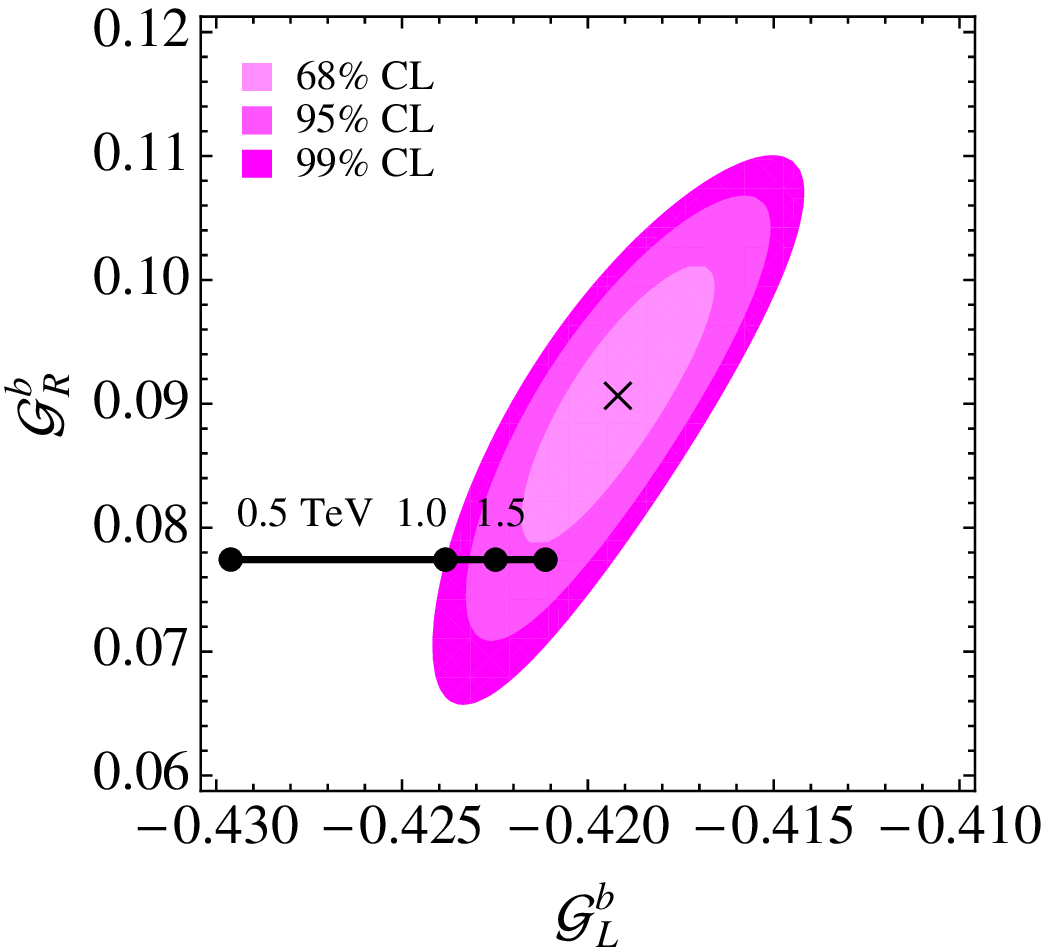}
\parbox{15.5cm}{\caption{\label{fig:gLbgRb} Regions of 68\%, 95\%, and
    99\% probability in the ${\cal G}_L^b$--${\cal G}_R^b$ plane. The
    black dot without label is the SM expectation for the reference
    point and the black cross represents the best-fit solution. The
    black line and the remaining dots indicate the predictions in the
    flavor non-universal axigluon model. They have been obtained for
    $\theta = 15^\circ$ and $M_G = 0.5 \, {\rm TeV}, 1.0 \, {\rm TeV},
    1.5 \, {\rm TeV}$.}}
\end{center}
\end{figure}

The possible size of the one-loop new-physics corrections to the
effective couplings ${\cal G}^b_{L,R}$ is shown in \Fig{fig:gLbgRb},
which displays the regions of $68\%$, $95\%$, and $99\%$~CL obtained
from a global fit to the bottom-quark POs in (\ref{eq:bPOsexp}). The
predictions in the flavor non-universal axigluon model are
superimposed in black. The shown points correspond to $\theta =
15^\circ$ and $M_G = 0.5 \, {\rm TeV}, 1.0 \, {\rm TeV}, 1.5 \, {\rm
  TeV}$. We see that the axigluon contributions drive ${\cal G}_L^b$
to smaller values with respect to the SM reference point (black dot),
while ${\cal G}_R^b$ remains essentially unaffected.\footnote{The
  corrections to ${\cal G}_R^b$ are, as anticipated, positive but so
  small that they are not visible in the figure.} For smaller (larger)
values of the mixing angle $\theta$, the corrections to ${\cal G}_L^b$
are more (less) pronounced due to the presence of the factor $\big (
g_L^h \big )^2 = \cot^2 \theta$ in the effective coupling
\eq{eq:calGbapprox}. Changing $\theta$ has only a minor effect on
${\cal G}_R^b$. This implies that in the considered axigluon model the
quality of the global fit to the bottom-quark POs does not improve
with respect to the SM. In particular, the best-fit values ${\cal
  G}_L^b = -0.41910$ and ${\cal G}_R^b = 0.091044$ (black cross)
cannot be obtained in the flavor non-universal axigluon model. The
constraints arising from the $Z \to b \bar b$ couplings on the
$M_G\hspace{0.5mm}$--$\hspace{0.5mm} \theta$ plane will be analyzed in
\Sec{sec:global}.

\begin{figure}[!t]
\begin{center}
\vspace{-1cm}
\includegraphics[height=7.5cm]{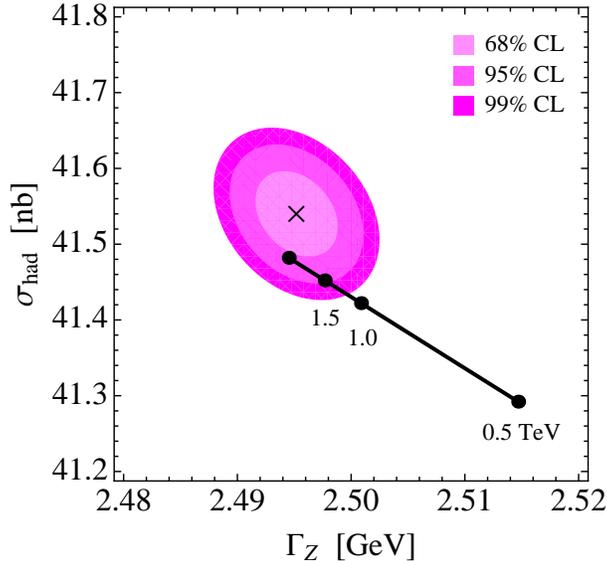}
\parbox{15.5cm}{\caption{\label{fig:gamZsighad} Regions of 68\%, 95\%,
    and 99\% probability in the $\Gamma_Z$--$\sigma_{\rm had}$
    plane. The black dot without label is the SM expectation for the
    reference point and the black cross represents the best-fit
    solution. The black line and the remaining dots indicate the
    predictions in the flavor non-universal axigluon model. They have
    been obtained for $\theta = 15^\circ$ and $M_G = 0.5 \, {\rm TeV},
    1.0 \, {\rm TeV}, 1.5 \, {\rm TeV}$.}}
\end{center}
\end{figure}

The roles of the total width $\Gamma_Z$ of the $Z$ boson and the
hadronic pole cross-section $\sigma_{\rm had}$ turn out to be even
more important in constraining the parameter space of the considered
axigluon model than the bottom-quark POs alone. This is due to the
fact that the former observables are more sensitive to the altered
light-quark couplings.  The relevant quantities can be written as
\beq \label{eq:GammaZsigmahad}
\Gamma_Z = \Gamma_{\rm lep} + \Gamma_{\rm inv} + \Gamma_{\rm had} \,,
\qquad \sigma_{\rm had} = \frac{12\pi}{M_Z^2} \frac{\Gamma_{e}
  \hspace{0.25mm} \Gamma_{\rm had}}{\Gamma_Z^2} \,,
\eeq
where $\Gamma_{\rm lep}$, $\Gamma_{\rm inv}$, and $\Gamma_{e}$ denote
the partial decay width to leptons, the invisible width from $Z$
decays to neutrinos, and the partial width for electrons,
respectively. In our numerical analysis, we will set these partial
widths to their SM values, $(\Gamma_{\rm lep})_{\rm SM} = 251.733 \,
{\rm MeV}$, $(\Gamma_{\rm inv})_{\rm SM} = 501.579 \, {\rm MeV}$, and
$(\Gamma_{e})_{\rm SM} = 83.975 \, {\rm MeV}$
\cite{Arbuzov:2005ma}. This is an excellent approximation, since
axigluon corrections affect these observables first at the three-loop
level. The hadronic width $\Gamma_{\rm had} = \sum_{q \neq t} \,
\Gamma_{q}$ is given by the sum over all quark final states with mass
smaller than $M_Z$. The partial $Z$-decay widths to quarks $\Gamma_q$
themselves are defined inclusively, \ie, they contain factorizable QED
and QCD final-state corrections encoded in the radiator factors
$R^q_{V,A}$ as well as non-factorizable radiative corrections
parametrized by $\Delta_{{\rm EW}/{\rm QCD}}^q$. Explicitly, one has
\beq \label{eq:Gammaqq}
\Gamma_{q} = N_c \hspace{0.5mm} \frac{G_F \hspace{0.25mm} M_Z^3}{6
  \sqrt{2} \pi} \, \Big ( R_V^q \, |{\cal G}_L^q + {\cal G}_R^q |^2 +
R_A^q \, |{\cal G}_L^q - {\cal G}_R^q |^2 \Big ) + \Delta_{{\rm
    EW}/{\rm QCD}}^q \,,
\eeq 
where $N_c =3$ is the number of colors and $G_F = 1.16637 \cdot
10^{-5} \GeV^{-2}$ is the Fermi constant.  The radiator factors are
$R_V^u = 1.03972$, $R_A^u = 1.04678$, $R_V^{d,s} = 1.03911$,
$R_A^{d,s} = 1.03205$, $R_V^c = 1.03974$, $R_A^c = 1.04651$, $R_V^b =
1.03970$, and $R_A^b = 1.02544$, while the relevant non-factorizable
corrections amount to $\Delta_{{\rm EW}/{\rm QCD}}^{u,c} =-0.113 \,
{\rm MeV}$, $\Delta_{{\rm EW}/{\rm QCD}}^{d,s} =-0.160 \, {\rm MeV}$,
and $\Delta_{{\rm EW}/{\rm QCD}}^{b} =-0.040 \, {\rm MeV}$
\cite{Arbuzov:2005ma}.  Evaluating the relations
\eq{eq:GammaZsigmahad} and \eq{eq:Gammaqq} using the input detailed
above leads to the following SM predictions
\beq \label{eq:GZSHSM}
\big ( \Gamma_Z \big)_{\rm SM} = (2.4945 \pm 0.0007 )\GeV \,, \qquad \big (
\sigma_{\rm had} \big)_{\rm SM} = (41.482 \pm 0.006)\, {\rm nb} \,.
\eeq 
The corresponding experimental extractions and their correlation read
\cite{LEPEWWG:2005ema}
\beq \label{eq:ZtotsHad}
  \begin{array}{l}
    \big ( \Gamma_Z \big)_{\rm exp} = (2.4952 \pm 0.0023) \GeV \,, \\[1.5mm] 
    \big ( \sigma_{\rm had} \big)_{\rm exp} = (41.540 \pm 0.037) \, {\rm nb} \,, 
  \end{array}
  \qquad 
  \rho = \begin{pmatrix}
    1.00 \, & \, -0.30 \\ 
    \, -0.30 & \, 1.00 
  \end{pmatrix} .  
\eeq 
Notice that, while the results for the total width $\Gamma_Z$ shows
agreement with the SM expectation \eq{eq:GZSHSM} within errors, the
experimental value of $\sigma_{\rm had}$ is above the SM value by
about $+1.6 \sigma$. After $A_{\rm FB}^b$ this is the largest
deviation in the global electroweak fit.  The principal dependence of
$\sigma_{\rm had}$ is on the number of light neutrino generations as
encoded in $\Gamma_{\rm inv}$, which is constant and equal to three in
the SM and in the considered axigluon model.\footnote{We assume that
  the mass of the fourth-generation neutrino is larger than the 95\%
  CL LEP II bound, which amounts to $m_{\nu_4} > 101.5 \, {\rm GeV}$
  \cite{Achard:2001qw}.}

The regions of $68\%$, $95\%$, and $99\%$~CL in the
$\Gamma_Z$--$\sigma_{\rm had}$ plane resulting from \eq{eq:ZtotsHad}
are shown in \Fig{fig:gamZsighad}. The predictions in the flavor
non-universal axigluon model are indicated in black. The displayed
points correspond to $\theta = 15^\circ$ and $M_G = 0.5 \, {\rm TeV},
1.0 \, {\rm TeV}, 1.5 \, {\rm TeV}$. We see that the axigluon
contributions lead to a correlated shift in $\Gamma_Z$ and
$\sigma_{\rm had}$ to higher and lower values, respectively. For
smaller (larger) values of the mixing angle $\theta$, the corrections
to both observables are more (less) pronounced, but the form of the
correlation remains unchanged.  This tells us that in the flavor
non-universal axigluon model the quality of the global fit to
$\Gamma_Z$ and $\sigma_{\rm had}$ does not improve relative to the
SM. The constraints arising from $\Gamma_Z$ and $\sigma_{\rm had}$
will be combined with those stemming from the bottom-quark POs in
\Sec{sec:global}. 

\begin{figure}[!t]
\begin{center}
\vspace{-1cm}
\includegraphics[width=15cm]{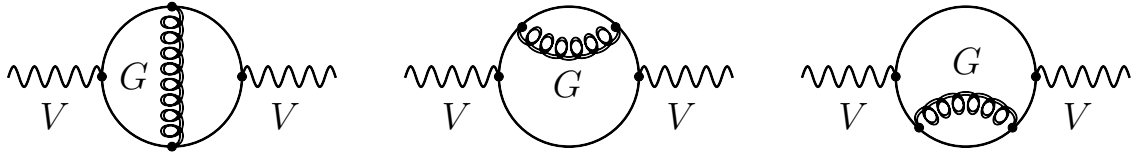}
\begin{picture}(0,0)(0,0)
\put(-223.5,22){\large $G$}
\put(-72.5,31){\large $G$}
\put(-387.5,25){\large $G$}
\put(-417.5,10){\large $V$}
\put(-330.5,10){\large $V$}
\put(-265.5,10){\large $V$}
\put(-180.5,10){\large $V$}
\put(-115.5,10){\large $V$}
\put(-30.5,10){\large $V$}
\end{picture}
\end{center}
\begin{center} 
  \parbox{15.5cm}{\caption{\label{fig:oblique} Two-loop axigluon
      contributions to the oblique parameters. In the case of the $WW$
      selfenergy ($VV = WW$) one has top and bottom quarks in the
      loop, while in the case of the $ZZ$ and $Z \gamma$ selfenergies
      ($VV = ZZ, Z\gamma$) only top quarks appear as intermediate
      states.}}
\end{center}
\end{figure}

We now turn our attention to the axigluon corrections to the oblique
(or Peskin-Takeuchi) parameters $S$, $T$, and $U$
\cite{Peskin:1991sw}. They measure deviations from the electroweak
radiative corrections in the SM induced by universal new-physics
effects, \ie, those entering through vacuum polarization diagrams. The
three parameters are defined as shifts relative to a fixed set of SM
values, so that $S$, $T$, and $U$ are identical to zero at that
point. The relevant set of oblique corrections is
\begin{equation} \label{eq:STUdef}
\begin{split}
    S & = \frac{16 \hspace{0.5mm} \pi s^2_w c^2_w}{e^2} \left[ \,
      \Pi_{ZZ}^{\hspace{0.25mm} \prime}(0)+ \frac{s^2_w-c^2_w}{s_w
        c_w} \, \Pi_{Z\gamma}^{\hspace{0.25mm}
        \prime}(0)-\Pi_{\gamma\gamma}^{\hspace{0.25mm}
        \prime}(0) \, \right] ,\\
    T & = \frac{4 \hspace{0.5mm} \pi}{e^2 c^2_w M_Z^2} \, \Big
    [\Pi_{WW}(0)-c^2_w \, \Pi_{ZZ}(0) -2 \, s_w c_w \, \Pi_{Z\gamma}(0)-
    s^2_w \, \Pi_{\gamma\gamma}(0) \Big]\,,\\
    U & = \frac{16 \hspace{0.5mm} \pi s^2_w}{e^2} \, \Big
    [\Pi_{WW}^{\hspace{0.25mm} \prime}(0)-c^2_w \,
    \Pi_{ZZ}^{\hspace{0.25mm} \prime}(0) -2 \, s_w c_w \,
    \Pi_{Z\gamma}^{\hspace{0.25mm} \prime}(0) -s^2_w \,
    \Pi_{\gamma\gamma}^{\hspace{0.25mm} \prime}(0) \Big ] \,,
\end{split}
\end{equation} 
where $\Pi_{VV} (p^2) = \Pi_{VV} (0) + p^2 \,
\Pi_{VV}^{\hspace{0.25mm} \prime} (0) + {\cal O} (p^4)$ with $VV = WW,
ZZ, \gamma\gamma, Z\gamma$ denotes the transversal part of the vacuum
polarization tensor of the corresponding selfenergy and $s_w$ and
$c_w$ are the sine and cosine of the weak mixing angle,
respectively. Notice that gauge invariance guarantees that
$\Pi_{\gamma \gamma}(0)=0$ to all orders in perturbation theory. 

The two-loop Feynman diagrams contributing to $\Pi_{VV} (p^2)$ in the
axigluon model are shown in \Fig{fig:oblique}. Notice that the mass
splitting between the top and the bottom quark provides the source of
isospin breaking necessary to generate a non-zero value for $T$ even
so the axigluon physics is isospin preserving. Similarly,
non-vanishing contributions to $S$ and $U$ are generated at the
two-loop level. In the leading-logarithmic approximation, we find the
following expressions
\begin{gather} 
  S \approx S_4 + \frac{2}{9 \pi \hspace{0.25mm}} \, \frac{
    \alpha_s}{4 \pi} \, C_F N_c \, \big [ (g_L^h)^2 + 2
  \hspace{0.25mm} (g_R^h)^2 \big ] \, \frac{m_t^2}{M_G^2} \, \ln^2
  \left ( \frac{m_t^2}{M_G^2}
  \right ) \,, \nonumber \\[-3mm] \label{eq:STapprox} \\[-3mm] 
  T \approx T_4 + \frac{m_t^2}{8 \pi \hspace{0.25mm} s_w^2 c_w^2
    \hspace{0.25mm} M_Z^2} \, \frac{ \alpha_s}{4 \pi} \, C_F N_c \,
  \big [ (g_L^h)^2 + 2 \hspace{0.25mm} (g_R^h)^2 \big ] \,
  \frac{m_t^2}{M_G^2} \, \ln^2 \left ( \frac{m_t^2}{M_G^2} \right )
  \,, \nonumber
\end{gather}
where $N_c = 3$ is the number of colors and the mass of the bottom
quark has been neglected.  Notice that the axigluon contribution to
$T$ is larger by a factor of $9/(16 \hspace{0.5mm} s_w^2 c_w^2)
\hspace{0.5mm} m_t^2/M_Z^2 \approx 11.7$ than the one to $S$. Since
the parameter $U$ is suppressed with respect to $T$ by an additional
factor of $M_Z^2/M_G^2$, we neither quote its analytic result nor use
this parameter when constraining the
$M_G\hspace{0.5mm}$--$\hspace{0.5mm}\theta$ plane. We remark that the
leading-logarithmic two-loop corrections in \eq{eq:STapprox} can be
found by contracting the axigluon propagator to a four-quark
interaction and considering only the first diagram in
\Fig{fig:oblique}. What concerns $T$, our leading-logarithmic result
agrees with the one presented in \cite{Burdman:1999us}, while the
expression for $S$ is, to the best of our knowledge, new. The full
two-loop results for the set of relevant oblique corrections are given
in \App{app:oblique}. Although the two-loop axigluon contributions to
$S$ and $T$ involve double logarithms, it turns out that numerically
the leading-logarithmic approximation is not an excellent
approximation for the relevant axigluon masses, so that in our
numerical analysis we will employ the complete two-loop expressions.

\begin{figure}[!t]
\begin{center}
\vspace{-1cm}
\includegraphics[height=7.5cm]{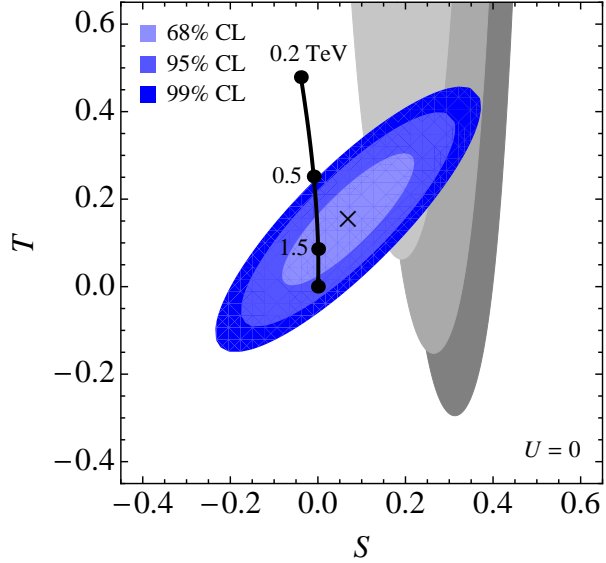}
\parbox{15.5cm}{\caption{\label{fig:ST} Regions of 68\%, 95\%, and
    99\% probability in the $S$--$T$ plane. The black dot without
    label is the SM point and the black cross represents the best-fit
    solution. The black line and the remaining dots indicate the
    contribution of a flavor non-universal axigluon. They have been
    obtained for $\theta = 15^\circ$ and $M_G = 0.2 \TeV, 0.5 \TeV,
    1.5 \TeV$. The gray areas illustrate the predicted regions for the
    fourth-generation contributions to $S$ and $T$ for three different
    values of the Higgs-boson mass. See text for further details.}}
\end{center}
\end{figure}

The terms $S_4$ and $T_4$ appearing in \eq{eq:STapprox} denote the
one-loop contributions of the fourth generation of sequential fermions
to the $S$ and $T$ parameters. Neglecting inter-generational mixing
with the additional generation, these corrections take the well-known
form \cite{Peskin:1991sw, He:2001tp}
\begin{gather} 
  S_4 = \frac{1}{6 \pi} \, \Big [ \, N_c \, {\cal S} (m_{u_4}, m_{d_4}) +
    {\cal S} (m_{\nu_4}, m_{\ell_4}) \, \Big ] \,,
  \nonumber \\
  T_4 = \frac{1}{16 \pi \hspace{0.25mm} s_w^2 c_w^2 \hspace{0.25mm}
    M_Z^2} \, \Big [ \, N_c \, {\cal T} (m_{u_4},m_{d_4}) + {\cal T}
  (m_{\nu_4},m_{\ell_4}) \, \Big ] \,,  \label{eq:S4T4}
\end{gather}
with 
\beq \label{eq:fT}
{\cal S} (m_1,m_2) = 1 - 2 \hspace{0.25mm} Y \hspace{0.25mm} \ln \left
  ( \frac{m_1^2}{m_2^2} \right ) \,, \quad {\cal T} (m_1, m_2) = m_1^2
+ m_2^2 - 2 \, \frac{m_1^2 \hspace{0.25mm} m_2^2}{m_1^2-m_2^2} \ln
\left ( \frac{m_1^2}{m_2^2} \right ) \,.
\eeq
Here $Y =1/6$ ($Y = -1/2$) denotes the hypercharge of the left-handed
doublet of quarks (leptons).  Notice that choosing the masses of the
fourth-generation quarks and leptons to be degenerate, which
corresponds to the isospin limit, one is left with the finite
correction $S_4 = 2/(3\pi)$ to the $S$ parameter while the
contribution $T_4$ vanishes identically.  This value of $S$ can be
reduced by splitting the multiplets such that $m_{u_4}/m_{d_4} > 1$
and $m_{\nu_4}/m_{\ell_4} < 1$, which in turn leads to a positive
shift in $T$ that is quadratic in the mass splittings. Given that
${\cal T} (m_1, m_2) = {\cal T} (m_2, m_1)$ the possible values of
$S_4$ and $T_4$ thus form an upright parabola-like shaped region in
the $S$--$T$ plane with a vertex at $\big (2/(3 \pi), 0 \big)$.

The experimental 68\% CL bounds on the $S$ and $T$ parameters,
corrected to the present world average of the top-quark mass
\cite{Group:1900yx}, and their correlation matrix are given by
\cite{LEPEWWG:2005ema}
\beq\label{eq:STexp}
   \begin{array}{l}
    S = 0.07 \pm 0.10 \,, \\[0.5mm] 
    T = 0.15 \pm 0.10 \,,
   \end{array} \qquad 
   \rho = \begin{pmatrix} 1.00~ & 0.85 \\ 
                          0.85~ & 1.00 
          \end{pmatrix} .
\eeq
The regions of 68\%, 95\%, and 99\% probability in the $S$--$T$ plane
are shown in Figure~\ref{fig:ST}. Notice that in the global fit to the
LEP and SLC measurements the parameter $U$ is set to zero. The
$S$--$T$ error ellipses show that there are no large unexpected
electroweak radiative corrections from physics beyond the SM, as the
values of the oblique parameters are in agreement with zero. The
contributions of a flavor non-universal axigluon to $S$ and $T$ are
indicated by the black line and points. The shown points have been
obtained for the value of the mixing angle fixed to $\theta =
15^\circ$ and the three different axigluon masses $M_G = 0.2 \TeV$,
$0.5 \TeV$, $1.5 \TeV$.  From the panel as well as the second formula
in \eq{eq:STapprox}, we see that the axigluon corrections to $T$ are
strictly positive. In contrast, the axigluon contribution to $S$
changes from positive to negative sign when the axigluon mass is
decreased. This feature arises from single logarithms and constant
terms not shown in \eq{eq:STapprox} and takes place at $M_G = 1.25 \,
\TeV$ for the specific value $\theta=15^\circ$ of the mixing angle.
For comparison the grayish areas in the latter figure show the
possible predictions for $S_4$ and $T_4$, assuming $m_h = 100 \GeV$
(light gray), $400 \GeV$ (gray), and $1000 \GeV$ (dark gray). The
shifts in $S$ and $T$ due to a Higgs-boson mass different from our
reference value $m_h^{\rm ref} = 150 \GeV$ are given to
leading-logarithmic accuracy by $\Delta S = 1/(12 \pi) \, \ln
\hspace{0.25mm} (m_h^2/(m_h^{\rm ref})^2)$ and $\Delta T = -3/(16 \pi
\hspace{0.25mm} c_w^2) \, \ln \hspace{0.25mm} (m_h^2/(m_h^{\rm
  ref})^2)$ \cite{Peskin:1991sw}.  For each value of the Higgs-boson
mass the masses of the fourth-generation fermions have been freely
varied in the ranges $m_{u_4} \in [311, 600] \GeV$, $m_{d_4} \in [372,
600] \GeV$, $m_{\nu_4} \in [101.5, 600] \GeV$, and $m_{\ell_4} \in
[101.9, 600] \GeV$. From the plot it is immediately clear that it is
not possible to obtain a model-independent bound on $M_G$ and $\theta$
from the oblique parameters in the case of a flavor non-universal
axigluon. In fact, one has to distinguish three different
cases. First, the value of $T_4$ is below the $S$--$T$ ellipse. In
this case one can derive a two-side limit on $M_G$ as a function of
$\theta$. Second, the prediction for $T_4$ lies inside the ellipse,
which allows one to obtain a lower bound on the axigluon mass for any
value of the mixing angle. Third, the prediction for $T_4$ is above
the $S$--$T$ ellipse. In this case a flavor non-universal axigluon is
at variance with the constraints imposed by the oblique
corrections. For illustrative purposes, we will in \Sec{sec:global}
determine the allowed region in the
$M_G\hspace{0.5mm}$--$\hspace{0.5mm}\theta$ plane for the flavor
non-universal axigluon model assuming the specific values $S_4 = 0.15$
and $T_4 = 0.19$. For a Higgs boson with a mass $m_h \in [100, 200]
\GeV$ such values are obtained for the mass splittings $m_{\ell_4} -
m_{\nu_4} \approx 70 \GeV$ and $m_{u_4} - m_{d_4} \approx \big (1 +
1/3 \, \ln \, (m_h^2/(m_h^{\rm ref})^2) \big ) \, 45 \GeV$ with $m_h^{\rm
  ref} = 150 \GeV$. A glimpse at \Fig{fig:ST} tells us that the
corresponding point in the $S-T$ plane lies within the 68\% CL of the
global fit to the oblique parameters.

\section{Collider Observables}
\label{sec:collider}

In this section we will discuss the direct bounds on massive
color-octet bosons following from the high-$p_T$ experiments performed
at the Tevatron and the LHC. In particular, we will consider
top-antitop quark as well as dijet production. In the former case we
will combine the available experimental information on the total
inclusive $t \bar t$ cross section \cite{CDFnotetot, D0notetot}, the
invariant mass spectrum \cite{Aaltonen:2009iz, Bridgeman:2008zz}, and
the differential forward-backward asymmetry \cite{Aaltonen:2011kc}. In
the latter case we explore the constraining power of the searches for
dijet resonances \cite{Khachatryan:2010jd, Collaboration:2011aj,
  Aaltonen:2008dn, Collaboration:2010bc} and the measurements of the
dijet angular distributions \cite{Khachatryan:2011as,
  Collaboration:2011aj, Abazov:2009mh, Collaboration:2010eza}. Like
before, we will neglect possible effects from flavor-changing axigluon
couplings when computing the relevant collider observables. See
\cite{Han:2010rf, Bai:2011ed, Hewett:2011wz} for other comprehensive
studies of color-octet resonances at hadron colliders similar to ours
in spirit.

At the Tevatron $t \bar t$ pairs are produced in collisions of protons
and antiprotons at an CM energy of $\sqrt{s} = 1.96 \TeV$. Within the
model at hand the hadronic process receives Born-level contributions
from quark-antiquark annihilation $q \bar q \to t \bar t$, associated
with tree-level exchange of an axigluon in the $s$ channel. On the
other hand, the gluon-fusion channel $gg \to t \bar t$ does not
contribute at Born level. This result is an important consequence of
the gauge invariance of the axigluon model, which forces the $gg
\hspace{0.25mm} G$ vertex to vanish at tree level. The absence of the
$gg \hspace{0.25mm} G$ tree-level interactions is a feature of many
models with extra massive bosons in the adjoint representation of
$SU(3)_c\hspace{0.25mm}$.\footnote{In the case of warped extra
  dimensions the absence of the coupling of two gluons to the massive
  spin-one octet, \ie, the KK gluon, is an artifact of the
  orthonormality of gauge-boson wave functions \cite{Lillie:2007yh,
    Randall:2001gb, Bauer:2010iq}. In addition, all gluon-axigluon
  vertices with an odd number of axigluons can be forbidden by strong
  parity \cite{Bagger:1987fz}.} Beyond tree level or through
non-renormalizable operators of dimension six and higher, such a
coupling can however be induced. Since the resulting effects are,
compared to the contributions from $q \bar q \to t \bar t$, suppressed
by two powers of the new-physics scale and possibly a loop factor, we
will not consider contributions to top-quark pair production from
gluon fusion in the following.

Constraints on new color-octet resonances in $t \bar t$ production
arise from the Tevatron measurements of the total cross section
$\sigma_s\hspace{0.5mm}$, the invariant mass spectrum $d
\sigma_s/dM_{t \bar t} \hspace{0.25mm}$, the forward-backward
asymmetry $A_{\rm FB}^t$, and its distribution $A_{\rm FB}^t(M_{t\bar
  t})$. Since the last bin of the available CDF measurement of the
differential cross section, \ie, $M_{t \bar t} \in [0.8, 1.4] \TeV$,
is most sensitive to the presence of new degrees of freedom with
masses of order TeV, we will restrict our attention to this range of
invariant masses when calculating $d \sigma_s/dM_{t \bar t}$.
Recently the CDF collaboration has measured the forward-backward
asymmetry in two bins of the $t\bar t$ invariant mass, separated at
$M_{t\bar t} = 0.45 \TeV$.  While in the low bin the observed
asymmetry agrees with the SM expectation as well as with zero within
errors, the result for the upper bin lies more than $3 \sigma$ above
the NLO QCD prediction. Similar to the symmetric spectrum, the region
of high $M_{t\bar t}$ for the asymmetry is expected to be most
sensitive to heavy new physics. In our global analysis of the $t \bar
t$ observables, we include both bins of the differential asymmetry,
treating the two measurements as fully uncorrelated. This should be a
good approximation, since at the moment the measurement of $A_{\rm
  FB}^t(M_{t\bar t})$ is statistically limited.

The quantities of interest can all be obtained from the double
differential $t \bar t$ cross section,
\beq \label{eq:d2sigmadmdc}
\frac{d^2\sigma}{d M_{t\bar t} \hspace{0.5mm} d \cos \hat \theta} = \left (
  \frac{d^2\sigma}{d M_{t\bar t} \hspace{0.5mm} d \cos \hat \theta} \right
)_{\rm SM} + \left ( \frac{d^2\sigma}{d M_{t\bar t} \hspace{0.5mm} d
    \cos \hat \theta} \right )_{G} \,,
\eeq 
by integrating over the invariant mass $M_{t\bar t}$ and the cosine of
the scattering angle $\hat \theta$ of the top quark in the partonic CM
frame. Here the labels SM and $G$ denote the SM and axigluon
contributions. In particular, the symmetric and asymmetric mass spectra
are calculated via
\beq \label{eq:dssdsa}
\begin{split}
  \hspace{10mm} &\frac{d\sigma_s}{dM_{t\bar t}} = \int_{-1}^1 \! d
  \cos
  \hat \theta \, \frac{d^2\sigma}{dM_{t\bar t} \hspace{0.5mm} d\cos \hat \theta} \,, \\
  & \hspace{-2.25cm} \frac{d\sigma_a}{dM_{t\bar t}} = \int_{0}^1 \! d
  \cos \hat \theta \, \frac{d^2\sigma}{dM_{t\bar t} \hspace{0.5mm}
    d\cos \hat \theta} - \int_{-1}^0 \! d \cos \hat \theta \,
  \frac{d^2\sigma}{dM_{t\bar t} \hspace{0.5mm} d\cos \hat \theta}\,,
\end{split}
\eeq 
while the forward-backward asymmetry takes the form $A^t_{\rm FB} =
\sigma_a/\sigma_s$. An analog definition holds, of course, in the case
of the $M_{t \bar t}$ distribution $A_{\rm FB}^t (M_{t \bar t})$ of
the asymmetry.

The SM contributions to \eq{eq:d2sigmadmdc} and \eq{eq:dssdsa} are
computed at NLO using {\tt MCFM} \cite{MCFM}, which for what concerns
$t \bar t$ production is based on the seminal work
\cite{Nason:1987xz}.  For the charge-symmetric observables we employ
{\tt MSTW2008NLO} parton distribution functions (PDFs)
\cite{Martin:2009iq} with the strong coupling constant
$\alpha_s=0.120$ as an input, corresponding to $\alpha_s(m_t)=0.109$
at two-loop accuracy. Since the charge-asymmetric terms are generated
first at the one-loop level, they are computed using {\tt MSTW2008LO}
PDFs and normalized to the LO symmetric cross section.  The
expectations for the $t\bar t$ observables in the SM along with the
corresponding measurements are collected in
Table~\ref{tab:ttobservables}.  The shown theoretical errors are due
to renormalization and factorization scale variations $\mu_r = \mu_f
\in [m_t/2,2m_t]$ as well as PDF uncertainties. We also mention that
the symmetric SM cross section at LO, integrated over the low- and
high-mass bin amounts to $(\sigma_s)^{M_{t\bar t} < 0.45\TeV} = 4.04
\, {\rm pb}$ and $(\sigma_s)^{M_{t\bar t} > 0.45 \TeV} = 2.58 \, {\rm
  pb}$. The axigluon effects in the $t\bar t$ observables are computed
by convoluting the well-known matrix element squared for
quark-antiquark annihilation induced by axigluon exchange
\cite{Frampton:2009rk, Cao:2010zb} with {\tt MSTW2008LO} PDFs and
fixing $\mu_r = \mu_f = m_t = 173.3 \GeV$. The relevant value of the
strong coupling constant is $\alpha_s(M_Z) = 0.139$, which translates
into $\alpha_s(m_t) = 0.126$ using LO RG running.

\begin{table}[!t]
\begin{center}
\begin{tabular}{|c|c|cc|}
  \hline
  Observable & SM & \hspace{10mm} Measurement & \\
  \hline
  $\sigma_s$ & $(6.73^{+0.52}_{-0.80})\,\rm pb$ & $(7.50\pm 0.48)\,\rm pb$ & 
  \cite{CDFnotetot}\\[1mm]
  $(d\sigma_s/dM_{t\bar t})^{M_{t \bar t} \in [0.8, 1.4] \TeV}$ & 
  $(0.061^{+0.012}_{-0.006})$ fb/GeV & 
  $(0.068\pm 0.034)$ fb/GeV & \cite{Aaltonen:2009iz}\\[1mm]
  $\big (A_{\rm FB}^{t} (M_{t\bar t}) \big )^{M_{t\bar t}<0.45\TeV}$ & 
  $(7.0^{+1.0}_{-0.8})\%$ 
  &  $(-11.6\pm 15.3)\%$ & 
  \cite{Aaltonen:2011kc}\\[1mm]
  $\big (A_{\rm FB}^{t} (M_{t\bar t}) \big )^{M_{t\bar t}>0.45\TeV}$ & 
  $(11.0^{+1.2}_{-1.3})\%$ & $(47.5\pm 11.2)\%$ & \cite{Aaltonen:2011kc}\\
\hline
\end{tabular}
\end{center}
\begin{center}
  \parbox{15.5cm}{\caption{\label{tab:ttobservables} SM expectations
      and measurements of the $t \bar t$ observables entering our
      analysis. The numbers given for the forward-backward asymmetries
      correspond to the partonic CM frame.}}
\end{center}
\end{table}

From \Tab{tab:ttobservables} we infer that the measurements of the
charge-symmetric observables agree fairly well with their SM
predictions, whereas the SM expectation for the asymmetry in the high
$M_{t \bar t}$ bin is significantly lower than its measured value. In
order to fit the data, an axigluon ought to yield positive
contributions to $\big(A_{\rm FB}^{t} (M_{t\bar t})\big)^{>} =
\big(A_{\rm FB}^{t} (M_{t\bar t})\big)^{M_{t\bar t} > 0.45 \TeV}$,
while leaving $\sigma_s$, $d\sigma_s/dM_{t\bar t}$, and $\big(A_{\rm
  FB}^{t} (M_{t\bar t})\big)^{<} = \big(A_{\rm FB}^{t} (M_{t\bar
  t})\big)^{M_{t\bar t} < 0.45 \TeV}$ essentially unaffected. As we
will see in \Sec{sec:global}, this leads to a generic tension in the
global fit, since the latter observables prefer the new degrees of
freedom to be heavy, whereas the asymmetry in the region $M_{t \bar t}
> 0.45 \TeV$ would like to have a low new-physics scale.

Any massive color-octet boson that couples to light $q \bar q$ pairs
is also subject to constraints arising from narrow resonance searches
in dijet production. Employing an integrated luminosity of only $3.1
\, {\rm pb}^{-1}$ \cite{Collaboration:2010bc} and $2.9 \, {\rm
  pb}^{-1}$ \cite{Khachatryan:2010jd}, ATLAS and CMS have already
surpassed the previous most stringent mass limits by CDF
\cite{Aaltonen:2008dn}.  Very recently, ATLAS has presented improved
95\% CL upper limits on the product of the resonant production cross
section ($\sigma$), branching fraction ($\cal B$), and acceptance
($A$) for an axigluon coupling decaying democratically to all quark
flavors \cite{Collaboration:2011aj}.  We will use the latter results,
which are based on a data set of $36 \, {\rm pb}^{-1}$, to derive the
relevant constraints on the allowed parameter space in the
non-universal axigluon model.

The tree-level cross section for resonant axigluon production receives
only corrections from $q \bar q \to G$, and can thus be written as 
\beq \label{eq:sigmaqqG}
\sigma = \sum_{q} \,\ff_{q \bar q} (M_G^2/s, \mu_f) \,\,
\frac{C_F}{N_c} \hspace{0.25mm} \frac{2\pi^2\alpha_s}{s} \,
\big((g_L^q)^2+(g_R^q)^2\big)\,,
\eeq
where the sum extends over the light quark flavors $q = u,d,s,c,b$ and
$\sqrt{s} = 7 \TeV$. The parton luminosity functions,
\beq \label{eq:luminosities}
\ff_{ij} (\tau,\mu_f) = \frac{2}{1+\delta_{ij}}
\int_{\tau}^1\frac{dx}{x}\,f_{i/p}(x,\mu_f)\,f_{j/p}(\tau/x,\mu_f)\,.
\eeq
are evaluated at the parton CM energy corresponding to the resonant
production of the axigluon, \ie, $\tau = M_G^2/s$. They are obtained
from a convolution of the universal non-perturbative PDFs $f_{i/p}(x,
\mu_f)$, which describe the probability of finding the parton $i$ in
the proton with longitudinal momentum fraction $x$. In our analysis we
employ {\tt MSTW2008LO} PDFs with the renormalization and
factorization scales set to $\mu_r = \mu_f = M_G$.

The branching ratio for the decay of a heavy spin-one color octet into
a pair of light quarks reads ${\cal B}(G\rightarrow q\bar q) =
\Gamma_q/\Gamma_G$. The total width in the case of the non-universal
axigluon is given by the sum $\Gamma_G = \sum_q \Gamma_q + \sum_Q
\Gamma_Q$ of light- and heavy-quark contributions, where $Q=t,u_4,d_4$
($Q = t$) in the case of an flavor non-universal (universal)
axigluon. Neglecting the masses of the light quarks, the partial
tree-level decay rates are 
\beq \label{eq:GammaG}
\begin{split}
  & \hspace{3.75cm} \Gamma_q = \frac{\alpha_s \hspace{0.25mm} T_F}{6}
  \, M_G \, \Big [ (g_L^\ell)^2 + (g_R^\ell)^2 \Big ] ,  \\
  & \Gamma_Q = \frac{\alpha_s \hspace{0.25mm} T_F}{6} \, M_G \,
  \sqrt{1 - \frac{4m_Q^2}{M_G^2}} \, \left [ \left ( (g_L^h)^2 +
      (g_R^h)^2 \hspace{0.5mm} \right ) \left ( 1 -
      \frac{m_Q^2}{M_G^2} \right ) + 6 \hspace{0.5mm} g_L^h
    \hspace{0.5mm} g_R^h \, \frac{m_Q^2}{M_G^2} \hspace{0.5mm} \right
  ] ,
\end{split}
\eeq
where $T_F = 1/2$ and $m_Q$ denotes the mass of the heavy quarks in
the final state. Since dijet resonance searches are based on the
narrow-width approximation, they do not apply if the total decay width
of the resonance significantly exceeds the mass resolution of the
detector. The article \cite{Collaboration:2011aj} lacks the explicit
information for which value of $\Gamma_G/M_G$ the ATLAS axigluon
analysis becomes inapplicable.\footnote{The ATLAS 95\% CL exclusion
  $M_G \in [0.6,2.1] \TeV$ applies to a flavor-universal axigluon that
  has QCD-like couplings to quarks. Such a color-octet boson has a
  total width of around $8\%$ to $9\%$ of its mass.} Yet, the
presented model-independent limits on Gaussian resonances, which cover
the ranges $M_G \in [0.6, 4.0] \TeV$ and $\Gamma_G/M_G \in [3, 15]\%$,
suggest that it should at least apply to cases where the ratio of the
resonance width and its mass is below $15\%$. In order to illustrate
the importance of knowing the applicability of the narrow-width
approximation, we will consider three different cases $\Gamma_G/M_G <
10\%, 15\%$, and $20 \%$.  As we will see in \Sec{sec:global}, these
three benchmark scenarios restrict the parameter space to the region
where $\theta \gtrsim 40^{\circ}, 30^{\circ}$, and $25^{\circ}$,
respectively.  To calculate the total decay width in the non-universal
axigluon model, we fix the fourth-generation quark masses to $m_{u_4}
= 311 \GeV$ \cite{Lister:2008is} and $m_{d_4} = 372 \GeV$
\cite{Aaltonen:2011vr}. The model dependence introduced by this choice
is small.

\begin{figure}[!t]
\begin{center}
\vspace{-1cm}
\includegraphics[height=7.5cm]{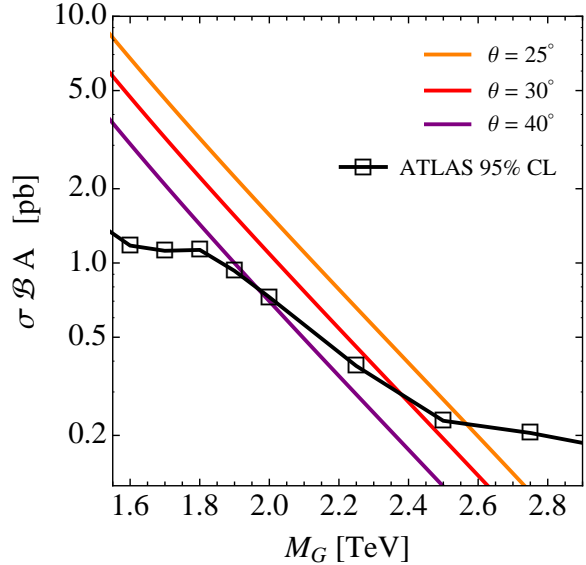}
\parbox{15.5cm}{\caption{\label{fig:ATLAS} Predictions for resonant
    axigluon production as a function of $M_G$ for three different
    values of the mixing angle. The orange, red, and purple line
    corresponds to $\theta = 25^\circ, 30^\circ$, and $40^\circ$,
    respectively. The black line represents the ATLAS $95\%$ CL upper
    limit on $\sigma \hspace{0.5mm} \mathcal{B} \hspace{0.5mm} A$ for
    resonances decaying to $q \bar q$.}}
\end{center}
\end{figure}

In order to utilize the 95\% CL upper limit on $\sigma \hspace{0.5mm}
{\cal B} \hspace{0.5mm} A$ from ATLAS, we have to correct for
perturbative QCD radiation, non-perturbative effects (\ie,
hadronization and multi-parton interactions), and the detector
acceptance. We achieve this by rescaling our result for the partonic
production cross section $q\bar q \to G \to q \bar q$ by
\beq \label{eq:Rscaling} 
\begin{split}
R & = \frac{\left (\sigma \hspace{0.5mm} {\cal B} \hspace{0.5mm}
    A\right)_{\rm axigluon}^{\rm ATLAS}}{\big (\sigma \hspace{0.5mm}
  {\cal B} (G \to q \bar q) \big ) \big |^{Q\neq u_4,
    d_4}_{g_{L,R}^{\ell,h} = \pm 1}} \\&  = 0.54 - \frac{0.09}{2.4
  \TeV} \; (M_G - 0.6 \TeV) + \left [ \frac{0.99}{2.4 \TeV} \; (M_G -
  0.6 \TeV) \right ]^2 \,,
\end{split}
\eeq 
where the numerator is given by the axigluon prediction of ATLAS,
while the denominator is calculated at the partonic level using
\eq{eq:sigmaqqG} and \eq{eq:GammaG}, including only three generations
of quarks, and employing $g_{L}^{\ell,h} = 1$ and $g_{R}^{\ell,h} =
-1$. Notice that \eq{eq:Rscaling} implicitly assumes that the
acceptance depends only on $M_G$ but not on $\Gamma_G$. We have
explicitly checked that using the limit on a simplified Gaussian
signal model presented in \cite{Collaboration:2011aj}, which
incorporates the dependence on $\Gamma_G$, essentially leads to the
same exclusions as the rescaling procedure described above.

In \Fig{fig:ATLAS} we compare the $95\%$ CL upper bound on $\sigma
\hspace{0.5mm} {\cal B} \hspace{0.5mm} A$ obtained by ATLAS (black
line) to our theory predictions for resonant axigluon production
employing the three values $\theta = 25^\circ, 30^\circ$, and
$40^\circ$ (orange, red, and purple lines) for the mixing angle. Cross
sections above the black curve are disfavored by the data. From the
intersection of the theory predictions with the ATLAS limit, we derive
the following 95\% CL bounds $M_G > 2.5 \TeV, 2.4 \TeV$, and $2.0
\TeV$. In \Sec{sec:global} we will compare the constraints on the
$M_G\hspace{0.5mm}$--$\hspace{0.5mm}\theta$ plane following from the
recent narrow resonance search at ATLAS with the other available
direct constraints.

A second quantity of interest for what concerns dijet production is
the jet angular distribution. The differential cross section for a
pair of jets with invariant mass $M_{jj}$ produced at an angle $\hat
\theta$ to the beam direction in the jet-jet CM frame, can be written
in the following way 
\beq \label{eq:dijetdouble} 
\frac{d^2 \sigma}{d M_{jj} \hspace{0.25mm} d \cos \hat \theta} =
\frac{M_{jj}}{s} \sum_{i,j} \ff_{ij} (M_{jj}^2/s,\mu_f) \,
\frac{d\sigma_{ij}}{d \cos \hat \theta} \;,
\eeq
with 
\beq \label{eq:dsigmaij}
\frac{d\sigma_{ij}}{d \cos \hat \theta} = \frac{1}{32 \pi M_{jj}^2} \,
\sum_{k,l} \, \overline{\sum} \, \big | {\cal M}
(ij \to kl) \big |^2 \, \frac{1}{1+\delta_{kl}}  \,.
\eeq
Here ${\cal M} (ij \to kl)$ denotes the matrix element for the
scattering of the incoming partons $i,j=q,\bar{q},g$ into the outgoing
partons $k,l$. The color and spin indices in \eq{eq:dsigmaij} are
averaged (summed) over initial (final) states as indicated by the
symbol $\overline \sum$. The expressions for the tree-level matrix
elements squared appearing in QCD can be found in \cite{Ellis:1991qj}.

Compared to the narrow resonance searches discussed before, the dijet
angular distribution \eq{eq:dijetdouble} has the salient advantage
that it also constrains broad $s$-channel resonances. This is due to
the fact that the dominant channels in QCD dijet production have the
familiar Rutherford scattering behavior $d \sigma_{ij}/d \cos \hat
\theta \propto 1/\sin^4 \hspace{0.25mm} (\hat \theta/2)$ at small
angle $\hat \theta$, which is characteristic for $t$-channel exchange
of a massless spin-one boson. In order to remove the Rutherford
singularity, one usually considers the dijet cross sections
differential in
\beq \label{eq:chi}
\chi = \frac{1+|\cos\hat \theta|}{1-|\cos \hat \theta|} \,.
\eeq 
In the small angle limit, \ie, $\chi \to \infty$, the partonic
differential QCD cross section then behaves as $d \sigma_{ij}/d \chi
\propto {\rm const}$. Relative to the QCD background, the production
of a heavy resonance leads to additional hard scattering and hence
more jets perpendicular to the beam. In turn one expects a deviation
from the QCD prediction in form of an enhanced activity of
high-energetic jets in the central region of the detector.  If the
angular distributions receive contributions from the presence of a
heavy degree of freedom, one should see an excess of events in
$d\sigma_{ij}/d\chi$ for $\chi \to 1$ and large $M_{jj}$ with respect
to the (almost) flat QCD spectrum.

The shape of the normalized dijet angular distribution $1/\sigma
\hspace{0.5mm} d\sigma/d\chi$ has been studied both at the Tevatron
\cite{Abazov:2009mh} and LHC \cite{Khachatryan:2011as,
  Collaboration:2011aj, Collaboration:2010eza} and found to be in good
agreement with the SM prediction. These results put stringent
constraints on any non-standard scattering mechanism leading to jet
pairs, including scenarios where the new particles are too heavy to be
produced directly. In models where the new-physics scale $\Lambda$ is
much larger than the CM energy of the colliding partons, the exchange
of new particles is most commonly described in terms of effective
four-quark contact interactions (CIs). These consist out of products
of left-handed color-singlet quark currents \cite{Eichten:1983hw,
  Lane:1996gr},
\beq \label{eq:HLL}
{\cal H}^{\rm CI}_{4q} = \frac{2\pi \hspace{0.25mm} \xi}{\Lambda^2} \,
\sum_{i} \, (\bar q_{i L} \hspace{0.25mm} \gamma^\mu \hspace{0.25mm}
q_{i L}) (\bar q_{i L} \hspace{0.25mm} \gamma_\mu \hspace{0.25mm} q_{i
  L}) \,.
\eeq
Here $i$ is a flavor index and $\xi$ determines whether the
interference between the new-physics and the SM contributions is
destructive ($\xi = +1$) or constructive ($\xi = -1$). The currently
most stringent 95\% CL bound on this type of four-quark interactions
amounts to {$\Lambda > 9.5 \TeV$} for $\xi = +1$ and is based on $36
\, {\rm pb}^{-1}$ of $\sqrt{s} = 7 \TeV$ data collected by ATLAS
\cite{Collaboration:2011aj}.

\begin{figure}[!t]
\begin{center}
\vspace{-1cm}
\includegraphics[height=7.5cm]{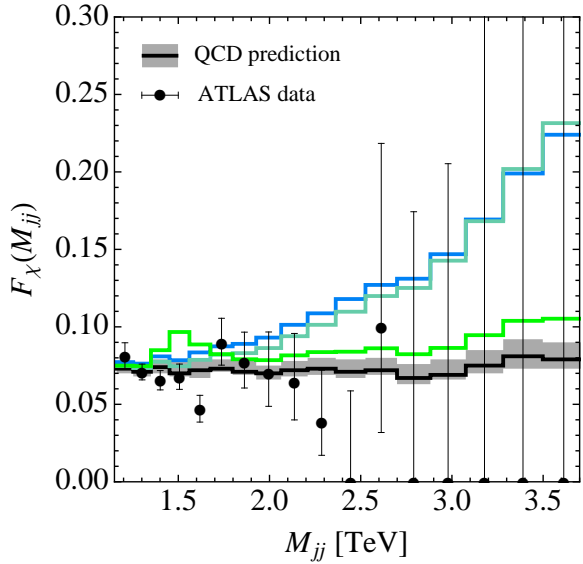}  
\parbox{15.5cm}{\caption{\label{fig:Fchi} Ratio $F_\chi (M_{jj})$ as a
    function of $M_{jj}$. The black line and gray band represent the
    central value of the QCD prediction with its total error. The
    ATLAS data points with their statistical uncertainties are
    overlaid as black error bars. The blue, aquamarine, and green
    lines correspond to QCD plus the effects of a flavor non-universal
    axigluon with $M_G = 1.5 \TeV$ and $\theta = 45^\circ$ for three
    different theoretical treatments. The first two curves are
    obtained in an effective theory, while the third one corresponds
    to the exact result. The latter prediction employs an axigluon
    propagator with a Breit-Wigner form and $\Gamma_G/M_G = 10\%$. See
    text for further details.}}
\end{center}
\end{figure}

The ATLAS collaboration determines the bound on the scale $\Lambda$
entering \eq{eq:HLL} by studying the fraction of centrally produced
dijets versus the total number of observed events for a specified
dijet mass range $M_{jj} \in [M_{jj}^{\text{min}},
M_{jj}^{\text{max}}]$. Specifically, ATLAS measures
\beq \label{eq:Fchi}
F_{\chi}(M_{jj}) = \frac{\sigma(\chi <
  3.32,[M_{jj}^{\text{min}},M_{jj}^{\text{max}}])}{\sigma(\chi <
  30,[M_{jj}^{\text{min}},M_{jj}^{\text{max}}])}\,,
\eeq
in 27 different bins of the dijet invariant mass.  Utilizing the same
amount of data the CMS collaboration obtains the weaker bound $\Lambda
> 5.6 \TeV$ \cite{Khachatryan:2011as}. The strong constraint by ATLAS
is due to the fact that relative to the expected QCD background there
are too few observed central events for $M_{jj}$ around $1.6 \TeV$ and
above $2.2 \TeV$. This feature is illustrated in \Fig{fig:Fchi}, which
shows the central value of the QCD prediction of $F_{\chi} (M_{jj})$
(black line) and its total error (gray band) in comparison to the
ATLAS data points with their statistical uncertainties (black error
bars).  To eliminate the possible bias from the downward statistical
fluctuations, we will employ the expected limit of $\Lambda > 5.7
\TeV$ from the ATLAS collaboration. This more conservative exclusion
is comparable to an alternative calculation by ATLAS using Bayesian
statistics as well as the aforementioned bound derived by CMS.

Translating the lower limit on $\Lambda$ into bounds on the parameter
space of the axigluon model is complicated by two features. First, the
chiral structure of the axigluon interactions is richer than the one
of CIs encoded in \eq{eq:HLL}. Second, in the parts of the phase space
with $M_G \lesssim M_{jj}$ a calculation of \eq{eq:Fchi} based on
dimension-six operators is not applicable, since the relevant momentum
transfer is comparable or larger than the mass of the new state
produced in the scattering.  We begin by commenting on the first
issue. Integrating out the heavy degrees of freedom from the
Lagrangian in \eq{eq:quarkaxiSU2} leads to the following effective
tree-level Hamiltonian
\beq \label{eq:HG4q} 
{\cal H}^{G}_{4q} = \frac{4 \pi \alpha_s}{M_G^2} \, \sum_{i,j}
\sum_{P,Q} \, g_P^\ell \hspace{0.5mm} g_Q^\ell \, (\bar q_{i P}
\hspace{0.25mm} \gamma^\mu T^a \hspace{0.25mm} q_{i P}) (\bar q_{j Q}
\hspace{0.25mm} \gamma_\mu T^a \hspace{0,25mm} q_{j Q}) \,,
\eeq 
where the sum over $P,Q =L,R$ includes both left- and right-handed
chiral fields. By calculating the partonic differential cross section
for the interference of the effective interactions \eq{eq:HLL} and
\eq{eq:HG4q} with the SM and with themselves, one obtains for the
dominant channel $uu\to uu$ in the former case
\beq \label{eq:suuLL} 
\left ( \frac{d\sigma_{uu}}{d\chi} \right )_{\rm CI} = \frac{4\pi
  \alpha_s}{9} \, \frac{\xi}{\Lambda^2} \, \frac{1}{\chi} \, + \,
{\cal O} (1/\Lambda^4) \,, \eeq while in the latter case one finds
\beq \label{eq:suuG} \left ( \frac{d\sigma_{uu}}{d\chi} \right )_G =
\frac{4\pi\alpha_s}{9} \, \frac{\alpha_s}{3M_G^2} \, \frac{1}{\chi}
\left [ \left ( (g_L^\ell)^2 + (g_R^\ell)^2 \right ) + 3
  \hspace{0.25mm} g_L^\ell g_R^\ell \,
  \frac{1-\chi+\chi^2}{(1+\chi)^2} \right ] \, + \, {\cal O} (1/M_G^4)
\,.
\eeq 
From \eq{eq:suuG} one observes that the axigluon contribution
proportional to the product $g_L^\ell \hspace{0.25mm} g_R^\ell$ is
sensitive to the relative sign between the couplings to left- and
right-handed quarks. It furthermore changes the shape of the angular
distribution with respect to \eq{eq:HLL}, which involves only
left-handed quarks. This feature is illustrated by the aquamarine
(blue) curve in \Fig{fig:Fchi}, which corresponds to the prediction for
$F_{\chi} (M_{jj})$ in the effective theory \eq{eq:HG4q} neglecting
(including) contributions proportional to $g_L^\ell \hspace{0.25mm}
g_R^\ell$. Both curves have been obtained for $M_G = 1.5 \TeV$ and
$\theta = 45^\circ$. We see that the exact effective theory result
predicts less central activity in all bins apart from the last two
than the one where the contributions proportional to $g_L^\ell
\hspace{0.25mm} g_R^\ell$ have been left out. This implies that the
actual limits on $M_G$ obtained from \eq{eq:HG4q} are weaker than the
ones derived by simply rescaling the bound on $\Lambda$ by a factor of
$\big ( \alpha_s \hspace{0.25mm} \big ((g_L^\ell)^2+ (g_R^\ell)^2 \big
)/3 \big)^{1/2}$.

In order to discuss the range of validity of the effective theory
formalism, we have also performed the exact calculation of $F_{\chi}
(M_{jj})$ for a flavor non-universal axigluon of mass $M_G$ and width
$\Gamma_G$. The exact axigluon result assuming $M_G = 1.5 \TeV$,
$\theta = 45^\circ$, and $\Gamma_G/M_G = 10\%$ is displayed by the
green curve in \Fig{fig:Fchi}. The corresponding matrix elements are
reported in \App{app:dijet}. The first noticeable feature of the exact
prediction is the relatively wide peak centered at $M_{jj} \approx 1.5
\TeV$ corresponding to the onset of resonant axigluon production. In
addition, one observes that in contrast to the effective theory
result, which scales as $M_{jj}^2/M_G^2$ for $M_{jj} \gg M_G$, the
exact $F_{\chi} (M_{jj})$ distribution has an almost flat tail for
large dijet masses. The different high-$M_{jj}$ behavior of the
predictions is readily understood by realizing that the propagator
associated with $s$-channel axigluon exchange is approximated by
$-1/M_G^2$ in the effective theory, while it behaves like $1/M_{jj}^2$
in the full theory. In view of the dissimilarity between the exact and
the effective theory result in large parts of the phase space, we will
use in \Sec{sec:global} the exact prediction for the dijet angular
distribution to determine the allowed parameter space in the
$M_G\hspace{0.5mm}$--$\hspace{0.5mm}\theta$ plane.

Before closing this section, let us finally spend some words on the
actual calculation of the dijet angular distributions. Our starting
point is the Born-level computation of the QCD expectation for
$F_{\chi} (M_{jj})$ using {\tt MSTW2008LO} PDFs with $\mu_r = \mu_f =
M_{jj}$.\footnote{In our calculation we respect the rapidity cuts
  imposed by ATLAS.  Effects from cuts on the dijet rapidity boost
  $y_B = (y_{j_1}+y_{j_2})/2$ largely cancel in the ratio
  $F_{\chi}(M_{jj})$, which is designed to be sensitive to the
  rapidity difference $|y_{j_1}-y_{j_2}| = \ln\chi$.  Neglecting the
  ATLAS cut on $y_B$ leaves the constraints on the axigluon parameters
  essentially unchanged.} This result is then multiplied by the
bin-wise $K$-factors computed by the ATLAS collaboration
\cite{Collaboration:2011aj} to obtain a reshaped spectrum that
includes corrections originating from NLO matrix elements. The
detector resolution as well as possible additional non-perturbative
QCD effects are incorporated by calculating the ratio between the
latter result and the central value of the ATLAS QCD
prediction. Numerically, we find that the bin-wise rescaling factor
determined in this way lies in the narrow range $[0.79, 0.91]$ for the
relevant kinematic region. Our prediction for $F_{\chi} (M_{jj})$ is
then obtained by adding the QCD and axigluon results and multiplying
the resulting expression by the rescaling factor. Following ATLAS, we
apply $K$-factors only to the QCD part of the dijet angular
distribution.\footnote{The NLO QCD corrections to dijet production
  induced by the CIs \eq{eq:HLL} have been calculated in
  \cite{Gao:2011ha} and shown to lower the exclusion limit on
  $\Lambda$ by around $10\%$. Our analysis neglects NLO corrections in
  the interference of the axigluon contributions with QCD and with
  themselves.} Notice that applying the same rescaling factor to both
the QCD and axigluon result is based on the assumption that the event
detection and the effects of the Monte Carlo shower depend only on the
invariant mass $M_{jj}$ of the dijet final state, but not on the
precise form of the new-physics signal. In order to determine to which
extend this assumption is justified would require to perform a
dedicated simulation of $F_\chi (M_{jj})$ including detector effects
and parton showering. However, such an analysis is beyond the scope of
this work.

\section{Numerical Analysis}
\label{sec:global}

In this section we present the constraints on the parameter space of
axigluon models and their look-a-likes that arise from flavor physics,
EWPOs, and collider observables. Throughout our analysis, we will
require that the couplings of heavy color-octet bosons to quarks
remain perturbative. In the flavor non-universal axigluon model, this
requirement is fulfilled if the mixing angle lies within $\theta \in
[15^\circ, 45^\circ]$. Notice that in this parameter region the
axigluon forms a distinct resonance, satisfying $\Gamma_G/M_G \ll 1$,
and that the bounds stemming from fermion condensation
\cite{Chivukula:2010fk} are also avoided.

\begin{figure}[!t]
\begin{center}
\includegraphics[height=7.5cm]{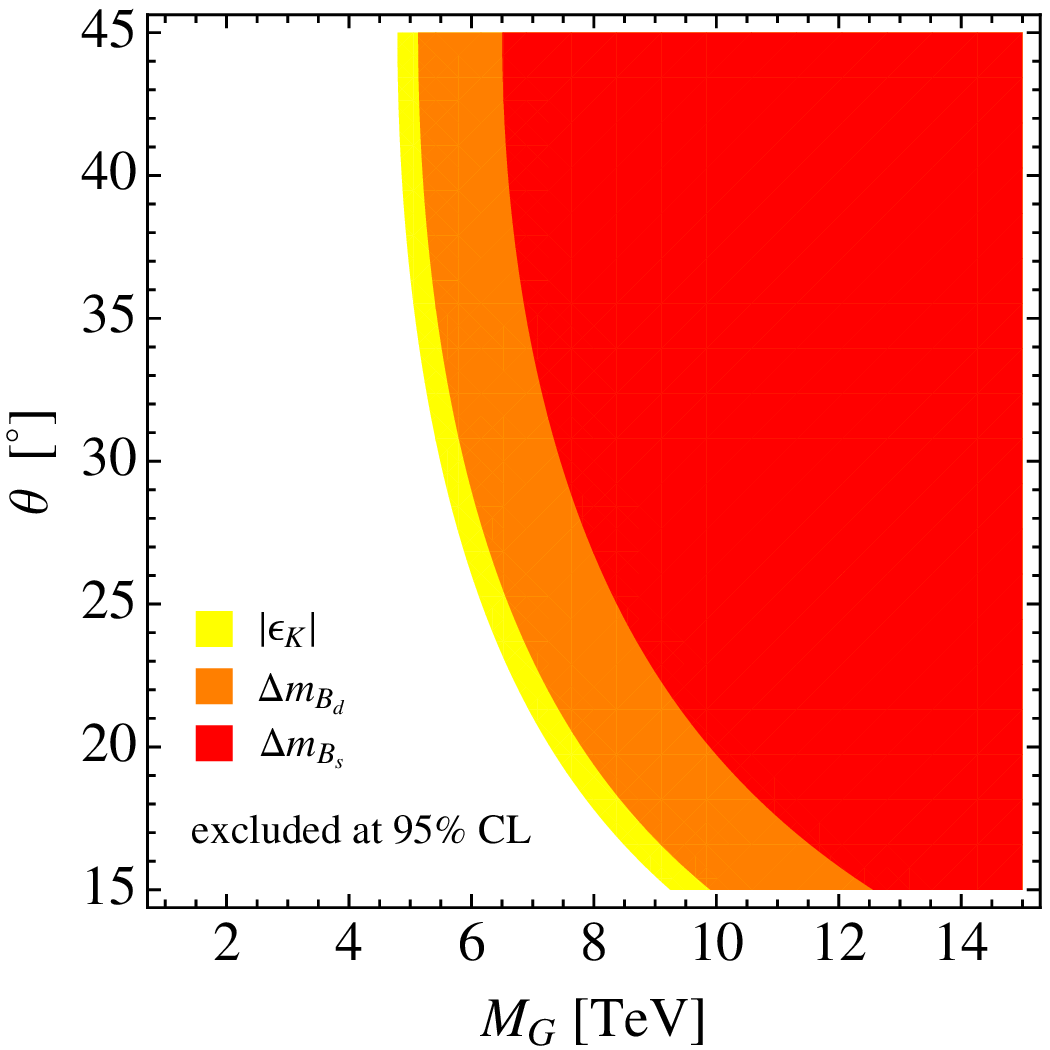}
\qquad 
\includegraphics[height=7.5cm]{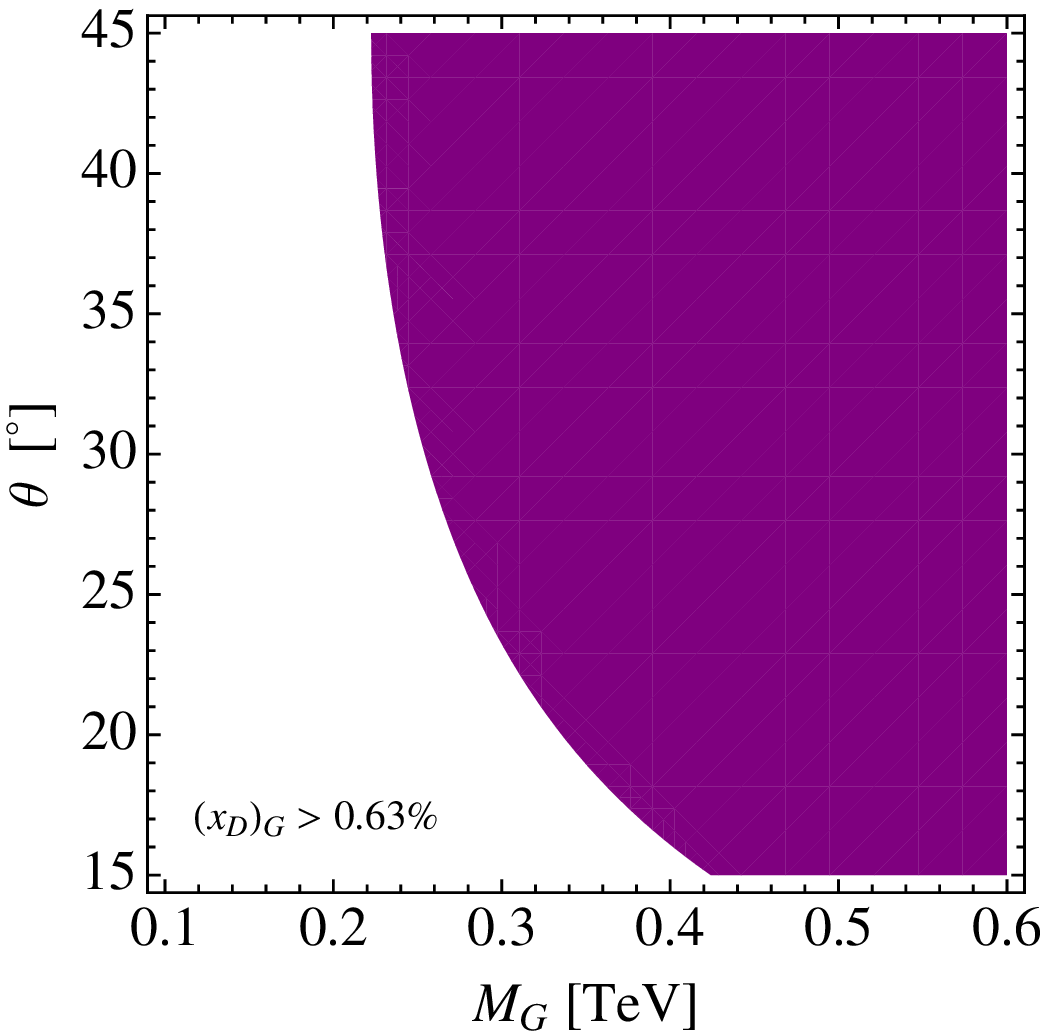}
\end{center}
\begin{center} 
  \parbox{15.5cm}{\caption{\label{fig:flavorbounds} Constraints in the
      $M_G\hspace{0.5mm}$--$\hspace{0.5mm}\theta$ plane imposed by the
      measured amount of flavor violation in the $\Delta F =2$
      sector. The left (right) panel corresponds to the scenario of
      up-type (down-type) flavor alignment in the non-universal
      axigluon model. The regions of parameter space colored white are
      disfavored by the existing experimental data.}}
\end{center}
\end{figure}

We begin our survey in the flavor sector.  A detailed discussion
similar to ours has been presented recently in \cite{Bai:2011ed}.  In
\Fig{fig:flavorbounds}, we show the constraints on the non-universal
axigluon model that follow from neutral meson mixing.  The left panel
corresponds to flavor alignment in the up-type quark sector (\ie, $U_u
= 1$), while the figure on the right-hand side reflects the situation
in the case of down-type quark flavor alignment (\ie, $U_d = 1$). We
recall that in both cases the resulting FCNCs are purely left-handed
\eq{eq:Ods} and governed by the CKM matrix \eq{eq:HeffDS}. From the
yellow, orange, and red regions in the left plot we infer that in the
case of up-type quark flavor alignment, the most stringent bound
arises from $\Delta m_{B_s}$.  The $95\%$ CL limit reads 
\beq \label{eq:downbound}
M_G \, > \, \frac{|g_L^h-g_L^\ell|}{2} \; 6.4 \TeV = \frac{6.4
  \TeV}{\sin \left ( 2 \theta \right )} \, > \, 6.4 \TeV \,,
\eeq 
where in the last step we have inserted the analytic expressions for
the left-handed couplings as given in \eq{eq:glgh}.  The 95\% CL upper
limits arising from $|\epsilon_K|$ and $\Delta m_{B_d}$ are weaker
than the bound quoted in \eq{eq:downbound} and simply obtained by
replacing $6.4 \TeV$ with $4.8 \TeV$ and $5.0 \TeV$,
respectively.\footnote{If the total errors in \eq{eq:CM} were enlarged
  by a factor of 2, the new-physics scales entering the numerator
  \eq{eq:downbound} would read $2.8 \TeV$ in the case of
  $|\epsilon_K|$ and $\Delta m_{B_d}$ and $4.0 \TeV$ in the case of
  $\Delta m_{B_s}$.} Notice that in the presence of new CP phases
and/or additional effective operators, the bounds on $M_G$ following
from $\Delta F = 2$ processes in the down-type quark sector are
typically even stronger than the numbers given above. This implies
that a flavor non-universal axigluon possessing generic couplings to
quarks is experimentally ruled out, unless its mass lies in the
multi-TeV range.

Yet, if the underlying theory that determines the pattern of flavor
breaking is left unspecified, it is possible to confine FCNC effects
to the up-type quark sector by aligning the axigluon couplings to
down-type quarks. The purple region in the right panel of
\Fig{fig:flavorbounds} shows that in this case the bounds imposed by
flavor physics are very weak. Requiring that the axigluon
contributions to the $D$-meson mixing parameter $x_D$ do not exceed
the measured value, \ie, $(x_D)_G < 0.63\%$, we derive the following
constraint
\beq \label{eq:xDbound}
M_G \, > \, \frac{|g_L^h-g_L^\ell|}{2} \; 0.22 \TeV = \frac{0.22
  \TeV}{\sin \left ( 2 \theta \right )} \, > \, 0.22 \TeV \,.
\eeq 
We emphasize that since the latter bound has been obtained by
minimizing the constraints from flavor violation in the $\Delta F = 2$
sector, it is a firm (though quite loose) exclusion limit that has to
be satisfied by any flavor non-universal model with extra massive
bosons in the adjoint representation of
$SU(3)_c\hspace{0.25mm}$. Obviously, flavor universal scenarios are
not subject to any flavor constraint by construction.

\begin{figure}[!t]
\begin{center}
\includegraphics[height=7.5cm]{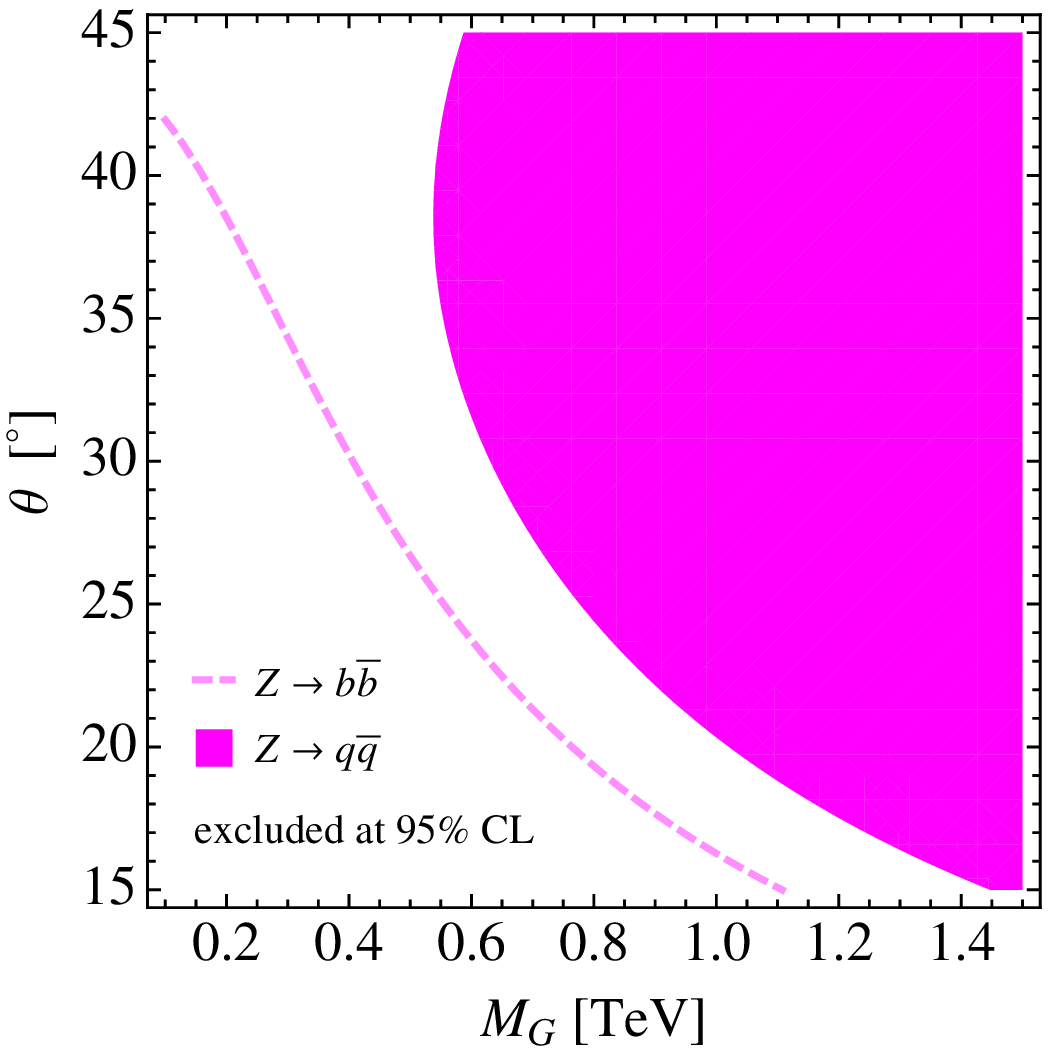}
\qquad 
\includegraphics[height=7.5cm]{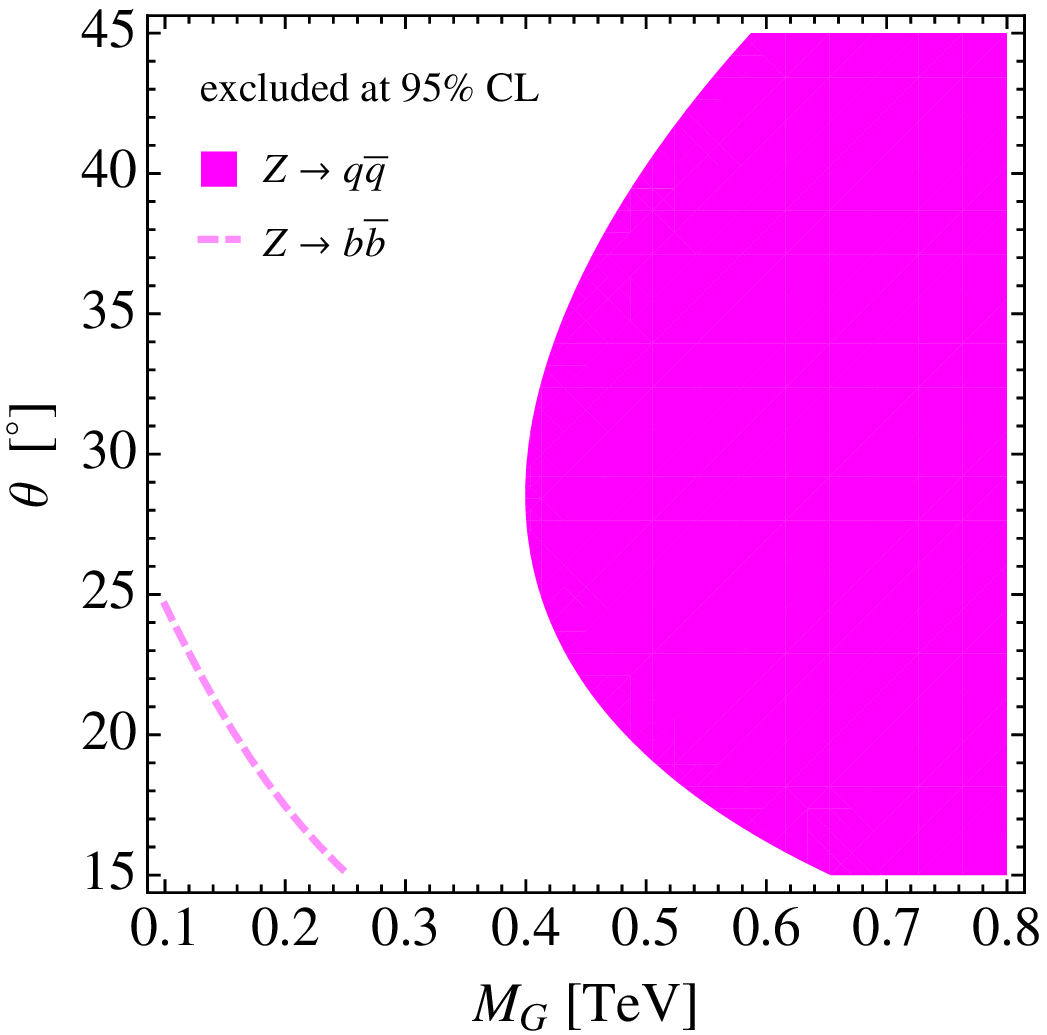}
\end{center}
\begin{center} 
  \parbox{15.5cm}{\caption{\label{fig:ew} Constraints in the
      $M_G\hspace{0.5mm}$--$\hspace{0.5mm}\theta$ plane arising from
      the precision measurements in the $Z \to q \bar q$ sector. The
      left (right) panel corresponds to the flavor non-universal
      (universal) axigluon model. The magenta colored regions of
      parameter space are preferred by the global $\chi^2$ fit to
      $R_b$, $A_b$, $A_{\rm FB}^b$, $\Gamma_Z$, and $\sigma_{\rm
        had}$. Parameter sets below the dashed magenta curve are
      disfavored by the bottom-quark POs alone.}}
\end{center}
\end{figure}

We now turn our attention to the constraints from the bottom-quark POs
\eq{eq:bPOtheory}.  The restrictions imposed by a combination of the
corresponding measurements \eq{eq:bPOsexp} are indicated by the dashed
magenta curve in the left and right panels of \Fig{fig:ew} for the
case of the flavor non-universal and universal axigluon, respectively.
The $M_G \hspace{0.25mm}$--$\hspace{0.25mm} \theta$ regions below the
lines are disfavored at 95\% CL. By comparing the two plots, one first
observes that the $Z \to b \bar b$ bound is significantly weaker for a
flavor universal axigluon than for a flavor non-universal axigluon.
This feature arises since, due to the smallness of the ratios $\big
({\cal G}_R^u/{\cal G}_L^u \big )^2 \approx 1/5$ and $\big ({\cal
  G}_R^d/{\cal G}_L^d \big)^2 \approx 1/30$ of $Z$-boson couplings,
the dominant observable $R_b$ measures approximately the difference
$(g_L^h)^2 - (g_L^\ell)^2$.  This combination of axigluon couplings is
equal to zero in the flavor universal case, which implies that for
fixed $\theta$ the constraints on $M_G$ are less severe for a flavor
universal than for a non-universal axigluon. A second feature that is
clearly visible in the plots is that $Z \to b \bar b$ does not provide
a sound constraint in the limit $\theta \to 45^\circ$.  This behavior
is readily understood by recalling that the axigluon effects in $R_b$
vanish in this limit in both variants of the model. Notice further
that the less constraining asymmetries $A_b$ and $A_{\rm FB}^b$ are to
a good approximation proportional to $(g^h_L)^2 - (g^h_R)^2$.  This
combination also tends to zero for $\theta \to 45^{\circ}$, which
further impairs the restrictive power of the bottom-quark POs.

\begin{figure}[!t]
\begin{center}
\includegraphics[height=7.5cm]{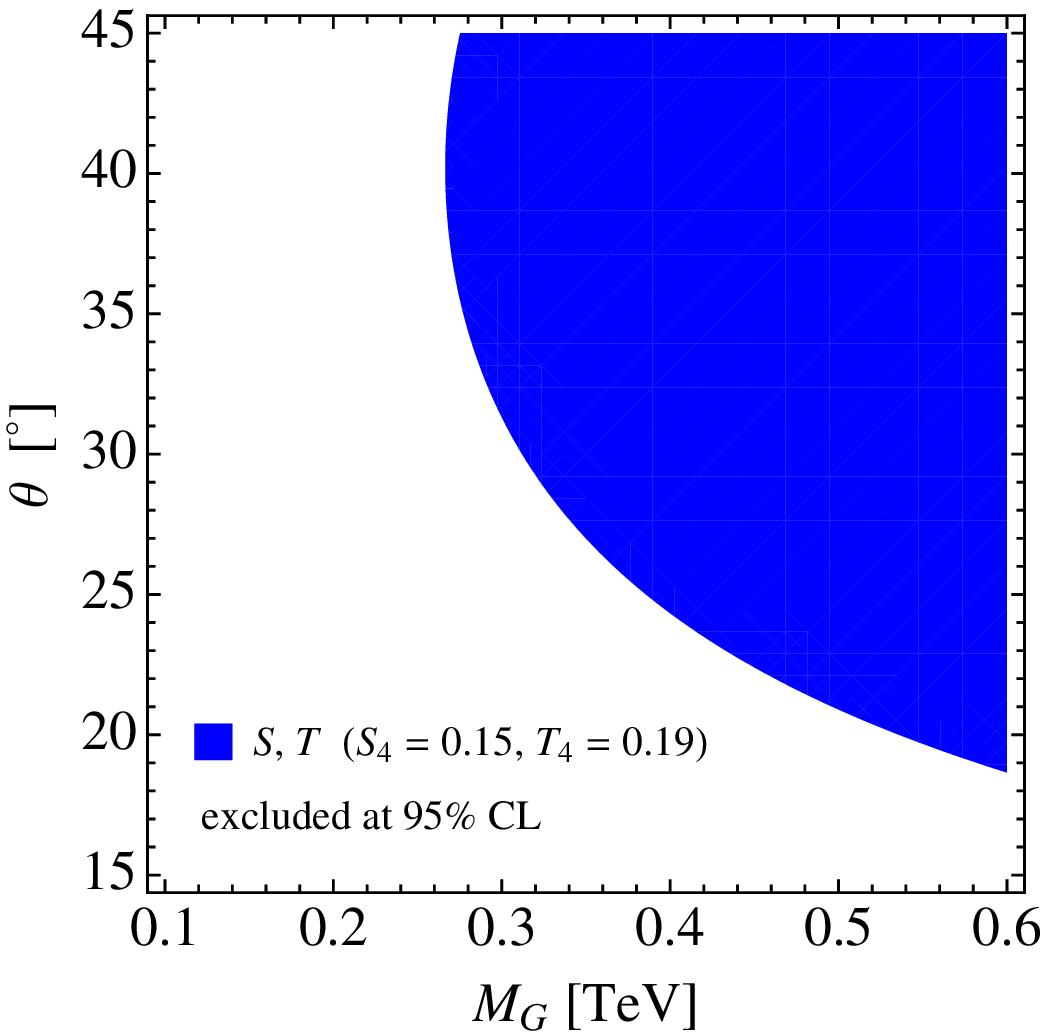}
\qquad 
\includegraphics[height=7.5cm]{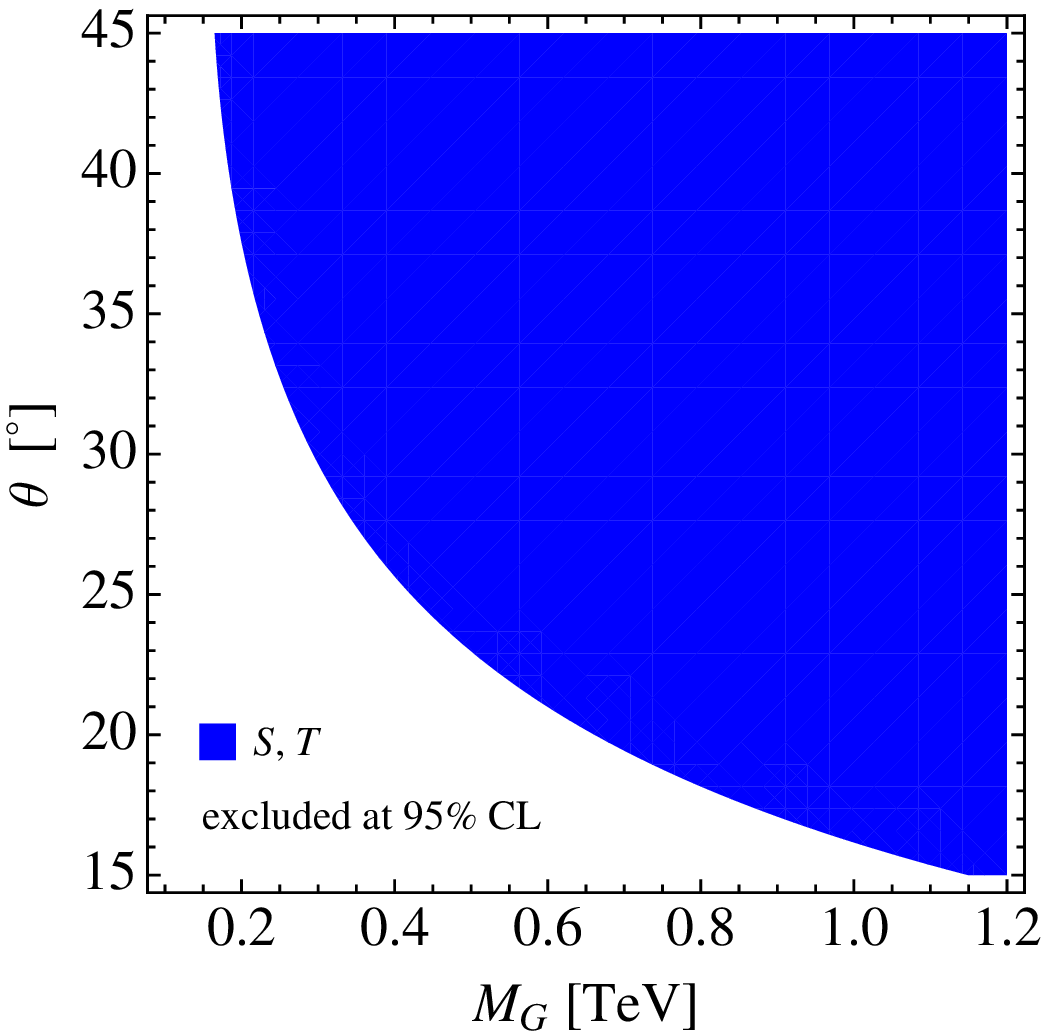}

\end{center}
\begin{center} 
  \parbox{15.5cm}{\caption{\label{fig:STplane} Constraints in
      the $M_G\hspace{0.5mm}$--$\hspace{0.5mm}\theta$ plane due to a
      combined fit to the oblique parameters $S$ and $T$. The
      left panel shows the results for the case of a flavor
      non-universal axigluon assuming $S_4 = 0.15$ and $T_4 = 0.19$,
      while the right plot represents the case of a flavor universal
      axigluon. The parameter space preferred by the fit is colored
      blue.}}
\end{center}
\end{figure}

While the restrictions imposed by $R_b$, $A_b$, and $A_{\rm FB}^b$ can
thus be avoided, combining them with $\Gamma_Z$ and $\sigma_{\rm
  had}$, as defined in \eq{eq:GammaZsigmahad}, turns out to lead to
non-trivial constraints on the axigluon mass for any value of the
mixing angle. From a combination of the whole set of $Z \to q \bar q$
observables, we derive in the case of the flavor non-universal
axigluon model the following 95\% CL limit
\beq \label{eq:Zaxinon}
M_G \, > \, \left ( 0.67 \, \sqrt{\tan^2 \theta + 0.40 \, \cot^2
    \theta - 0.35} - 0.10 \right ) \! \TeV \, > \, 0.54 \TeV \,.
\eeq
In the case of the flavor universal axigluon, we obtain instead
\beq \label{eq:Zaxiuni}
M_G \, > \, \left ( 0.71 \, \sqrt{\tan^2 \theta + 0.09 \, \cot^2
    \theta +0.07 } -0.18 \right ) \! \TeV \, > \, 0.40 \TeV \,.
\eeq
For $\theta = 45^\circ$ the inequalities \eq{eq:Zaxinon} and
\eq{eq:Zaxiuni} imply $M_G > 0.58 \TeV$ and $M_G > 0.59 \TeV$,
respectively.  The parameter regions allowed by $Z \to q \bar q$ at
95\% CL are shown as magenta areas in the left and right panels of
\Fig{fig:ew}. The panels show clearly that new color-octet bosons with
masses below the electroweak scale and QCD-like axigluon couplings to
all quark species are not compatible with the LEP and SLC measurements
of the $Z q\bar q$ couplings. Of course, the $Z \to q \bar q$
constraints can be weakened by assuming that the presence of the
axigluon only affects the couplings of a single light quark. 

We now discuss the constraints on the
$M_G\hspace{0.5mm}$--$\hspace{0.5mm}\theta$ plane that emerge from the
oblique parameters $S$ and $T$. We again distinguish between a flavor
non-universal and a universal axigluon. From \Sec{sec:ewpo} we know
already that in the former case a model-independent bound on $M_G$ and
$\theta$ cannot be derived, since the fourth-generation contributions
$S_4$ and $T_4$ can always be chosen in such a way as to compensate
for the corrections associated with the axigluon.  Studying the
constraints for a specific set of fourth-generation contributions to
$S$ and $T$ is nevertheless a useful exercise to get a feeling of how
stringent the bounds on the
$M_G\hspace{0.5mm}$--$\hspace{0.5mm}\theta$ plane can be.  Taking $S_4
= 0.15$ and $T_4 = 0.19$ as motivated in \Sec{sec:ewpo}, we obtain the
following 95\% CL bound
\beq \label{eq:STaxinon}
M_G > \left ( 1.68 \, \sqrt{\tan^2 \theta + 0.50 \, \cot^2 \theta +
    56.3} - 12.5 \right ) \! \TeV \, > \, 0.27 \TeV \,.
\eeq
This lower limit is displayed in the left plot in
\Fig{fig:STplane}. Since the axigluon contribution to $T$ is strictly
positive and larger by an order of magnitude than the one to $S$, a
stronger (weaker) bound is obtained if a value $T_4 > 0.19$ ($T_4 <
0.19$) inside the 68\% probability region of the $S$--$T$ ellipse is
chosen.  We conclude from this exercise that in the context of the
flavor non-universal axigluon the oblique parameters are typically
able to probe values of $M_G$ at and beyond the electroweak scale.

One can be more specific in the case of a flavor universal axigluon,
since the model is anomaly-free without the introduction of extra
matter. This feature allows one to derive a model-independent bound on
$M_G$ as a function of $\theta$ from $S$ and $T$. At 95\% CL we find
\beq \label{eq:STaxiuni}
M_G > \left ( 0.88 \, \sqrt{\tan^2 \theta + 1.23 \, \cot^2 \theta
    + 34.0} - 5.13 \right ) \! \TeV \, > \, 0.16 \TeV \,.
\eeq
This limit is shown in the right panel of \Fig{fig:STplane}.  For
$\theta= 45^{\circ}$ the latter inequality yields $M_G > 0.17\TeV$. We
see again that the oblique parameters, despite the fact that they are
first affected at the two-loop level by the presence of the axigluon,
rule out the possibility of masses $M_G$ significantly below the
electroweak scale. Notice however that the limits from $T$ can become
quite pesky, if the top quark couples very strongly to the massive
color-octet vector. In view of the enhancement of $(g_R^h)^2$ relative
to $(g_L^h)^2$ by a factor of 2 in the expression \eq{eq:STapprox},
this statement applies in particular to models where the right-handed
top quark is (fully) composite. In such a case the parameter $T$
receives unacceptably large positive corrections even for axigluon
masses in the TeV range.

\begin{figure}[!t]
\begin{center}
\includegraphics[height=7.5cm]{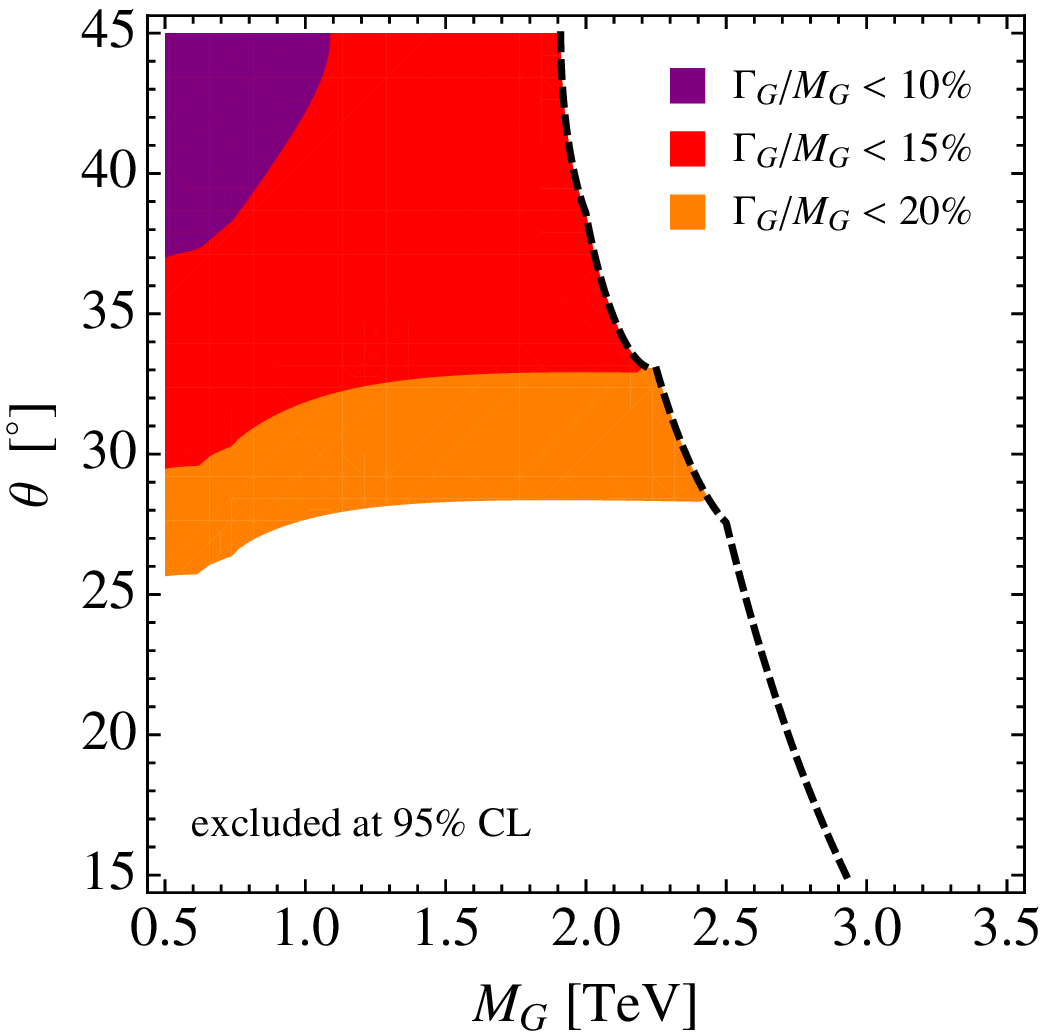}
\qquad
\includegraphics[height=7.5cm]{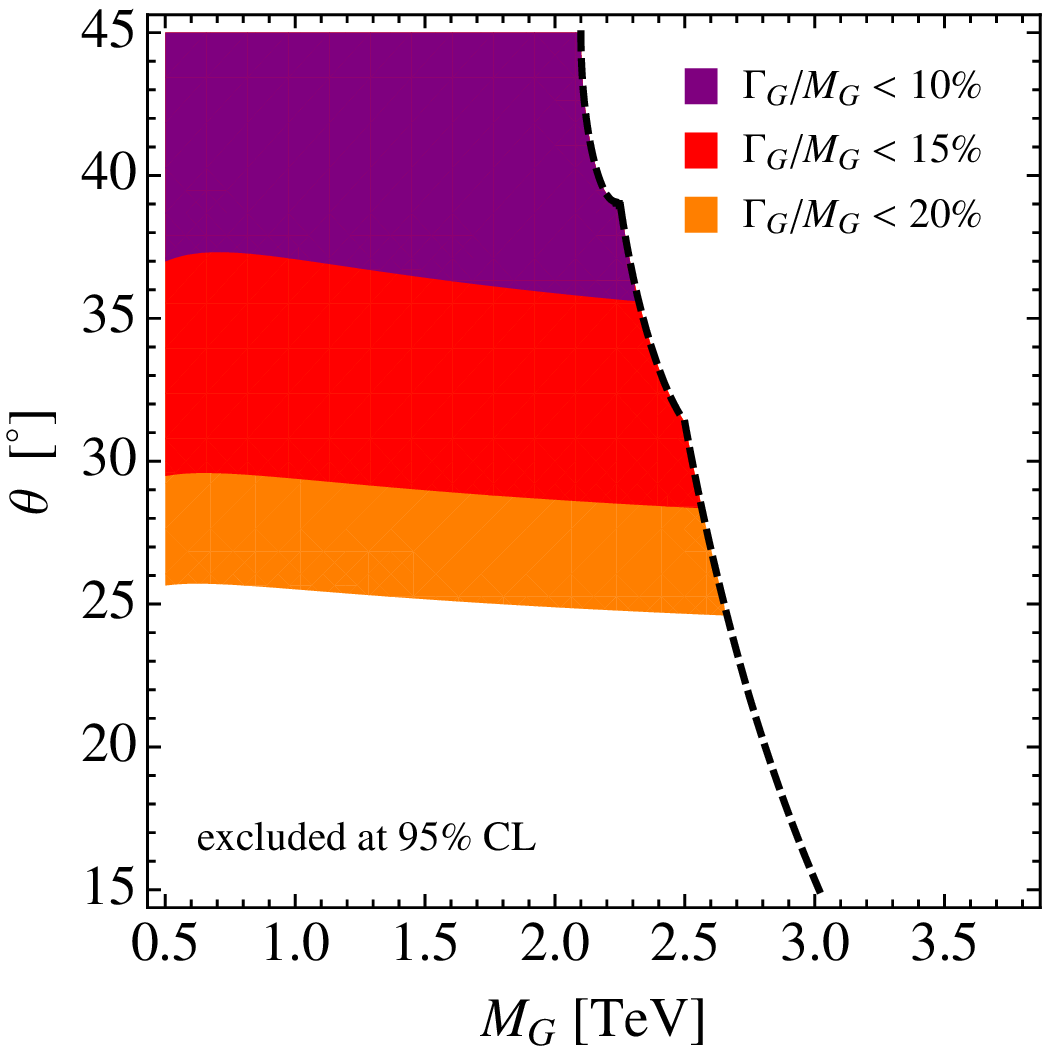}
\end{center}
\begin{center} 
  \parbox{15.5cm}{\caption{\label{fig:dijetres} Constraints from the
      latest ATLAS narrow resonance search in dijet production on the
      flavor non-universal (left) and universal (right) axigluon
      model. The $95\%$ CL contour is represented by the dashed black
      line. Excluded regions of parameter space are displayed in color
      for fixed relative axigluon decay widths $\Gamma_G/M_G < 10\%$
      (purple), $\Gamma_G/M_G < 15\%$ (red), and $\Gamma_G/M_G < 20\%$
      (orange).}}
\end{center}
\end{figure}

After this detailed discussion of the indirect probes of the axigluon
parameter space through high-intensity experiments, we now move to the
direct bounds stemming from the energy frontier. We start with the
constraints imposed by the narrow resonance searches in dijet
production. In \Fig{fig:dijetres} we display the $95\%$ CL upper
bounds in the $M_G\hspace{0.25mm}$--$\hspace{0.25mm}\theta$ plane for
the flavor non-universal (left) and universal (right) axigluon
model. The area to the left of the dashed black curve is disfavored by
the latest ATLAS measurement of the product $\sigma \hspace{0.25mm}
{\cal B} \hspace{0.25mm} A$ involving the resonant production cross
section, the branching fraction, and the acceptance.  As explained in
\Sec{sec:collider}, resonance search analyses are only applicable if
the total decay width $\Gamma_G$ is sufficiently small, so that the
axigluon forms a distinct resonance. The excluded regions are obtained
for fixed ratios $\Gamma_G/M_G < 10\%$ (purple), $\Gamma_G/M_G < 15\%$
(red), and $\Gamma_G/M_G < 20\%$ (orange), corresponding to mixing
angles of $\theta\gtrsim 40^{\circ}$, $30^{\circ}$, and $25^{\circ}$,
respectively.  The blank area features $\Gamma_G/M_G >
20\%$. Comparing the two panels, one observes that the bound on $M_G$
is weaker if the axigluon has flavor non-universal couplings. This is
due to the presence of the extra fermion generation, which lowers the
axigluon branching ratio into light quarks, and hence allows for a
bigger production cross section for fixed values of $\sigma
\hspace{0.5mm} {\cal B} \hspace{0.5mm} A$. Yet, a larger resonant
production cross section goes along with a lower axigluon mass.
Notice that the presence of the additional decay channels into $u_4
\bar u_4$ and $d_4 \bar d_4$ manifests itself even more clearly if on
considers the contours in the
$M_G\hspace{0.25mm}$--$\hspace{0.25mm}\theta$ plane corresponding to
constant values of $\Gamma_G/M_G$. We see that the resonance search is
applicable to a wider range of parameter space in the flavor universal
model. In the limit $\theta \to 45^{\circ}$, we find for a flavor
non-universal axigluon the following 95\% CL bound\footnote{Since the
  narrow resonance search of ATLAS includes only dijet events with
  $M_{jj} > 0.6 \TeV$, the mass window below $0.6 \TeV$ is strictly
  speaking also allowed by this measurement.  Older dijet searches
  \cite{Albajar:1988rs, Abe:1993it} extend the disfavored region down
  to around $0.15 \TeV$.}  \beq \label{eq:resnon} M_G > 1.9 \TeV \,,
\eeq while in the flavor universal case we obtain the somewhat
stronger constraint \beq \label{eq:resuni} M_G > 2.1 \TeV \,.  \eeq
The values of $\Gamma_G/M_G$ corresponding to these limits amount to
almost $11\%$ and a little bit more than $8\%$, respectively.
Although the publication \cite{Collaboration:2011aj} does not state
explicitly for which values of $\Gamma_G/M_G$ their axigluon analysis
is applicable, we conclude from the model-independent ATLAS limits on
Gaussian resonances, which cover relative widths up to $15\%$, that
the derived dijet resonance bounds \eq{eq:resnon} and \eq{eq:resuni}
are robust.  We also remark that the dependence of \eq{eq:resnon} on
the masses of the fourth-generation quarks is rather mild. If instead
of $m_{u_4} = 311 \GeV$ and $m_{d_4} = 372 \GeV$ one would use
$m_{u_4} = m_{d_4} = 600 \GeV$, the former bound would strengthen
slightly and read $M_G > 2.0\TeV$.

{

\begin{figure}[!t]
\begin{center}
\includegraphics[height=7.5cm]{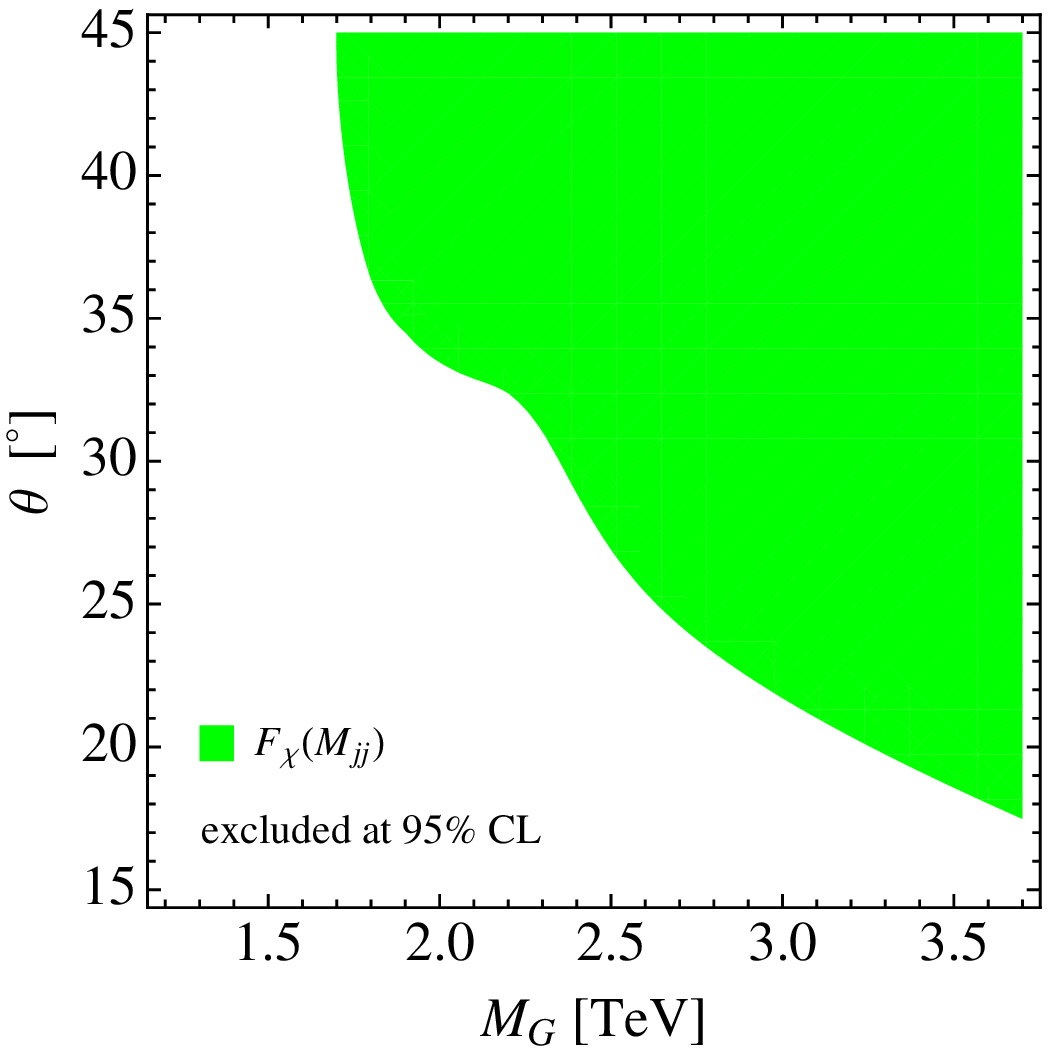}
\qquad
\includegraphics[height=7.5cm]{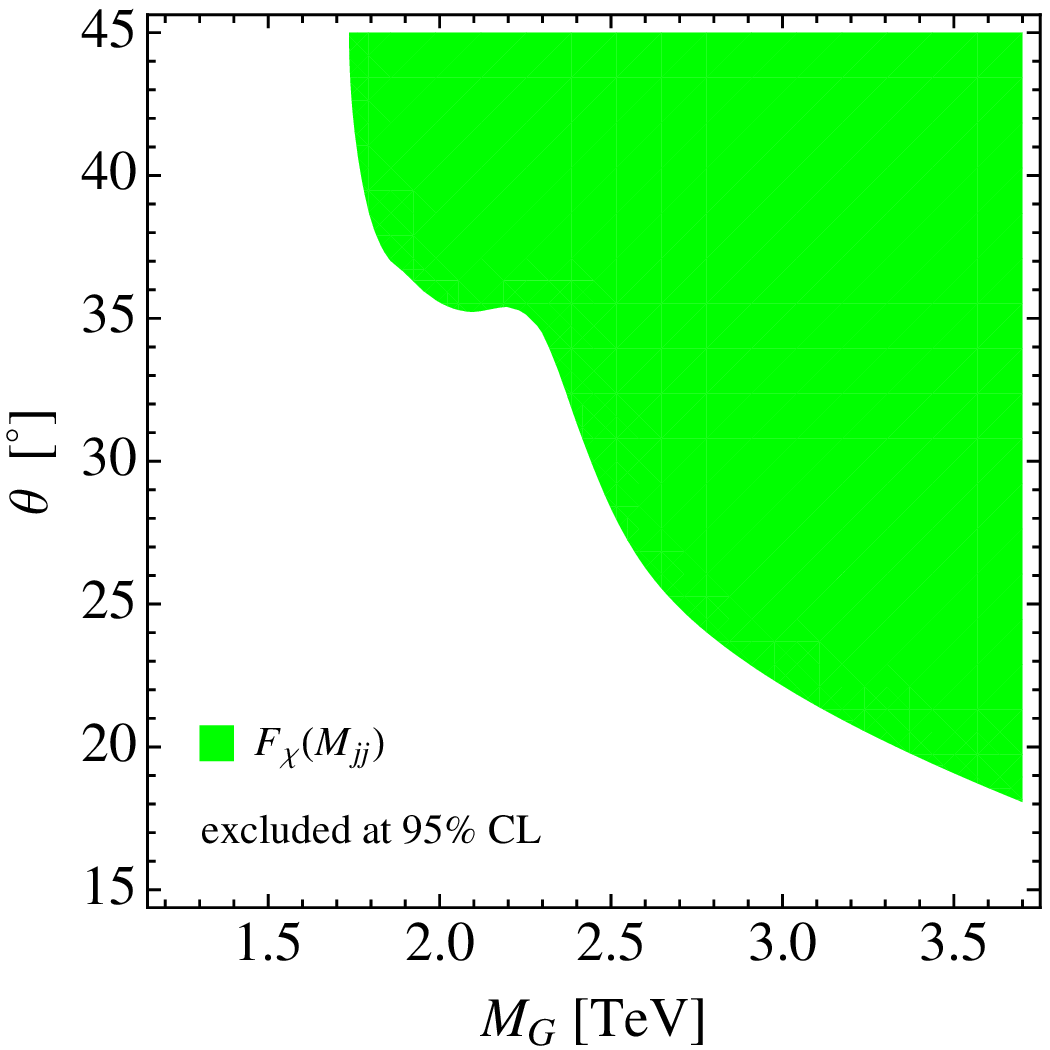}
\end{center}
\begin{center} 
  \parbox{15.5cm}{\caption{\label{fig:fchimgth}
      Constraints on $M_G\hspace{0.25mm}$--$\hspace{0.25mm}\theta$
      arising from the angular distribution in dijet production in the
      non-universal (left) and universal (right) axigluon model. The
      green areas show the parameter region in accordance with the
      ATLAS data at $95\%$ CL.}}
\end{center}
\end{figure}

The angular distribution in dijet production allows one to derive
constraints on the axigluon parameter space without restricting to the
narrow-width approximation. As described at the end of
\Sec{sec:collider}, we perform a fit of the axigluon prediction for the
centrality ratio $F_{\chi}(M_{jj})$ in \eq{eq:Fchi} by constructing a
likelihood following the procedure outlined in
\cite{Collaboration:2011aj}. Our analysis uses the ATLAS data for
$F_{\chi}(M_{jj})$ in $17$ different bins, covering dijet masses of
$M_{jj}\in[1.2,3.7]\TeV$. Due to the low statistics at high $M_{jj}$
the fit is however essentially driven by the bins below $2.5\TeV$. The
resulting constraints on the axigluon parameters in the
$M_G\hspace{0.25mm}$--$\hspace{0.25mm}\theta$ plane are shown in
\Fig{fig:fchimgth} for the non-universal (left) and universal (right)
axigluon model. The white areas are excluded at $95\%$ CL. Notice the
distortion of the exclusion contour at $M_G \approx 2.2\TeV$. The
constraints are strengthened in this region due to the strong decline
of the measured $F_{\chi}(M_{jj})$ spectrum starting at this value of
$M_{jj}\hspace{0.25mm}$, as displayed in \Fig{fig:Fchi}. In the limit
$\theta\to 45^{\circ}$, the $95\%$ CL bound reads\footnote{The quoted
  number is based on the expected ATLAS limit of $\Lambda > 5.7
  \TeV$. If instead the observed limit $\Lambda > 9.5 \TeV$ is used,
  one finds $M_G > 2.4 \TeV$.}  \beq \label{eq:ang} M_G > 1.7 \TeV \,,
\eeq for both the flavor non-universal and universal axigluon model.
Since the matrix elements for axigluon production are symmetric under
the exchange of left- and right-handed couplings, the difference
between the contours is due to the presence of the fourth generation
in the non-universal case only. The additional contributions to the
width broaden the resonance in the spectrum $F_{\chi}(M_{jj})$,
yielding a slightly weaker constraint. The dependence of \eq{eq:ang}
on the masses of the extra quarks themselves is however negligible.

We emphasize that the recent measurement of the dijet angular
distribution also puts constraints on axigluons with masses below
$0.5\TeV$, where no ATLAS data is available, since light axigluons
enhance the tail of the $F_\chi (M_{jj})$ distribution with respect to
the QCD expectation. This feature is clearly visible in the left panel
of \Fig{fig:updijetres}, which shows the axigluon predictions for $M_G
= 0.3 \TeV$ (red curve), $M_G = 0.6 \TeV$ (green curve), and $M_G =
1.2 \TeV$ (blue curve). All lines have been obtained assuming QCD-like
couplings and $\Gamma_G/M_G = 10\%$.  Performing a likelihood fit to
the set of the first 22 invariant mass bins,\footnote{We do not
  include the last 5 bins of the recent ATLAS measurement in the fit,
  because their central values are all zero and they have large
  statistical errors.}  we find that the parameter space with $M_G >
0.17 \TeV$ is disfavored at 95\% CL in the case of both the flavor
non-universal and universal axigluon. Axigluon masses down to $0.25
\TeV$ are also excluded directly by an earlier analysis of angular
distributions at the Tevatron \cite{Abazov:2009mh}. These results
indicate that color-octet resonances with QCD-like couplings to light
and heavy quarks and masses at and below the electroweak scale are in
general in conflict with existing constraints arising from angular
distributions in dijet production.
  
\begin{figure}[!t]
\begin{center}
\includegraphics[height=7.5cm]{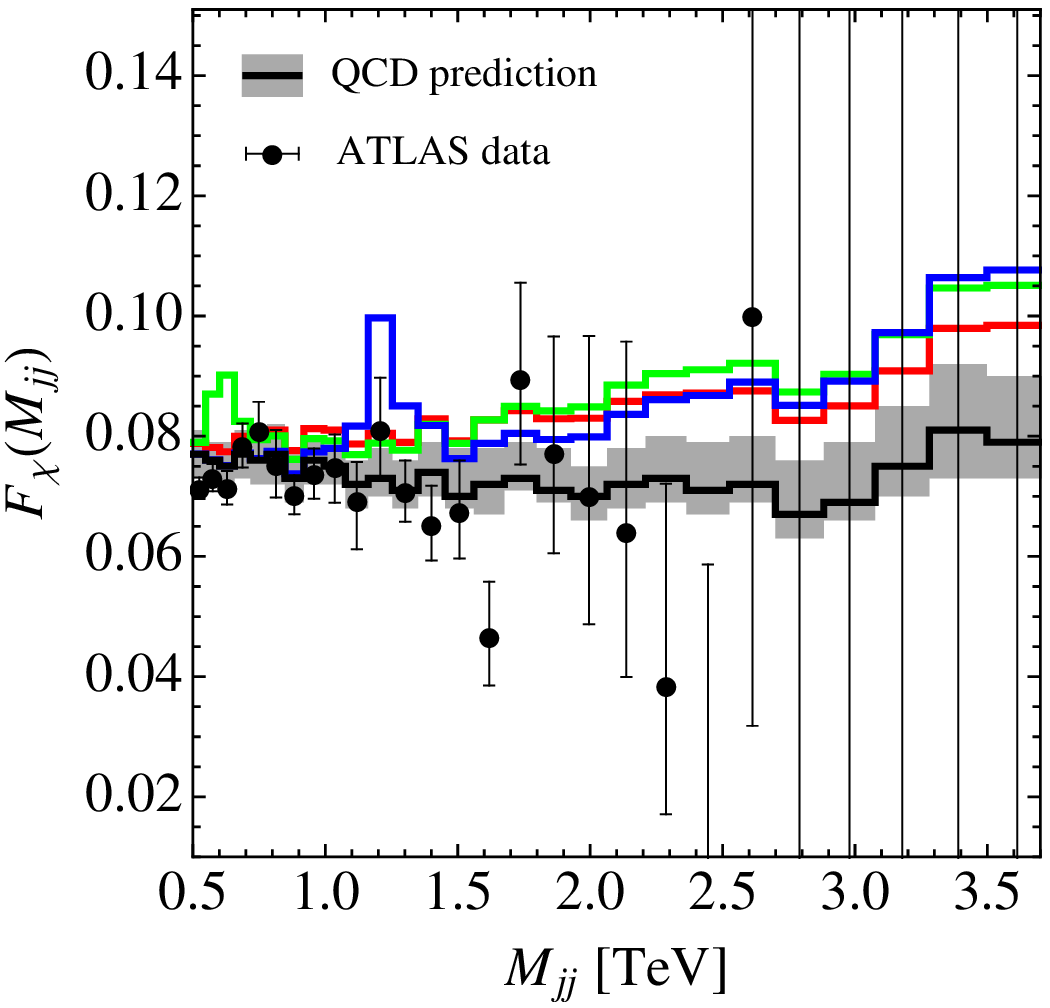}
\qquad
\includegraphics[height=7.5cm]{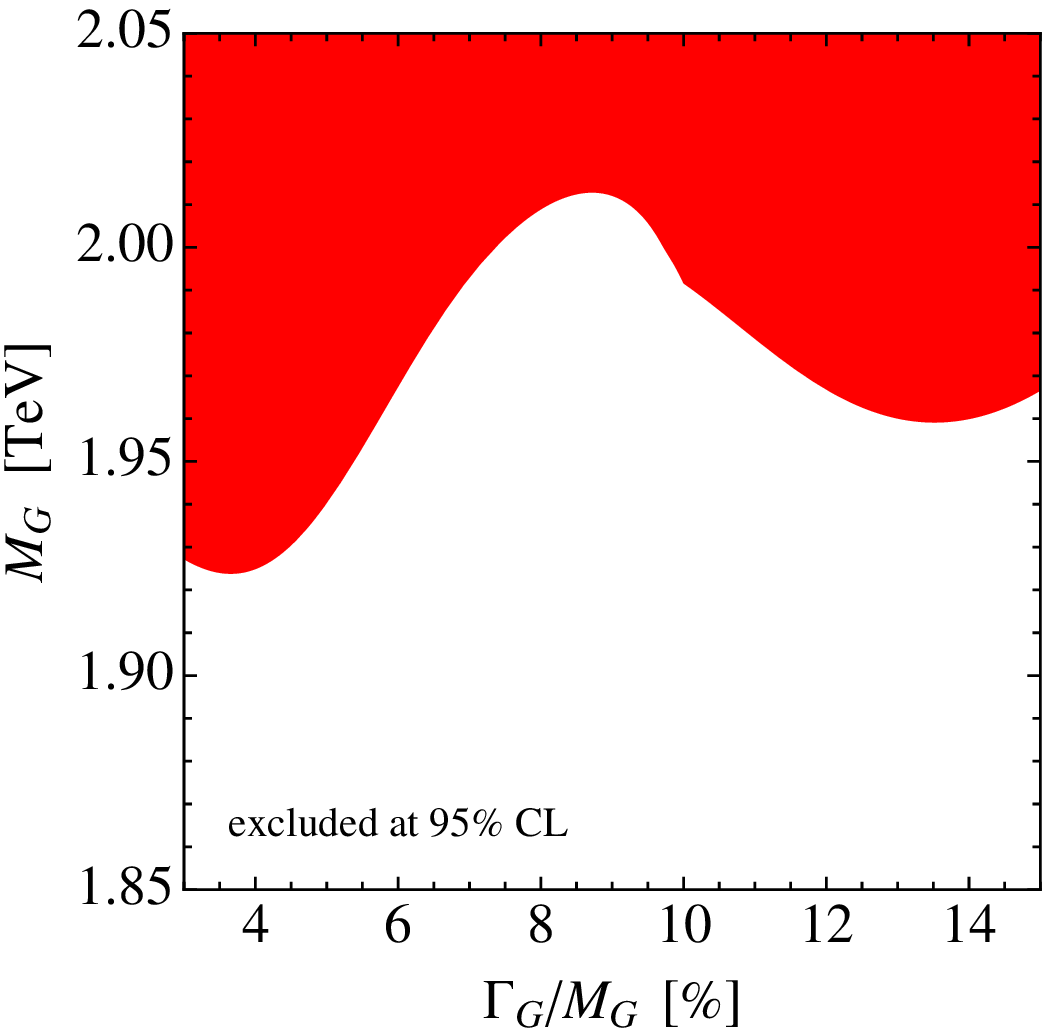}
\end{center}
\begin{center} 
  \parbox{15.5cm}{\caption{\label{fig:updijetres} Left: Ratio $F_\chi
      (M_{jj})$ as a function of $M_{jj}$ for three different
      axigluons with $M_G = 0.3 \TeV$ (red line), $M_G = 0.6 \TeV$
      (green line), and $M_G = 1.2 \TeV$ (blue line). The shown
      predictions correspond to QCD-like couplings and $\Gamma_G/M_G =
      10\%$. For comparison the QCD prediction with errors (black line
      and gray band) and the ATLAS measurement (black error bars) are
      also shown.  Right: Constraints from the latest ATLAS narrow
      resonance search in dijet production on a color-octet boson
      with relative width $\Gamma_G/M_G$ that only couples to up
      quarks. The blank region is excluded at 95\% CL for $g_L^u =
      g_R^u = 1$. See text for details.}}
\end{center}
\end{figure}
 
Let us also spend some words on how the bounds on the new-physics
scale $M_G$ change if dijet production can only proceed via $u \bar u
\to G$ scattering. First recall that it is more likely to find an up
quark than any other light quark in the proton. In fact, the up-quark
luminosity $\ff_{u\bar u} (M_G^2/s,\mu_f)$, which is a measure of this
probability, amounts to 60\% to 92\% of the total quark luminosity
$\sum_q \ff_{q\bar q} (M_G^2/s,\mu_f)$ for axigluon masses $M_G \in
[0.5, 4.0] \TeV$ and $pp$ collisions at $\sqrt{s} = 7 \TeV$. These
numbers imply that the sensitivity to $M_G$ is largely unchanged if
the axigluon has ${\cal O} (1)$ couplings to up quarks only. On the
right-hand side of \Fig{fig:updijetres}, we show the bound on $M_G$ as
a function of $\Gamma_G/M_G$ that arises from the model-independent
limits on resonances with Gaussian shape presented by ATLAS
\cite{Collaboration:2011aj}. The area allowed at 95\% CL is colored
red. One observes that the limit on $M_G$ depends only very weakly on
the relative width of the resonance and amounts to around $1.9 \TeV$
to $2.0 \TeV$ over the whole range of $\Gamma_G/M_G$.  The shown
exclusion region has been obtained assuming $g_L^u = g_R^u = 1$, but
translating it to other couplings just requires to perform a simple
rescaling by $\big ( (g_L^u)^2+(g_R^u)^2 \big )/2$. To give an
example, a color-octet resonance interacting only with right-handed up
quarks, a coupling strength of $g_R^u = 2$, and $\Gamma_G/M_G < 15\%$
is disfavored at 95\% CL if its mass is below $4.0 \TeV$. The
situation is different in the case of the dijet angular
distribution. For an axigluon that couples only to up quarks, the
effect on the centrality ratio $F_{\chi} (M_{jj})$ is significantly
reduced. In particular, the enhancement of the tail at high $M_{jj}$
is very mild, such that an axigluon with $M_G < 0.5\TeV$ cannot be
excluded by the LHC measurements.  Axigluons with a mass in the
accessibility range of the ATLAS detector are visible as a sharp
resonance in the $F_{\chi} (M_{jj})$ spectrum only if their relative
width does not exceed a few percent. Broader axigluon resonances
escape the constraints set by the angular distribution.

\begin{figure}[!t]
\begin{center}
\includegraphics[height=7.5cm]{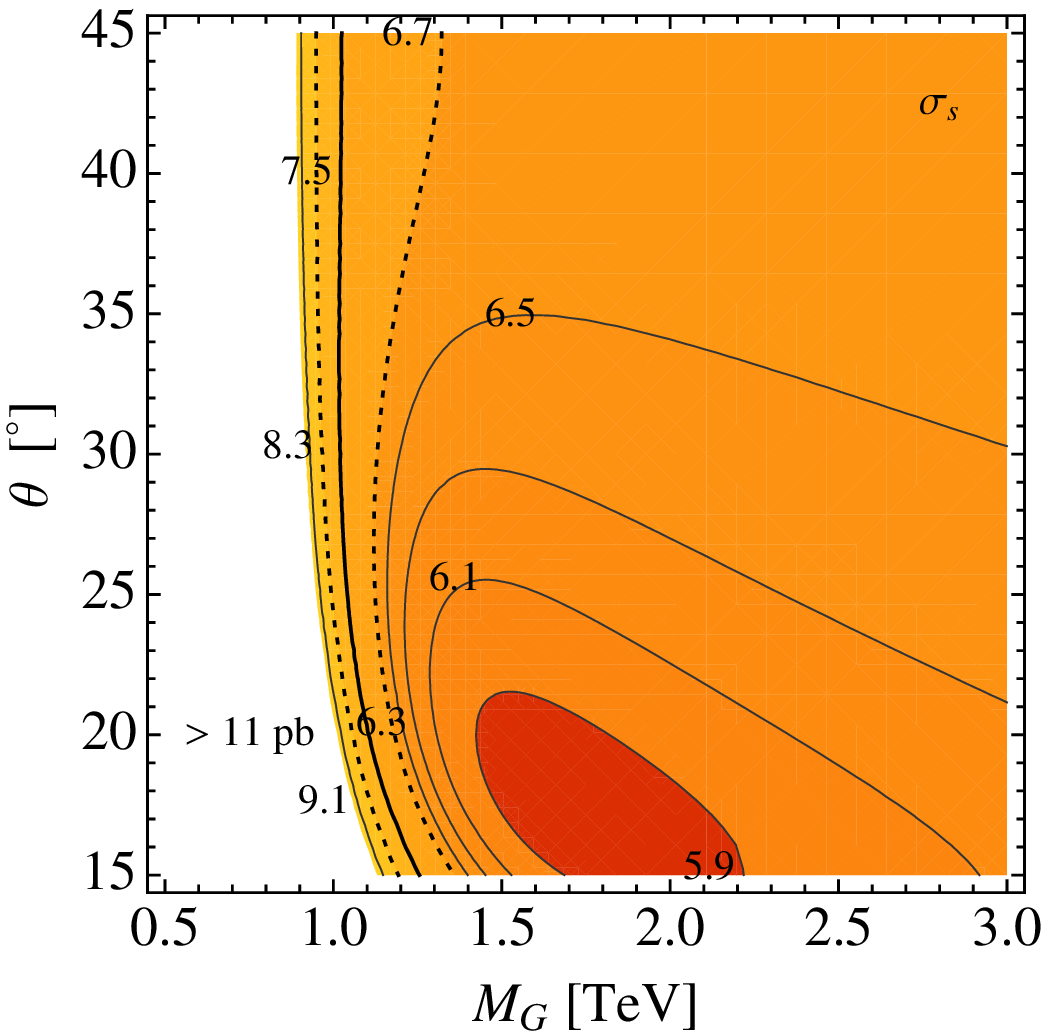}
\qquad 
\includegraphics[height=7.5cm]{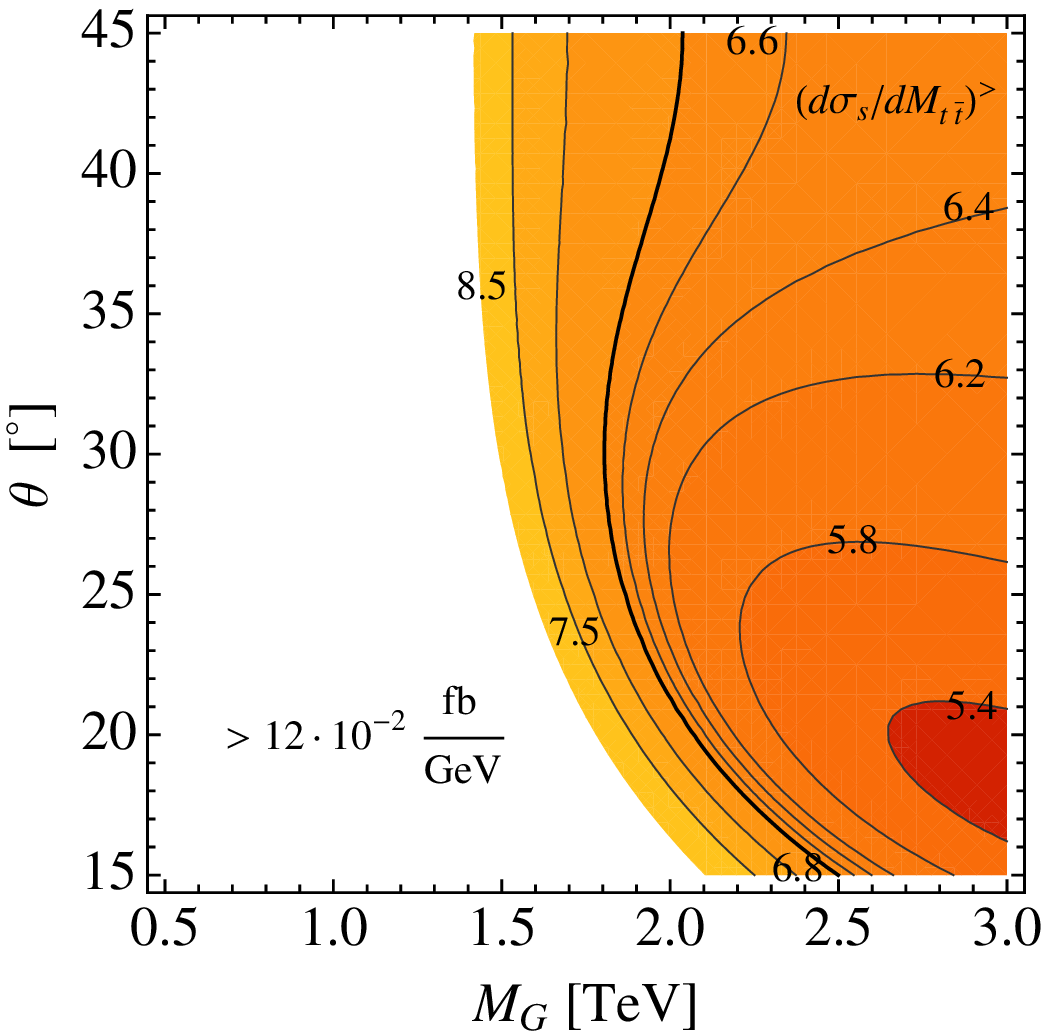}

\vspace{5mm}

\includegraphics[height=7.5cm]{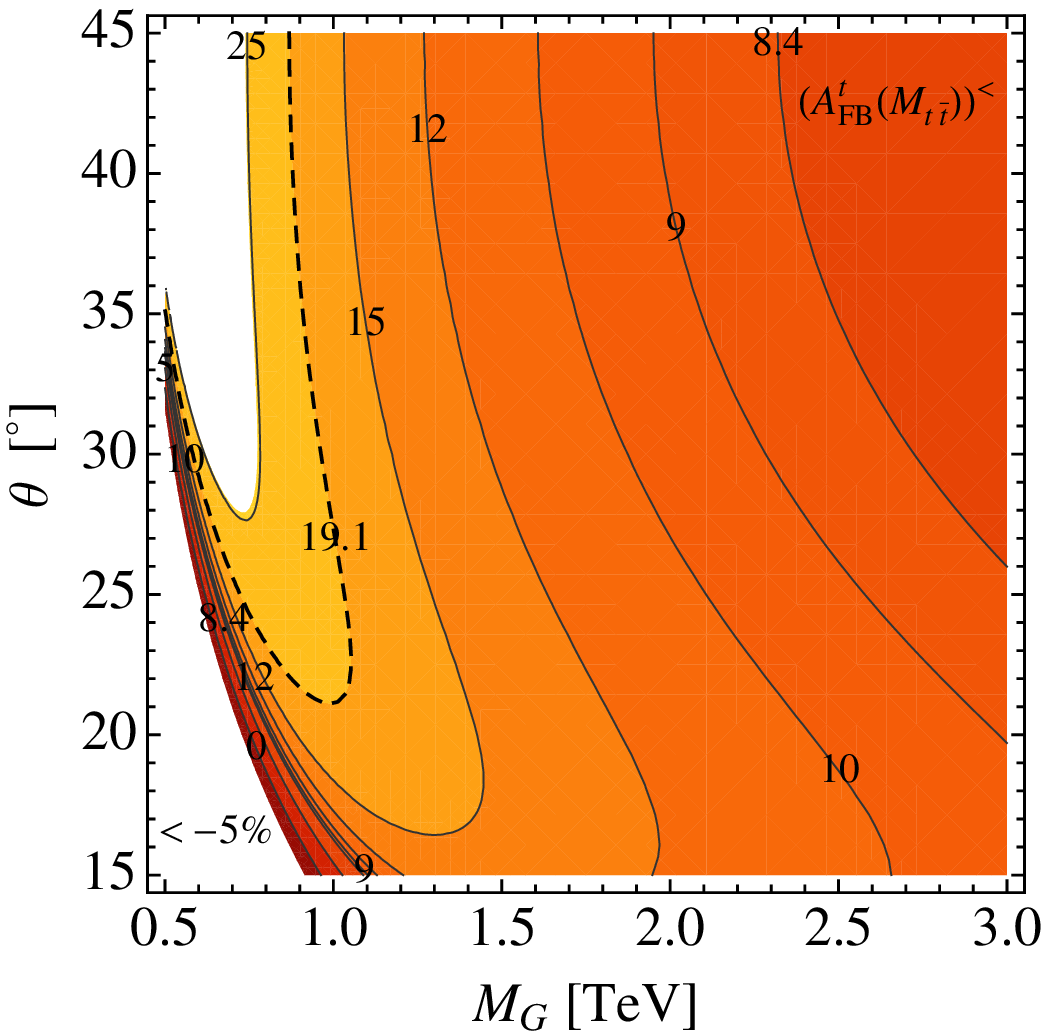}
\qquad 
\includegraphics[height=7.5cm]{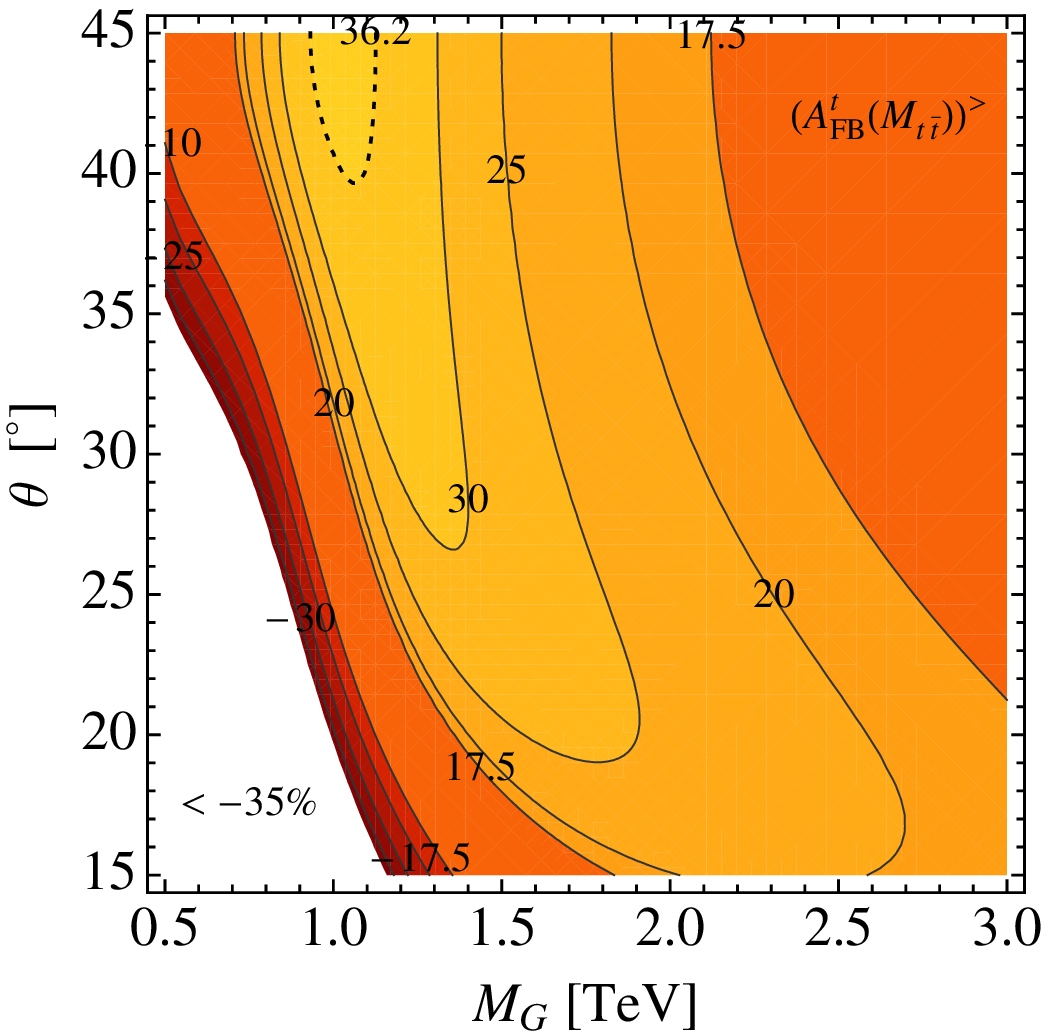}
\end{center}
\begin{center} 
  \parbox{15.5cm}{\caption{\label{fig:tt} Predictions for the
      observables in $t \bar t$ production obtained in the
      non-universal axigluon model. From the upper left to the lower
      right $\sigma_s$, $(d\sigma_s/dM_{t\bar t})^{>}$, $\big (A_{\rm
        FB}^t (M_{t\bar t}) \big)^{<}$, and $\big (A_{\rm FB}^t
      (M_{t\bar t}) \big)^{>}$ are shown. The central values of the
      measurements are displayed as solid black contours, while the
      $1\sigma$ ($2\sigma$) total error ranges are indicated by the
      dotted (dashed) lines. See text for further details.}}
\end{center}
\end{figure}

We finally explore the constraints on axigluons from top-antitop quark
production.  In \Fig{fig:tt} we compare the inclusive cross section
$\sigma_s$ and its high-$M_{t\bar t}$ bin, $(d\sigma_s/dM_{t\bar
  t})^{>} = (d\sigma_s/dM_{t\bar t})^{M_{t\bar t}\in [0.8,1.4]\,
  \text{TeV}}$, with the forward-backward asymmetry in two bins
separated at $M_{t\bar t} = 0.45\TeV$ (see
Table~\ref{tab:ttobservables} for the SM predictions and experimental
values of the relevant $t \bar t$ observables).  As anticipated in
\Sec{sec:collider}, the large measured values of the asymmetry are in
tension with the remaining observables, which are well described by
QCD.  This tension is most pronounced in the region of high $M_{t\bar
  t}$, which is most sensitive to contributions from new physics. From
the upper and lower right panels it becomes apparent that the
asymmetry $\big (A_{\text{FB}}^t(M_{t\bar t}) \big )^{>}$ favors an
axigluon with a mass around $1\TeV$, while the symmetric cross section
$(d\sigma_s/dM_{t\bar t})^{>}$ is in agreement with the measurement
within errors for $M_G\gtrsim 1.5\TeV$. Notice that the observable
$\big (A_{\text{FB}}^t(M_{t\bar t}) \big)^{>}$ prefers large mixing
angles $\theta$, corresponding to axigluon couplings that are mostly
axial-vector-like. This maximizes the contribution that arises from
the axigluon interference with the SM amplitude, proportional to
$-g_A^q \hspace{0.25mm} g_A^t$. Axial-vector-like couplings are also
preferred by $(d\sigma_s/dM_{t\bar t})^{>}$, as in this case the
axigluon-gluon interference is suppressed and, in addition, the
different terms that stem from the interference of the new-physics
contribution with itself tend to cancel. These features can be read
off from the model-independent formulas presented in
\App{app:ttobservables}. The asymmetry at low $M_{t\bar t}$, as shown
in the lower left panel, disfavors constructive axigluon
contributions, but is subject to large statistical errors, rendering
its restrictive power marginal at present. Finally, since the QCD
prediction is below the measurement, the total cross section displayed
on the upper left leaves space for moderate positive axigluon
contributions. Such moderate enhancements are in fact predicted in the
flavor non-universal model for values of $\theta$ close to maximal and
$M_G \gtrsim 1 \TeV$.

\begin{figure}[!t]
\begin{center}
\includegraphics[height=7.5cm]{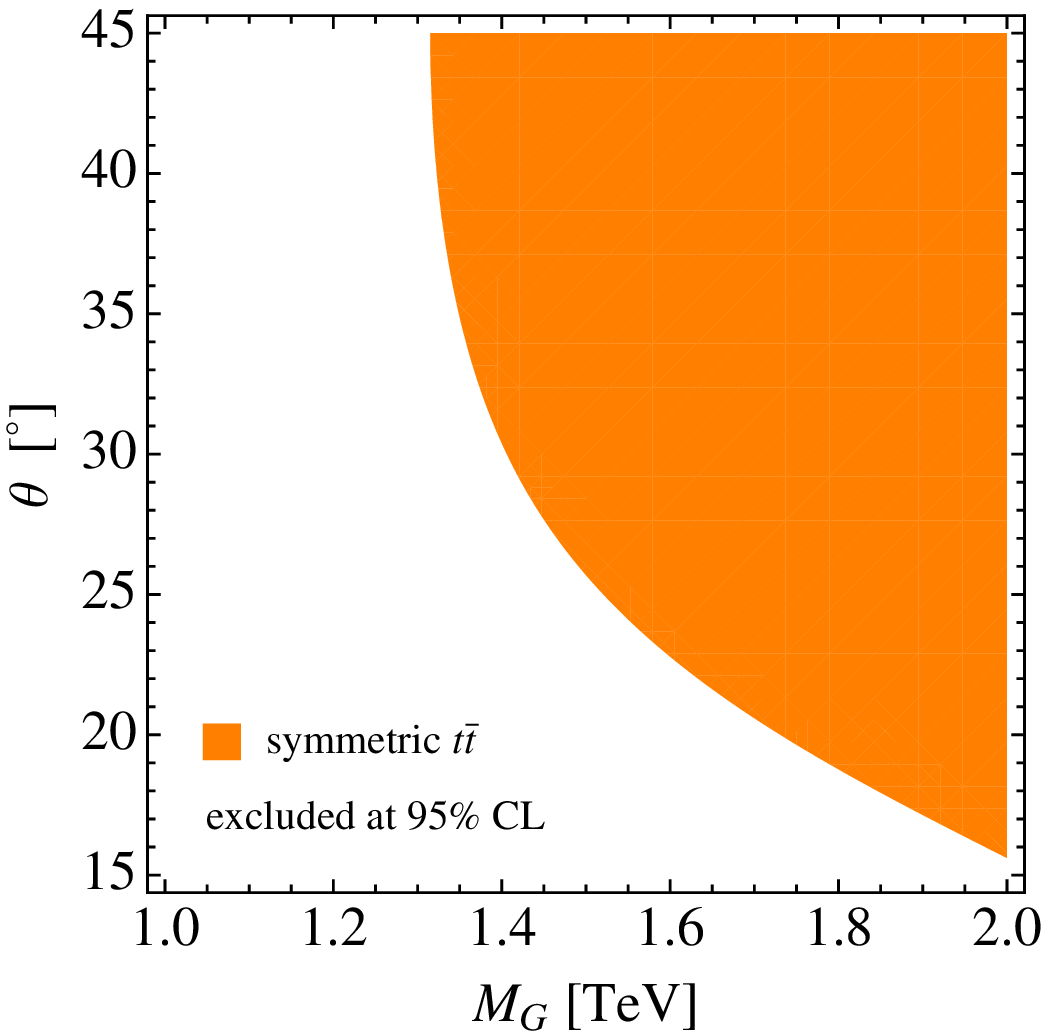}
\qquad
\includegraphics[height=7.5cm]{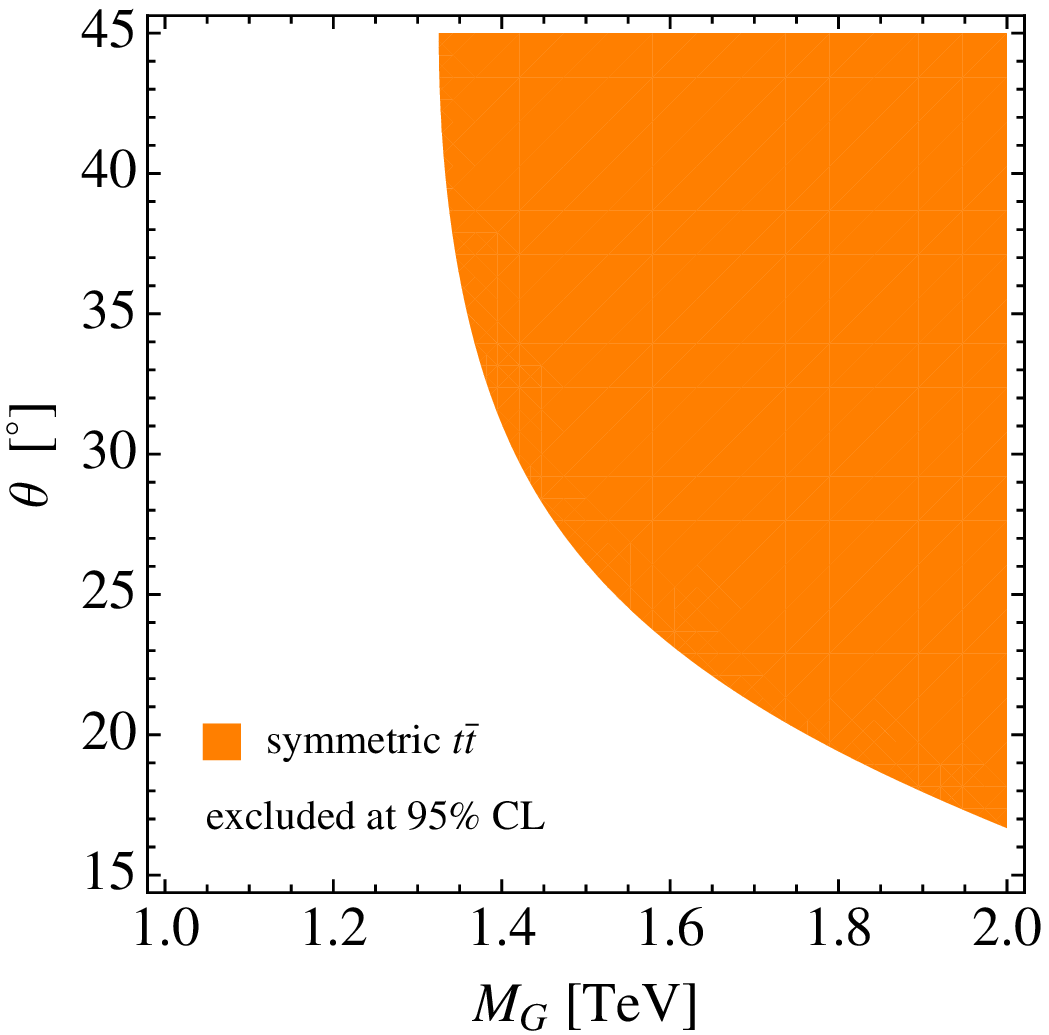}
\end{center}
\begin{center} 
  \parbox{15.5cm}{\caption{\label{fig:ttbounds1} Constraints in the
      $M_G\hspace{0.25mm}$--$\hspace{0.25mm}\theta$ plane imposed by
      the charge-symmetric observables in $t\bar t$ production in the
      flavor non-universal (left) and universal (right) axigluon
      model. The regions below the orange area are excluded at $95\%$
      CL. }}
\end{center}
\end{figure}

The constraints that derive from $t\bar t$ production are presented in
Figures \ref{fig:ttbounds1} and \ref{fig:ttbounds2}. Let us first
focus on the charge-symmetric observables, namely $\sigma_s$ and
$(d\sigma_s/dM_{t\bar t})^{>}$. The corresponding
$M_G\hspace{0.25mm}$--$\hspace{0.25mm}\theta$ planes are displayed in
the left and right panels of \Fig{fig:ttbounds1} for the case of a
flavor non-universal and universal axigluon, respectively. The regions
below the orange areas are excluded at $95\%$ CL.  For a given mixing
angle $\theta$, we derive the following mass bound on a non-universal
axigluon,
\beq \label{eq:ttsymnon} 
M_G > \left ( 0.45 + 0.47\,\tan \theta + 0.40 \, \cot \theta \right )
\! \TeV \, > \, 1.3 \TeV \,,
\eeq
while in the flavor universal case we obtain
\beq \label{eq:ttsymuni}
M_G > \left ( 0.27 + 0.61\,\tan \theta + 0.46 \, \cot \theta \right )
\! \TeV \, > \, 1.3 \TeV \,,
\eeq
at 95\% CL.  Notice that the axigluon contributions to the symmetric
$t\bar t$ cross section in both models differ only marginally as a
result of the changed axigluon width.  Due to the low sensitivity on
the width, the bound in \eq{eq:ttsymnon} is robust and to first
approximation independent from the masses of the extra quarks $u_4$
and $d_4$. We also emphasize that our fit to $\sigma_s$ and
$(d\sigma_s/dM_{t\bar t})^{>}$ excludes masses down to $0.2 \TeV$,
irrespectively of whether a flavor non-universal or universal axigluon
is considered.

\begin{figure}[!t]
\begin{center}
\includegraphics[height=7.5cm]{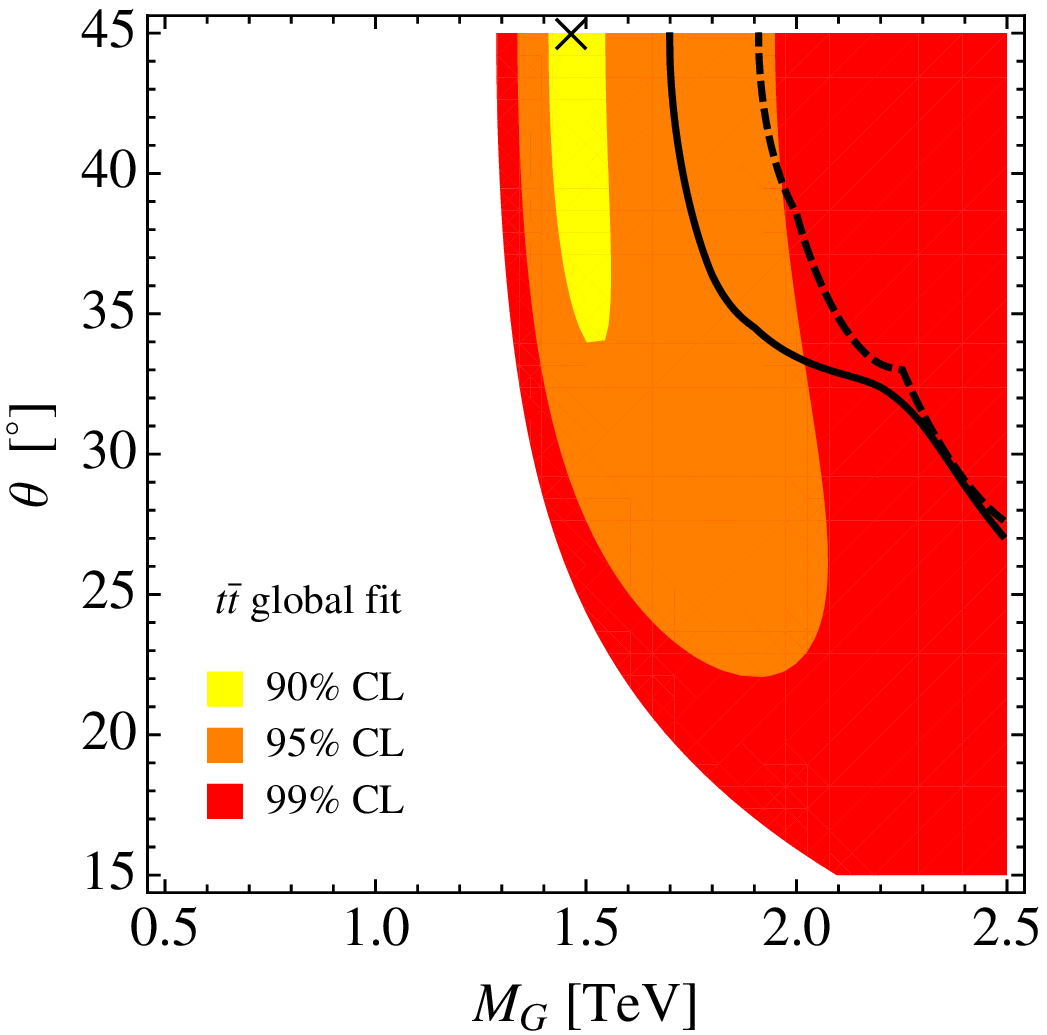}
\qquad
\includegraphics[height=7.5cm]{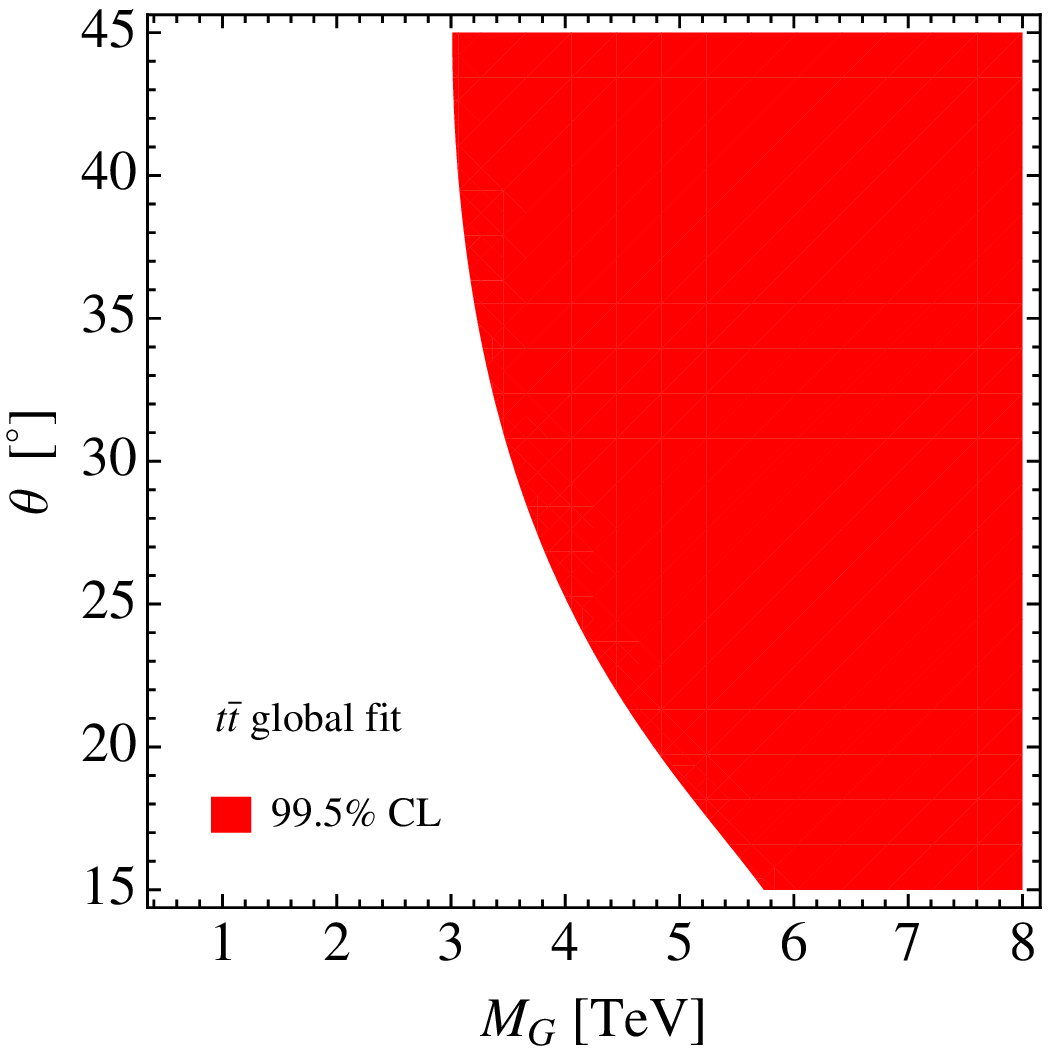}
\end{center}
\begin{center} 
  \parbox{15.5cm}{\caption{\label{fig:ttbounds2} Confidence regions in
      the $M_G\hspace{0.25mm}$--$\hspace{0.25mm}\theta$ plane
      resulting from a fit to the $t\bar t$ observables in the flavor
      non-universal (left) and universal (right) case. For comparison
      the constraints from the dijet resonance search (angular
      distribution) is also shown in the left panel as a dashed (solid)
      black line. The areas to the left of the curves are disfavored
      at 95\% CL.}}
\end{center}
\end{figure}

In order to gauge how the presence of an axigluon affects the overall
consistency of the data in the $t \bar t$ sector, we perform a
$\chi^2$ fit to the four observables $\sigma_s$, $(d\sigma_s/dM_{t\bar
  t})^{>}$, $\big (A_{\text{FB}}^t(M_{t\bar t}) \big)^{<}$, and $\big
(A_{\text{FB}}^t(M_{t\bar t}) \big )^{>}$. We treat the individual
measurements as fully uncorrelated.  The left (right) panel in
\Fig{fig:ttbounds2} shows the constraint in the
$M_G\hspace{0.25mm}$--$\theta\hspace{0.25mm}$ plane for the
non-universal (universal) axigluon model.  The white area of the
parameter space is excluded at $99\%$ CL ($99.5\%$ CL). The orange
(yellow) area in the left panel shows clearly that agreement with the
data at 95\% CL (90\% CL) is only achieved for a flavor non-universal
axigluon with a mass in the range of $1.3\TeV \lesssim M_G \lesssim
2.1\TeV$ ($1.4\TeV \lesssim M_G \lesssim 1.5\TeV$). In the global fit,
the lower bound on the axigluon mass is essentially determined by the
two charge-symmetric $t\bar t$ observables, while the upper limit
arises from the requirement to accommodate large values of $\big
(A_{\text{FB}}^t(M_{t\bar t}) \big )^{>}$. In fact, due to the tension
between the asymmetry in the high-$M_{t\bar t}$ bin with the remaining
observables, it is not possible to obtain a fit to the data with a CL
of 68\%. The best fit to the $t\bar t$ data, as indicated by the black
cross, is achieved for the axigluon parameters $M_G=1.5\TeV$ and
$\theta=45^{\circ}$. It has $\chi^2_G/\rm{ndf}=7.6/4$. Compared to the
goodness of the fit in the SM, $\chi^2_{\rm{SM}}/\rm{ndf}=13.1/4$, the
presence of a non-universal axigluon thus provides a significant
improvement from $99\%$~CL to $90\%$~CL. The central values of the
total cross section, the cross section in the highest $M_{t \bar t}$
bin, the total asymmetry in the parton frame, and the asymmetry at
large $M_{t \bar t}$ corresponding to the best-fit point are shown on
the left-hand side in \Fig{fig:bestfit}. We see that large positive
values of the $t \bar t$ asymmetries are indeed possible in the flavor
non-universal axigluon model without impairing the overall consistency
with the symmetric observables.  Relative to the SM, the shifts in
$\sigma_s$, $(d\sigma_s/dM_{t\bar t})^{>}$, $A_{\text{FB}}^t$, and
$\big (A_{\text{FB}}^t(M_{t\bar t}) \big)^{>}$ amount to around $1\%$,
$46\%$, $94\%$, and $133\%$.  For the best-fit point the central value
of the asymmetry in the low $M_{t \bar t}$ bin reads $\big (A_{\rm
  FB}^t (M_{t \bar t})\big )^{<} = 10.7 \%$. It is not included in the
figure.

\begin{figure}[!t]
\begin{center}
\includegraphics[height=5.5cm]{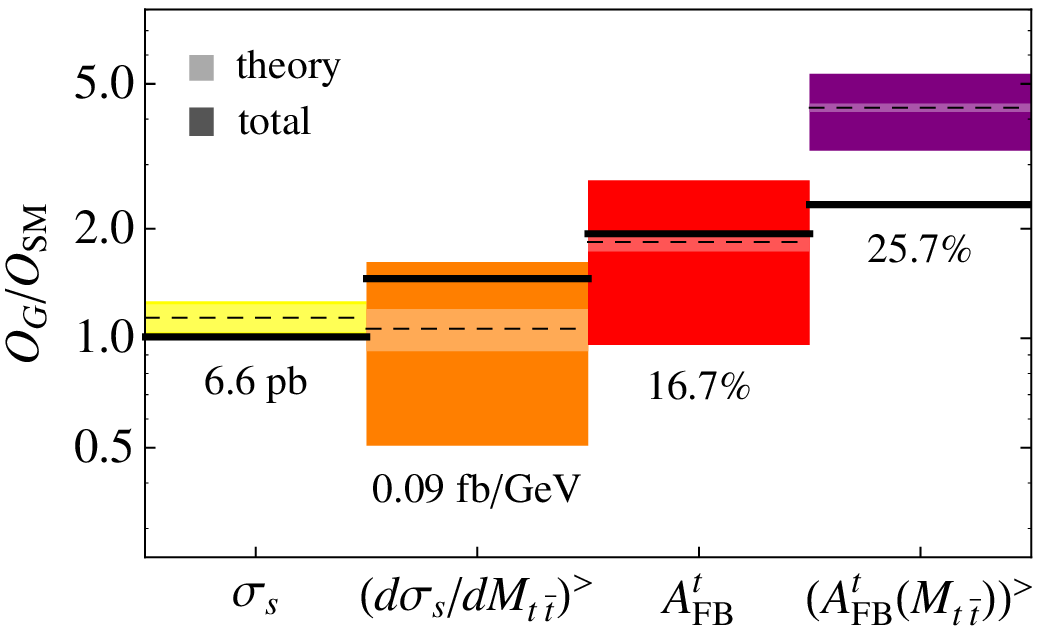}
\quad
\includegraphics[height=5.5cm]{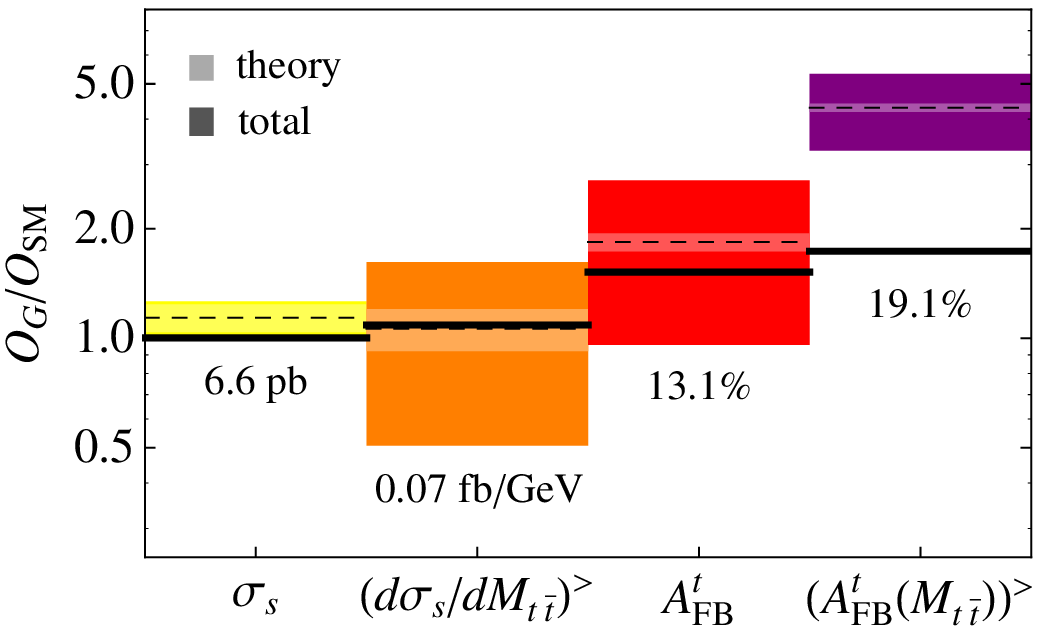}
\end{center}
\vspace{-10mm}
\begin{center} 
  \parbox{15.5cm}{\caption{\label{fig:bestfit} Predictions for the $t
      \bar t$ observables in the flavor non-universal axigluon model
      (thick black lines and numbers) corresponding to the best fit
      before (left) and after (right) including dijet constraints. For
      comparison the central values of the associated measurements
      (dashed black lines) are also shown. All predictions have been
      normalized to the SM expectations.  The light colored bars
      illustrate the theoretical uncertainties, while the combined
      experimental and theoretical errors as displayed in bright
      colors. See text for further details.}}
\end{center}
\end{figure}

The region of parameter space that is preferred by the $t\bar t$ data
in the case of a flavor non-universal axigluon is consistent with the
indirect constraints from flavor physics and EWPOs that have been
discussed above. Large parts of it are however disfavored by the
recent results on dijet production at the LHC. This feature is
illustrated by the dashed and solid black lines in the left panel of
\Fig{fig:ttbounds2} corresponding to the 95\%~CL limits from the
resonance search and the angular distribution measurement in dijet
production at ATLAS. These constraints exclude the best-fit solution
and allow only for a reduced compatibility with the $t \bar t$ data at
$95\%$~CL. Notice that, by taking into account the constraints from
dijet production, the axigluon mass is confined to the narrow range of
$1.9\TeV \lesssim M_G \lesssim 2.0\TeV$. The maximal axigluon effects
in the asymmetries are thus obtained for $M_G=1.9\TeV$ and
$\theta=45^{\circ}$.  The predictions for the relevant $t \bar t$
observables corresponding to this set of parameters are presented on
the right in \Fig{fig:bestfit}.  The quoted values correspond to
$\chi^2_G/\rm{ndf} = 9.3/4$, still yielding a considerably better fit
than within the SM.  With respect to the SM prediction the symmetric
observables $\sigma_s$ and $(d\sigma_s/dM_{t\bar t})^{>}$ change by
$0.2\%$ and $9\%$. Both the total charge asymmetry as well as the
prediction for the high-$M_{t \bar t}$ bin exhibit a large enhancement
of $52\%$ and $73\%$ relative to the SM.  It is important to realize
that, compared to the best-fit values, the asymmetries are not only
smaller, but also the value of the cross section in the highest $M_{t
  \bar t}$ has come down. This shows the strong positive correlation
between these observables.  The forward-backward asymmetry at low
$M_{t \bar t}$ amounts to $9.1\%$ for the latter set of parameters. We
conclude from these numbers, that an axigluon with QCD-like couplings
to all quarks, able to explain the top forward-backward asymmetry and
to simultaneously evade the bounds from dijet production, should have
a mass of $2\TeV$ and hence is expected to be soon discovered at the
LHC.

\begin{figure}[!t]
\begin{center}
\includegraphics[height=7.5cm]{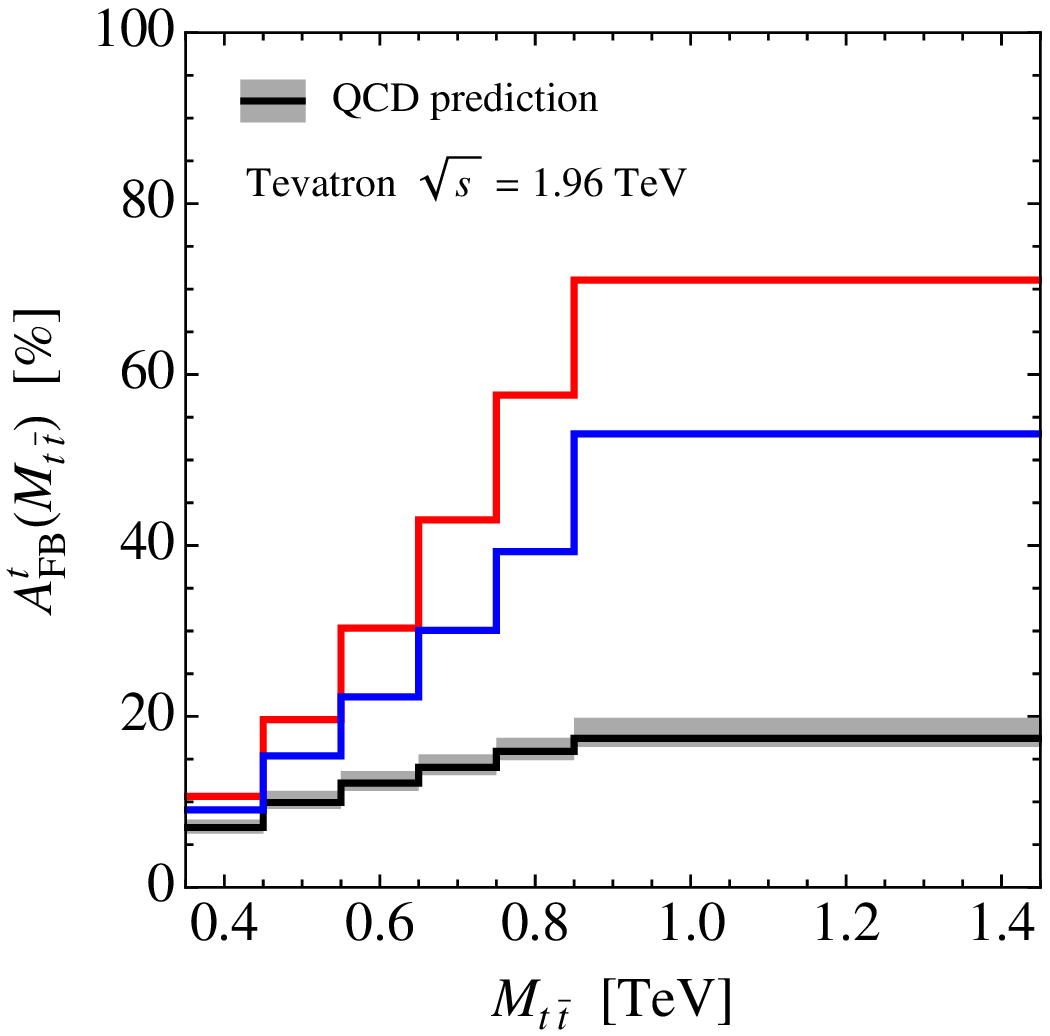}
\qquad 
\includegraphics[height=7.5cm]{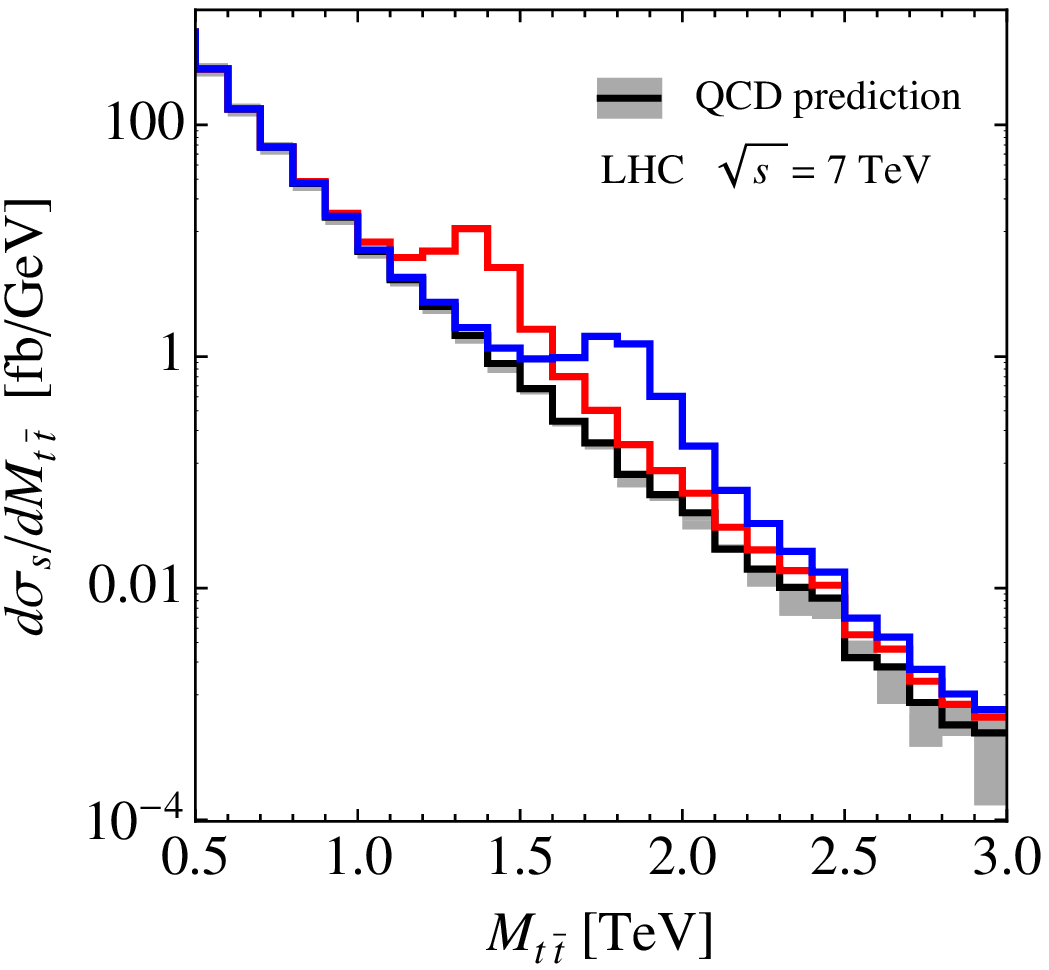}
\end{center}
\begin{center} 
  \parbox{15.5cm}{\caption{\label{fig:predictions} Left: Prediction
      for the $t \bar t$ forward-backward asymmetry at the Tevatron as
      a function of $M_{t \bar t}$. The red (blue) curve represents
      the result for a flavor non-universal axigluon with $\theta =
      45^\circ$ and $M_G = 1.5 \TeV$ ($M_G = 1.9 \TeV$). The central
      value and uncertainty of the QCD expectation are indicated by
      the black line and the gray band, respectively.  Right:
      Prediction for the $t \bar t$ cross section differential in
      $M_{t \bar t}$ for $pp$ collisions at $\sqrt{s} = 7 \TeV$. The
      red (blue) line corresponds to a flavor non-universal axigluon
      with $\theta = 45^\circ$ and $M_G = 1.5 \TeV$ ($M_G = 1.9
      \TeV$). For comparison the SM expectation with its uncertainty
      is also shown (black line and gray band).}}
\end{center}
\end{figure}

In the left panel of \Fig{fig:predictions}, we show the prediction for
the $M_{t \bar t}$ spectrum of the asymmetry at the Tevatron
corresponding to the best-fit solution before (red curve) and after
(blue curve) imposing the latest dijet constraints.  The QCD
prediction with its theoretical uncertainty, obtained by varying the
renormalization and factorization scales in the range $\mu_r = \mu_f
\in [m_t/2,2m_t]$, is displayed as a black line on top of a gray
band. We see that for the preferred axigluon parameters $A_{\rm FB}^t
(M_{t \bar t})$ is predicted to be strictly positive and a factor of
around 3 to 4 above the SM expectation for $M_{t \bar t} > 0.85
\TeV$. Notice that the former behavior is characteristic for the
presence of a heavy $s$-channel resonance with $g_A^q \hspace{0.5mm}
g_A^t < 0$. In fact, the interference term in $A_{\rm FB}^t (M_{t \bar
  t})$ only changes sign across a resonance, switching from positive
to negative (negative to positive) values in going from $M_{t\bar t}$
below to above the pole for $g_A^q \hspace{0.5mm} g_A^t < 0$ ($g_A^q
\hspace{0.5mm} g_A^t > 0$). Given the high sensitivity of the
differential asymmetry to the precise nature of the underlying
physics, a more precise measurement of this quantity should be a
primary goal of the Tevatron experiments and pursued with great vigor.
Further information on the possible origin of the anomaly in $A_{\rm
  FB}^t$ is expected to come in the near future from the measurement
of the distribution of the $t \bar t$ cross section at the LHC. On the
right of \Fig{fig:predictions}, we depict the prediction for
$d\sigma_s/dM_{t \bar t}$ corresponding to the best-fit solution
before (red curve) and after (blue curve) incorporating the ATLAS
dijet constraints. The QCD prediction including its scale variation is
displayed for comparison (black line and gray band). The relatively
wide axigluon resonances are clearly visible in the figure.  By
integrating over the range $M_{t \bar t} \in [1.0, 3.0] \TeV$, we find
that the presence of an axigluon with $\theta = 45^\circ$ and $M_G =
1.5 \TeV$ ($M_G = 1.9 \TeV$) leads to an enhancement of the tail of
the spectrum by a factor of around 2.5 (1.3).  From this observation,
we conclude that if the anomaly in $A_{\rm FB}^t$ persists and is due
to a new heavy color-octet resonance, then a notable enhancement of
the $t \bar t$ cross section with respect to the SM should be observed
at the LHC. The synergy/complementarity of Tevatron and LHC
measurements has also been stressed in \cite{Han:2010rf, Bai:2011ed,
  AguilarSaavedra:2011vw, Hewett:2011wz, AguilarSaavedra:2011hz}.

It is possible to circumvent the dijet constraints by relaxing the
conditions on the axigluon couplings present in the simplest chiral
color model. Since the $t\bar t$ observables are mostly sensitive to
the product of the couplings to the light quarks and the top quark, a
simple rescaling 
\beq \label{eq:rescaling} 
g_{L,R}^q \, \rightarrow \,\xi \,g_{L,R}^q\,,\qquad g_{L,R}^t \,
\rightarrow \, g_{L,R}^t/\xi \,,
\eeq 
leaves the goodness of the global fit essentially unchanged.  This
freedom can be exploited to suppress the couplings to light quarks
that govern effects in dijet production. In order to reach the
axigluon best-fit point for $t\bar t$ production around $M_G=1.5\TeV$
and simultaneously evade the bounds from resonance searches in dijet
production, one needs a suppression factor of $\xi \lesssim 0.85$.
For the minimal suppression the resulting couplings are hence
$g_L^q=-g_R^q \approx 0.85$ and $g_L^t=-g_R^t \approx -1.2$, which
implies that $\Gamma_G/M_G \approx 9\%$, if only SM quark decay modes
are open. For such a resonance, we find $A_{\rm FB}^t = 13.7\%$, $\big
(A_{\rm FB}^t (M_{t \bar t}) \big )^{<} = 10.7\%$, and $\big (A_{\rm
  FB}^t (M_{t \bar t}) \big )^{>} = 25.6\%$. The total cross section
is SM-like, while $(d \sigma_s/dM_{t \bar t})^{>} = 0.09 \, {\rm
  fb}/{\rm GeV}$ shows the previously mentioned characteristic
enhancement with respect to the QCD prediction. The corresponding
results for $A_{\rm FB}^t (M_{t \bar t})$ and $d \sigma_s/dM_{t\bar
  t}$ essentially resemble the blue curves in the left and right
panels of \Fig{fig:predictions}.  Relaxing the conditions further to
$|g_L^q|\neq|g_R^q|$ and/or $|g_L^t|\neq|g_R^t|$ does not improve the
fit to the $t\bar t$ data, since it enhances the effects on the
symmetric cross section for fixed contributions to the asymmetry. From
these general considerations we conclude that the predictions of the
asymmetries presented in the left panel of \Fig{fig:bestfit}
represent, in fact, the maximal values that can be obtained in any
model of new physics, where $t \bar t$ production receives the
dominant corrections from $s$-channel exchange of a color-octet
resonance.\footnote{It might be possible to lever out these arguments
  and evade the constraints from both $t \bar t$ and dijet production
  by introducing more than one color-octet boson with almost
  degenerate masses and arrange for the couplings so that the
  individual contributions in the symmetric observables cancel each
  other to a large extent.}

Let us finally also briefly comment on the global fit to the $t\bar t$
observables in presence of an axigluon with flavor universal couplings
to quarks. The allowed parameter region is displayed in the right
panel of \Fig{fig:ttbounds2}.  While the effects on the symmetric
observables are essentially the same as in the non-universal case, the
contributions to the asymmetric cross section now interfere
destructively with QCD, driving the asymmetry $\big
(A_{\text{FB}}^t(M_{t\bar t}) \big )^{>}$ below the SM value. A flavor
universal axigluon is therefore excluded at $99.5\%$ CL by the global
fit to the $t\bar t$ data unless its mass is above $3.0 \TeV$. Notice
also that the obtained fit is always worse than the one in the SM, and
that the $t \bar t$ observables provide a constraint that is notably
stronger than the bound that follows from dijet production as well as
from any indirect constraints.

\section{Conclusions}
\label{sec:concl}
\vspace{2mm}

In this article we have performed a comprehensive study of direct and
indirect constraints on massive color-octet vector bosons and their
impact on top-quark pair production.  Motivated by the observation of
large effects in asymmetric $t\bar t$ production at the Tevatron, we
have focused on axigluons with flavor non-universal couplings to
quarks in the framework of chiral color. However, it is
straightforward to apply our general results and formulas to other
strongly-coupled $s$-channel resonances with arbitrary couplings to
quarks.

We have pointed out that any model with a heavy color-octet boson that
possesses flavor non-universal couplings to quarks inevitably features
FCNC interactions at tree level. In the case of generic flavor
violation in the down-quark sector, we found that neutral meson mixing
rules out axigluons with masses below several $\TeV$.  Without
specifying an underlying theory that fixes the pattern of flavor
breaking, the FCNC effects can however be confined to the up-quark
sector and aligned such that they are of MFV type.  Minimal
constraints are thus derived from the mass difference in the neutral
$D$-meson sector and yield the rather weak bound of $M_G >
0.22\TeV$. This bound has to be fulfilled by any axigluon with flavor
non-universal couplings of QCD strength to quarks. In contrast, models
with flavor universality are not constrained by flavor physics at all.

The precision observables at the $Z$ pole are sensitive to one-loop
effects in the $Zq\bar q$ couplings associated to the virtual exchange
of massive color-octet bosons. Constraints from the bottom-quark POs
are largely avoided if $|g_L^b|=|g_L^q|$ and $|g_L^b|=|g_R^b|$.  In
the axigluon model these conditions are both fulfilled for a mixing
angle of $\theta = 45^{\circ}$. Axigluon effects in the total
$Z$-boson decay width $\Gamma_Z$ and the hadronic cross section
$\sigma_{\rm had}$, on the other hand, cannot be decoupled for any
choice of the mixing angle. From the combined fit to all five
$Z\rightarrow q\bar q$ observables, we derive the mass bounds $M_G >
0.54\TeV$ in the flavor non-universal and $M_G > 0.40\TeV$ in the
universal axigluon model. Colored resonances with QCD-like couplings
to all quark flavors and a mass around the electroweak scale are
therefore highly disfavored. In fact, the presence of an axigluon
worsens the quality of the global fit to the $Z q\bar q$ couplings
with respect to the SM and, in particular, does not allow to reduce
the long-standing discrepancy in the bottom-quark asymmetry $A_{\rm
  FB}^b$.

Heavy-gluon corrections affect the Peskin-Takeuchi parameters $S$ and
$T$ first at the two-loop level. Since the $T$ parameter measures
isospin violation, axigluon effects to it are enhanced relative to $S$
by a chiral factor $m_t^2/M_Z^2$. In the flavor non-universal axigluon
model, the presence of a sequential fourth generation of fermions
leads to non-zero one-loop effects to $S$ and $T$, making it
impossible to derive a model-independent bound on the parameter
space. For a given set of typical fourth generation parameters, it
however turns out that the oblique parameters probe axigluon masses at
and slightly above the scale of EWSB. In the case of an axigluon with
flavor universal QCD-like couplings, the absence of extra fermions
allows to derive a firm bound on the axigluon mass of $M_G >
0.16\TeV$. The limits from isospin violation can however become
competitive or even more restrictive than those arising from the other
indirect constraints, if the top quark couples very strongly to the
new colored boson.  In particular, in models where the right-handed
top quark is fully composite, $T$ can receive unacceptably large
positive corrections even for axigluon masses in the TeV range.

\begin{figure}[!t]
\begin{center}
\includegraphics[height=4cm]{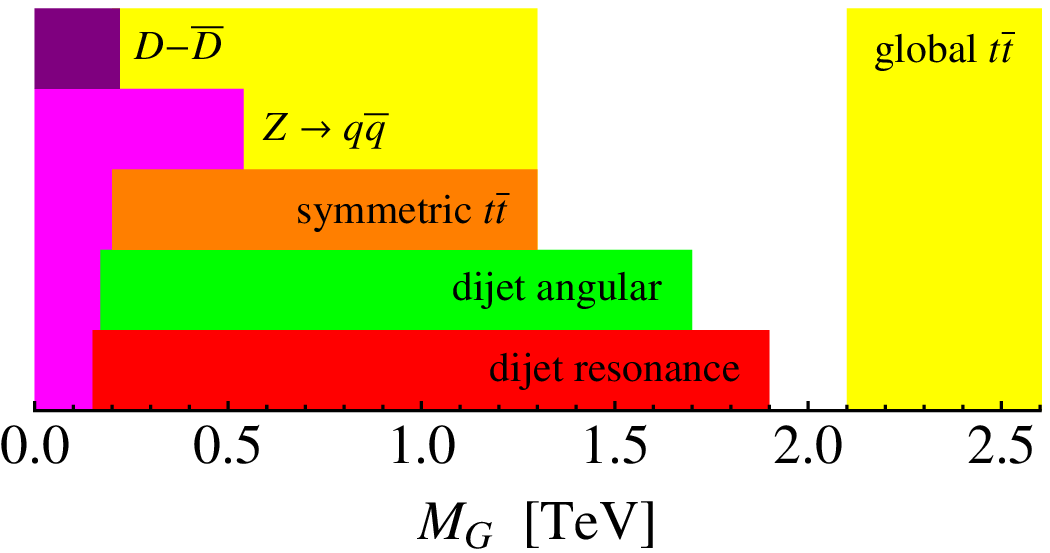}
\qquad
\includegraphics[height=4cm]{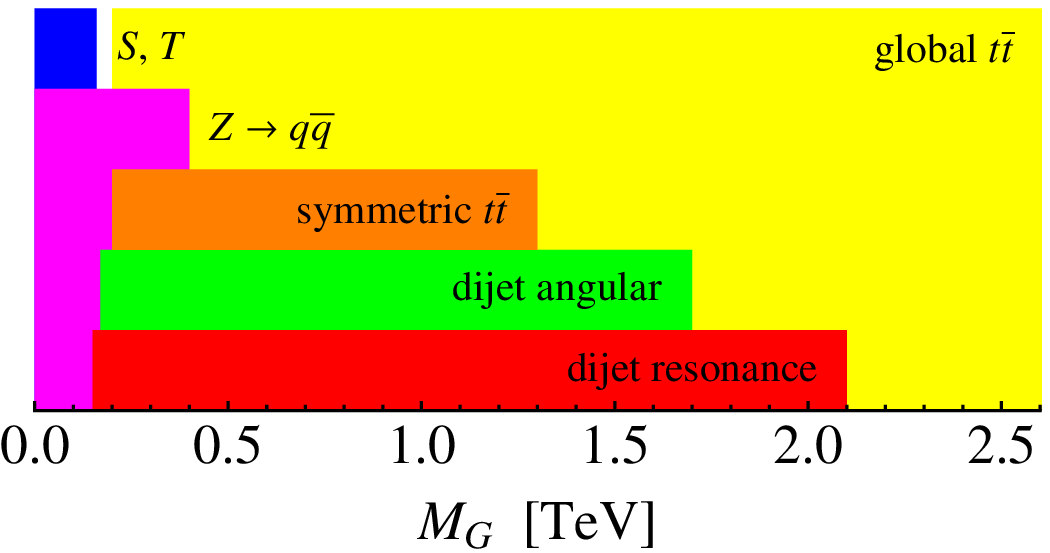}
\end{center}
\begin{center} 
  \parbox{15.5cm}{\caption{\label{fig:lego} Bounds on $M_G$ in the
      flavor non-universal (left) and universal (right) axigluon
      model. Colored regions are excluded at $95\%$ CL by $D$--$\bar
      D$ mixing (purple), $Z\rightarrow q\bar q$ (magenta), $S$ and
      $T$ (blue), $\sigma_s$ and $(d\sigma_s/dM_{t\bar t})^{>}$
      (orange), dijet angular distributions (green) and resonances
      searches (red) at the LHC, and by the global fit to the Tevatron
      $t\bar t$ data (yellow).}}
\end{center}
\end{figure}

The most stringent constraints on the axigluon parameter space stem
from the production of massive color-octet bosons decaying to dijets
at the LHC. Searches for narrow resonances in the dijet invariant mass
spectrum exclude the mass window $0.6\TeV < M_G < 1.9\TeV$ in the
flavor non-universal model. The lower bound is due to the limited
detector sensitivity to dijet events with small invariant masses
$M_{jj}$. Earlier resonance searches exclude axigluon masses down to
around $0.15\TeV$. In the flavor universal model, the axigluon width
$\Gamma_G$ is smaller due to the absence of the fourth generation,
which results in a stronger exclusion limit of $0.6 \TeV <M_G <
2.1\TeV$. These bounds are only relevant in the case of narrow
resonances with $\Gamma_G/M_G \lesssim 15\%$. Constraints from the
angular distribution of dijets are applicable for broad resonances as
well. In this context, we have shown that the inclusion of width
effects is important to obtain reliable bounds from the analysis of
the angular distribution, since the shape of the centrality ratio of
the jet pairs depends significantly on them. We found that, by probing
the enhancement of jets in the central region, one can exclude the
mass range of $0.17\TeV < M_G < 1.7\TeV$ for both the case of a flavor
non-universal and universal axigluon. The excluded range extends well
below the region $M_{jj} > 0.5 \TeV$ directly probed by ATLAS, because
light axigluons enhance the central activity also at $M_{jj} \gg M_G$.

We have finally studied the possible effects of a massive color-octet
vector boson in top-quark pair production. Our analysis shows that
there is a generic tension between having large effects in the
asymmetric and small effects in the symmetric $t \bar t$ cross
sections, because these observables are positively correlated. While
the anomalously large forward-backward asymmetry can in principle be
accommodated by a color octet with strong flavor non-universal
axial-vector couplings to quarks and a mass $M_G \approx 1\TeV$, such
a possibility is disfavored by a combined fit to the total cross
section $\sigma_s$ and its spectrum $(d\sigma_s/dM_{t\bar t})^{>}$ at
high invariant mass, which excludes masses in the range $0.2\TeV < M_G
< 1.3\TeV$, independently from flavor (non-)universality.  The tension
between the different observables also limits the goodness of a global
$t \bar t$ fit. For masses satisfying $1.3\TeV < M_G < 2.1\TeV$,
agreement with the Tevatron data is found at $95\%$ CL. The best fit
has $\chi^2/{\rm ndf} = 7.6/4$ (corresponding to 90\% CL) and is
achieved for a flavor non-universal axigluon with $M_G=1.5\TeV$ and
$\theta=45^{\circ}$.  This solution hence represents a significant
improvement with respect to the SM ($\chi^2/{\rm ndf} = 13.1/4$),
giving rise to asymmetries of $A_{\rm FB}^t=16.7\%$ and $\big (A_{\rm
  FB}^t (M_{t\bar t}) \big )^{>} = 25.7\%$. A flavor universal
axigluon, on the other hand, does not provide an acceptable fit to the
$t\bar t$ data, since it fails to yield large positive values of the
asymmetry.

The various mass constraints are summarized in \Fig{fig:lego} for the
flavor non-universal (left) and universal (right) axigluon model. In
the non-universal case the best-fit point for the $t\bar t$
observables at $M_G=1.5\TeV$ is at variance with the constraints
arising from dijet production, resulting in a narrow window of allowed
masses around $2\TeV$.  Axigluon effects in the $t\bar t$ asymmetry
are in this mass window reduced to at most $A_{\rm FB}^t=13.1\%$ and
$\big (A_{\rm FB}^t (M_{t \bar t}) \big )^{>} = 19.1\%$.  We have also
emphasized that the bounds from dijet production can be evaded by
relaxing the conditions on the axigluon couplings arising in models of
chiral color. For the rescaled couplings $g_L^q=-g_R^q \approx 0.85$
and $g_L^t=-g_R^t \approx -1.2$, the best-fit point to the $t\bar t$
data at $M_G=1.5\TeV$ becomes viable again.  It however turns out that
the fit cannot be further improved by allowing for
$|g_L^q|\neq|g_R^q|$ and/or $|g_L^t|\neq|g_R^t|$. These general
arguments imply that the presence of a single color-octet boson cannot
raise the prediction for the top-quark forward-backward asymmetry
$\big (A_{\rm FB}^t (M_{t \bar t}) \big )^{>}$ above $26\%$. In
conclusion, our findings suggest that a consistent explanation of the
Tevatron $t\bar t$ data by a massive color-octet boson predicts a
resonance of $1.5 \TeV$ to $2\TeV$ in the spectrum of top-quark pair
production at the LHC. Combined with improved Tevatron determinations
of the top-quark properties and an explicit measurement of the $t \bar
t$ charge asymmetry by ATLAS, CMS, and possibly LHCb, this should
allow to confirm or rule out the axigluon hypothesis in the near
future.

\subsubsection*{Acknowledgments}

It is a pleasure to thank Giulia Zanderighi for providing the analytic
result for one of the integrals needed for the calculations presented
in this article. Useful discussions and correspondence with Roberto
Barcelo, Mads Frandsen, Barbara J\"ager, Alex Kagan, J\"urgen
Rohrwild, and Jose Santiago are also acknowledged.  We are finally
grateful to Volker B\"uscher, Regina Demina, Sebastian Eckweiler,
Rhorry Gauld, Adam Gibson, Weina Ji, Lucia Masetti, Andrei Nomerotski,
and Christian Schmitt for helpful discussions about experimental
aspects of top-antitop and/or dijet production.

\begin{appendix}

\section{\boldmath Form Factors for $Z \to q \bar q$ \unboldmath}
\label{app:Zqq}

\renewcommand{\theequation}{A\arabic{equation}}
\setcounter{equation}{0}

Below we present the analytic expressions for the one-loop $Z \to q
\bar q$ form factor describing the virtual exchange of a color-octet
spin-one state. For light quarks ($q = u,d,s,c,b$) the given
expressions correspond to the limit of an on-shell $Z$ boson and
on-shell external quarks with vanishing mass, while in the case of the
top quark we evaluate the form factor at zero momentum
transfer. Throughout the calculation we will ignore possible
flavor-violating effects.

The one-loop axigluon corrections to the couplings of the $Z$ boson to
light quarks lead to a shift in the tree-level couplings $g_{P}^{q}$
with $P = L,R$. In the case of the bottom quark, the resulting
effective couplings can be written as
\beq \label{eq:calG}
{\cal G}_{P}^{b} = g_{P}^{b} \left [1 + \frac{\alpha_s}{4\pi} \; C_F
  \left ( g_P^h \right)^2 {\cal K} (x_Z) \right ] \,,
\eeq
where the real and imaginary parts of the complex form factor ${\cal
  K} (x_Z)$ take the following form
\beq \label{eq:RefImf}
\begin{split}
  {\rm Re} \hspace{0.5mm} {\cal K} (x_Z) \ & = -\frac{4+ 7 x_Z}{2
    x_Z}+ \frac{2+ 3 x_Z }{x_Z} \, \ln x_Z - \frac{2 \left (1 +
      x_Z\right )^2}{x_Z^2} \, \big [ \ln x_Z \, \ln (1+x_Z) +
  \text{Li}_2(-x_Z) \big ] \,, \\
  {\rm Im} \hspace{0.5mm} {\cal K} (x_Z) & = \left[-\frac{2 + 3
      x_Z}{x_Z} + \frac{2 \left (1 +x_Z \right )^2 }{x_Z^2} \, \ln
    (1+x_Z) \right] \pi \,.
\end{split}
\eeq
The above result agrees with the analytic expression given in
\cite{Carone:1994aa, Holdom:1995id}. Here $x_Z = M_Z^2/M_G^2$ and
$\text{Li}_2 (z) = -\int_0^z \!  dt \, \ln (1-t)/t$ denotes the
dilogarithm. We remark that in the decoupling limit $x_Z \to 0$, the
real and imaginary parts of the form factor ${\cal K} (x_Z)$ behave as
\beq \label{eq:ReKImKlim} 
{\rm Re} \hspace{0.5mm} {\cal K} (x_Z) \, \to \, x_Z \left(
  -\frac{2}{3} \, \ln x_Z + \frac{11}{9} \right) \,, \qquad {\rm Im}
\hspace{0.5mm} {\cal K} (x_Z) \, \to \, \frac{2\pi}{3} \, x_Z \,.
\eeq 
Notice finally that the imaginary components which are not
logarithmically enhanced, have been ignored in the discussion of the
$Z \to q \bar q$ constraints on the parameter space of the axigluon
model. In particular, the expression \eq{eq:calGbapprox} contains only
the leading-logarithmic corrections.

In the case of the top quark, we find instead 
\beq \label{eq:GAt}
{\cal G}_{P}^{t} = g_{P}^{t} \pm \frac{g}{c_w} \,
\frac{\alpha_s}{4\pi} \, C_F \left ( g_P^h \right )^2 \, G(x_t) \,,
\eeq 
where the plus (minus) sign applies in the case $P = L$ ($P=R$) and
$x_t = m_t^2/M_G^2$. The short-distance coefficient $G(x_t)$ is given
by
\beq \label{eq:Gfun}
G(x_t) = \frac{x_t}{1-x_t} + \frac{x_t}{(1-x_t)^2} \, \ln x_t \,.
\eeq
It obeys $G(x_t) \to x_t \left (\ln x_t + 1 \right )$ for $x_t \to 0$. 

\section{\boldmath Corrections to Oblique Parameters
    \unboldmath}
\label{app:oblique}

\renewcommand{\theequation}{B\arabic{equation}}
\setcounter{equation}{0}

In this appendix we give the analytic results for the two-loop
corrections to the Peskin-Takeuchi parameters $S$ and $T$ arising in
the considered axigluon model. The calculation has been performed in
the on-shell renormalization scheme and the bottom-quark mass has been
neglected throughout. Effects from the fourth generation of fermions,
needed to make the model anomaly-free, are ignored in the
following. Finally, flavor-violating effects that are strongly CKM
suppressed in the case of flavor alignment are neglected.

Our result for the $S$ parameter reads
\beq \label{eq:S}
S = \frac{2}{9 \pi \hspace{0.25mm}} \, \frac{ \alpha_s}{4 \pi} \, C_F
N_c \, \Big [ (g_L^h)^2 \, s_{LL} (x_t) + (g_R^h)^2 \, s_{RR} (x_t) +
g_L^h g_R^h \, s_{LR} (x_t) \Big ] \,,
\eeq
where the functions $s_{LL} (x_t)$, $s_{RR} (x_t)$, and $s_{LR} (x_t)$ are
given by
\beq  \label{eq:sLsRsLR}
\begin{split}
  s_{LL} (x_t) & = \frac{2 -21 x_t +80x_t^2-152x_t^3+60 x_t^4}{8
    \hspace{0.5mm} (1-4 x_t)^2 \hspace{0.25mm} x_t} \\ & \phantom{xx}
  + \frac{1 - 16 x_t +98 x_t^2 - 240 x_t^3 + 30 x_t^4 + 400 x_t^5 -
    120 x_t^6}{8 \hspace{0.5mm} (1- 4 x_t)^3 \hspace{0.25mm} x_t^2} \,
  \ln x_t \\ & \phantom{xx} + \frac{1-2 x_t }{4 x_t} \, \psi (x_t) +
  \frac{4x_t -29 x_t^2 +34 x_t^3 - 18 x_t^4 +60 x_t^5}{4
    \hspace{0.25mm} (1-4 x_t)^3} \; \phi \left(\frac{1}{4
      x_t}\right)\,, \\
  s_{RR} (x_t) & = \frac{2 -21 x_t +76x_t^2-136x_t^3+60 x_t^4}{8
    \hspace{0.5mm} (1-4 x_t)^2 \hspace{0.25mm} x_t} \\ & \phantom{xx}
  + \frac{1 - 16 x_t +98 x_t^2 - 240 x_t^3 + 6 x_t^4 + 496 x_t^5 - 120
    x_t^6}{8 \hspace{0.5mm} (1- 4 x_t)^3 \hspace{0.25mm} x_t^2} \, \ln
  x_t \\ & \phantom{xx} + \frac{1-2 x_t }{4 x_t} \, \psi (x_t) +
  \frac{8x_t -79 x_t^2 +230 x_t^3 - 258 x_t^4 +60 x_t^5}{4
    \hspace{0.25mm} (1-4 x_t)^3} \; \phi \left(\frac{1}{4
      x_t}\right) \,, \\
  s_{LR} (x_t) & = \frac{8 -79 x_t +178x_t^2-60x_t^3}{4 \hspace{0.5mm}
    (1-4 x_t)^2} \\ & \phantom{xx} + \frac{2 - 28 x_t +131 x_t^2 - 256
    x_t^3 + 118 x_t^4 + 60 x_t^5}{2 \hspace{0.5mm} (1- 4 x_t)^3
    \hspace{0.25mm} x_t} \, \ln x_t \\ & \phantom{xx} + 2
  \hspace{0.5mm} \psi (x_t) - \frac{9x_t^2 -42 x_t^3 +54 x_t^4 + 30
    x_t^5}{(1-4 x_t)^3} \; \phi \left(\frac{1}{4 x_t}\right) \,.
\end{split}
\eeq
The function $\psi(z)$ arises from the one-loop top-quark selfenergy
entering the counterterm contribution and takes the form
\beq \label{eq:psi} 
\psi (z) = \begin{cases} \displaystyle \frac{\sqrt{1 - 4 z}}{z} \,
  \tanh^{-1} \left(\sqrt{1 - 4 z}\right) \,, & z < 1/4 \,, \\[4mm]
  \displaystyle -\frac{\sqrt{4 z - 1}}{z} \left [ \, \tan
    ^{-1}\left(\frac{2 z-1}{\sqrt{4 z-1}}\right) + \cot
    ^{-1}\left(\sqrt{4 z-1}\right) \right ], & z > 1/4 \,. \end{cases}
\eeq
The function $\phi(z)$ stems from the two-loop scalar tadpole
with two different masses. The corresponding analytic expression reads
\cite{Davydychev:1992mt}
\bea \label{eq:phi}
\phi (z) = \begin{cases} \, 4\hspace{0.25mm} \displaystyle
  \sqrt{\frac{z}{1-z}} \, \text{Cl}_2 \left( 2 \sin^{-1}
    \left(\sqrt{z}\right)\right) , & z < 1 \,, \\[4mm]
  \frac{\displaystyle -4 \hspace{0.25mm} \text{Li}_2 \!
    \left(\frac{1}{2} \! \hspace{0.25mm} \left[1-\sqrt{1-
          z^{-1}}\right]\right)+2 \ln ^2\left(\frac{1}{2} \!
      \hspace{0.25mm} \left[1-\sqrt{1-z^{-1}}\right]\right) -\ln ^2(4
    z)+\frac{\pi ^2}{3}}{\displaystyle \sqrt{1- z^{-1}}} \, , & z > 1
  \,, \end{cases}
\eea
where $\text{Cl}_2 (z) = {\rm Im} \left [\text{Li}_2 (e^{iz})\right ]$
denotes the Clausen function.

The formulas in \eq{eq:sLsRsLR} are not very illuminating. We
therefore also give explicit expressions for the coefficient functions
in the limit $x_t \to 0$, corresponding to an infinitely heavy
axigluon. We obtain
\beq \label{eq:sLtRsLRexp}
\begin{split}
  s_{LL} (x_t) & \, \to \, x_t \left( \ln ^2 x_t + 5 \hspace{0.25mm} \ln x_t +
    \frac{\pi^2}{3} + \frac{7}{6} \right) , \\[2mm]
  s_{RR} (x_t) & \, \to \, x_t \left( 2 \ln ^2 x_t + 5 \hspace{0.25mm} \ln x_t
    + \frac{2}{3} \, \pi^2 + \frac{2}{3} \right) , \\[2mm]
  s_{LR} (x_t) & \, \to \, x_t \left ( -\frac{9}{2} \, \ln x_t -
    \frac{11}{4} \right) \, ,
\end{split}
\eeq
which implies that, if one is interested only in the
leading-logarithmic behavior for $x_t \to 0$, one simply has $s_{LL}
(x_t) \approx x_t \ln^2 x_t$, $s_{RR} (x_t) \approx 2 \hspace{0.25mm}
s_{LL} (x_t)$, and $s_{LR} (x_t) \approx 0$. These approximate
formulas have been used in \eq{eq:STapprox}.

In the case of the $T$ parameter, we find instead 
\beq \label{eq:T}
T = \frac{m_t^2}{8 \pi \hspace{0.25mm} s_w^2 c_w^2 \hspace{0.25mm}
  M_Z^2} \, \frac{ \alpha_s}{4 \pi} \, C_F N_c \, \Big [ (g_L^h)^2 \,
t_{LL} (x_t) + (g_R^h)^2 \, t_{RR} (x_t) + g_L^h g_R^h \, t_{LR} (x_t)
\Big ] \,,
\eeq
The coefficient functions $t_{LL} (x_t)$, $t_{RR} (x_t)$, and $t_{LR}
(x_t)$ are given by
\beq  \label{eq:tLtRtLR}
\begin{split}
  t_{LL} (x_t) & = -\frac{9 - 8 x_t}{2 x_t} + \frac{1 - x_t}{6 x_t^3}
  \, \pi^2 - \frac{1 - 20 x_t + 64 x_t^2 -24 x_t^3}{4 \hspace{0.5mm}(1
    - 4 x_t ) \hspace{0.25mm}x_t^2} \, \ln x_t + \frac{1 - x_t}{2
    x_t^3} \ln^2 x_t \\ & \phantom{xx} +\frac{2 - 7 x_t + 8 x_t^2 - 3
    x_t^3}{x_t^3} \, \text{Li}_2\left(\frac{x_t-1}{x_t}\right)
  -\frac{1 - 2 x_t}{2 x_t} \, \psi (x_t) \\ & \phantom{xx} + \frac{1 -
    12 x_t + 50 x_t^2 - 83 x_t^3 + 50 x_t^4 - 12 x_t^5}{2
    \hspace{0.5mm}(1 - 4 x_t) \hspace{0.25mm} x_t^3} \, \phi
  \left ( \frac{1}{4 x_t} \right) , \\
  t_{RR} (x_t) & = -\frac{3 - 4 x_t}{2 x_t} - \frac{1 - 12 x_t + 32
    x_t^2 -24 x_t^3}{4 \hspace{0.5mm} (1 - 4 x_t ) \hspace{0.25mm}
    x_t^2} \, \ln x_t -\frac{1 - 2 x_t + x_t^2}{x_t^2} \,
  \text{Li}_2\left(\frac{x_t-1}{x_t}\right) \\ & \phantom{xx} -\frac{1
    - 2 x_t}{2 x_t} \, \psi (x_t) - \frac{1 - 8 x_t +19 x_t^2 -18
    x_t^3+ 12 x_t^4}{2 \hspace{0.5mm} (1 - 4 x_t) \hspace{0.25mm}
    x_t^2} \, \phi \left (
    \frac{1}{4 x_t}\right) , \\
  t_{LR} (x_t) & = -7 + \frac{\pi^2}{6 x_t^2} - \frac{2 - 14 x_t + 12
    x_t^2 }{ (1 - 4 x_t ) \hspace{0.25mm} x_t} \, \ln x_t +
  \frac{\ln^2 x_t}{2x_t^2} + \frac{2 - 4x_t + 2x_t^2}{x_t^2} \,
  \text{Li}_2 \left(
    \frac{x_t-1}{x_t} \right) \\
  & \phantom{xx} -4\hspace{0.5mm} \psi (x_t) + \frac{1 - 10 x_t + 32
    x_t^2 - 32 x_t^3 + 24 x_t^4}{2 \hspace{0.5mm} (1 - 4 x_t)
    \hspace{0.25mm} x_t^2} \, \phi \left(\frac{1}{4 x_t} \right) .
\end{split} 
\eeq

In the limit $x_t \to 0$, the coefficients given in \eq{eq:tLtRtLR}
behave as
\beq \label{eq:tLtRtLRexp}
\begin{split}
  t_{LL} (x_t) & \, \to \, x_t \left( \ln ^2 x_t + \frac{7}{3}
    \ln x_t + \frac{\pi ^2}{3} + \frac{35}{36} \right) , \\
  t_{RR} (x_t) & \, \to \, x_t \left( 2 \ln ^2 x_t + \frac{11}{3}
    \ln x_t + \frac{2}{3} \, \pi^2 + \frac{5}{18} \right) , \\[2mm]
  t_{LR} (x_t) & \, \to \, x_t \left ( 6 \ln x_t - 1 \right) \, .
\end{split}
\eeq
so that for $x_t \to 0$ and leading-logarithmic accuracy, $t_{LL}
(x_t) \approx x_t \ln^2 x_t$, $t_{RR} (x_t) \approx 2 \hspace{0.5mm}
t_{LL} (x_t)$, and $t_{LR} (x_t) \approx 0$. The latter relations have
been employed in \eq{eq:STapprox}.

\section{\boldmath Contributions to $t \bar t$ Observables
    \unboldmath}
\label{app:ttobservables}

\renewcommand{\theequation}{C\arabic{equation}}
\setcounter{equation}{0}

Below we present approximate formulas that allow for a
model-independent global analysis of the $t \bar t$ observables
measured at the Tevatron. In terms of the following combinations
\beq \label{eq:gcombinations}
\begin{split}
  c_{gG}^{\pm} & = \left (g_L^\ell \pm g_R^\ell \right )
  \left (g_L^h \pm g_R^h \right ) \,, \\
  c_{GG}^{\pm} & = \big ((g_L^\ell)^2 \pm (g_R^\ell)^2 \big )
  \hspace{0.5mm} \big ((g_L^h)^2 \pm (g_R^h)^2 \big ) \,, \\
  c_{GG} & = \big ((g_L^\ell)^2 + (g_R^\ell)^2 \big ) \hspace{0.5mm}
  g_L^h \, g_R^h \,,
\end{split}
\eeq
of the coupling strengths \eq{eq:glgh}, we find 
\begin{align} \label{eq:ttobservables} 
  & \hspace{1cm} (\sigma_s)_G \approx \left \{ \left [ -\left (
        \frac{803 \hspace{-0.5mm} \GeV}{M_G} \right )^2 -\left (
        \frac{642 \hspace{-0.5mm} \GeV}{M_G} \right )^4 -\left (
        \frac{579 \hspace{-0.5mm} \GeV}{M_G} \right )^6 -\left (
        \frac{710 \hspace{-0.5mm} \GeV}{M_G} \right )^8 \right ]
    c_{gG}^+ \right.  \nonumber \\
  & \hspace{1cm} \phantom{xxx} + \left [ \left ( \frac{494
        \hspace{-0.5mm} \GeV}{M_G} \right )^4 +\left ( \frac{659
        \hspace{-0.5mm} \GeV}{M_G} \right )^6 -\left ( \frac{773
        \hspace{-0.5mm} \GeV}{M_G} \right )^8 +\left ( \frac{882
        \hspace{-0.5mm} \GeV}{M_G} \right )^{10} \right ] c_{GG}^+ \nonumber \\
  & \hspace{1cm} \phantom{xxx} \left. + \left [ \left ( \frac{462
          \hspace{-0.5mm} \GeV}{M_G} \right )^4 +\left ( \frac{556
          \hspace{-0.5mm} \GeV}{M_G} \right )^6 -\left ( \frac{553
          \hspace{-0.5mm} \GeV}{M_G} \right )^8 +\left ( \frac{734
          \hspace{-0.5mm} \GeV}{M_G} \right )^{10} \right ] c_{GG}
    \hspace{0.75mm} \right \} \hspace{0.5mm} {\rm pb} \,,
  \nonumber \\[2mm]
  & \hspace{-5mm} \left ( \frac{d\sigma_s}{d M_{t\bar t}} \right
  )_G^{M_{t \bar t} \in [0.8, 1.4] {\rm TeV}} \approx \left \{ \left [
      -\left ( \frac{181 \hspace{-0.5mm} \GeV}{M_G} \right )^2 -\left
        ( \frac{429 \hspace{-0.5mm} \GeV}{M_G} \right )^4 +\left (
        \frac{494 \hspace{-0.5mm} \GeV}{M_G} \right )^6 -\left (
        \frac{733 \hspace{-0.5mm} \GeV}{M_G} \right )^8 \right ]
    c_{gG}^+ \right.
  \nonumber \\
  & \phantom{xxxxxxxi}+ \left [ \left ( \frac{241\hspace{-0.5mm}
        \GeV}{M_G} \right )^4 +\left ( \frac{712\hspace{-0.5mm}
        \GeV}{M_G} \right )^6 -\left ( \frac{896 \hspace{-0.5mm}
        \GeV}{M_G} \right )^8 +\left ( \frac{955 \hspace{-0.5mm}
        \GeV}{M_G} \right )^{10} \right ] c_{GG}^+ \nonumber \\ &
  \phantom{xxxxxxxi} \left. + \left [ \left ( \frac{202
          \hspace{-0.5mm} \GeV}{M_G} \right )^4 +\left ( \frac{515
          \hspace{-0.5mm} \GeV}{M_G} \right )^6 -\left ( \frac{691
          \hspace{-0.5mm} \GeV}{M_G} \right )^8 +\left ( \frac{783
          \hspace{-0.5mm} \GeV}{M_G} \right )^{10} \right ] c_{GG}
    \hspace{0.75mm} \right \} \hspace{0.5mm} \frac{{\rm fb}}{\GeV} \,,
  \nonumber \\[2mm]
  & \left ( \frac{d\sigma_s}{d M_{t\bar t}} \right )_G^{M_{t \bar t} <
    0.45 {\rm TeV}} \approx \left \{ \left [ -\left ( \frac{1660
          \hspace{-0.5mm} \GeV}{M_G} \right )^2 -\left ( \frac{846
          \hspace{-0.5mm} \GeV}{M_G} \right )^4 +\left ( \frac{481
          \hspace{-0.5mm} \GeV}{M_G} \right )^6 -\left ( \frac{671
          \hspace{-0.5mm} \GeV}{M_G} \right )^8 \right ] c_{gG}^+
  \right. \nonumber \\
  & \phantom{xxxxxxxi}+ \left [ \left ( \frac{567\hspace{-0.5mm}
        \GeV}{M_G} \right )^4 +\left ( \frac{702\hspace{-0.5mm}
        \GeV}{M_G} \right )^6 -\left ( \frac{688 \hspace{-0.5mm}
        \GeV}{M_G} \right )^8 +\left ( \frac{674 \hspace{-0.5mm}
        \GeV}{M_G} \right )^{10} \right ] c_{GG}^+ \nonumber \\ &
  \phantom{xxxxxxxi} \left. + \left [ \left ( \frac{618
          \hspace{-0.5mm} \GeV}{M_G} \right )^4 +\left ( \frac{729
          \hspace{-0.5mm} \GeV}{M_G} \right )^6 -\left ( \frac{703
          \hspace{-0.5mm} \GeV}{M_G} \right )^8 +\left ( \frac{688
          \hspace{-0.5mm} \GeV}{M_G} \right )^{10} \right ] c_{GG}
    \hspace{0.75mm} \right \} \hspace{0.5mm} \frac{{\rm fb}}{\GeV} \,,
  \nonumber \\[2mm]
  & \left ( \frac{d\sigma_s}{d M_{t\bar t}} \right )_G^{M_{t \bar t} >
    0.45 {\rm TeV}} \approx \left \{ \left [ -\left ( \frac{487
          \hspace{-0.5mm} \GeV}{M_G} \right )^2 -\left ( \frac{536
          \hspace{-0.5mm} \GeV}{M_G} \right )^4 -\left ( \frac{522
          \hspace{-0.5mm} \GeV}{M_G} \right )^6 -\left ( \frac{673
          \hspace{-0.5mm} \GeV}{M_G} \right )^8 \right ] c_{gG}^+
  \right. \nonumber  \\
  & \phantom{xxxxxxxi}+ \left [ \left ( \frac{418\hspace{-0.5mm}
        \GeV}{M_G} \right )^4 +\left ( \frac{609\hspace{-0.5mm}
        \GeV}{M_G} \right )^6 -\left ( \frac{735 \hspace{-0.5mm}
        \GeV}{M_G} \right )^8 +\left ( \frac{847 \hspace{-0.5mm}
        \GeV}{M_G} \right )^{10} \right ] c_{GG}^+ \nonumber \\ &
  \phantom{xxxxxxxi} \left. + \left [ \left ( \frac{365
          \hspace{-0.5mm} \GeV}{M_G} \right )^4 +\left ( \frac{500
          \hspace{-0.5mm} \GeV}{M_G} \right )^6 -\left ( \frac{535
          \hspace{-0.5mm} \GeV}{M_G} \right )^8 +\left ( \frac{704
          \hspace{-0.5mm} \GeV}{M_G} \right )^{10} \right ] c_{GG}
    \hspace{0.75mm} \right \} \hspace{0.5mm} \frac{{\rm fb}}{\GeV} \,,
  \nonumber \\[2mm]
  & \hspace{1cm} (\sigma_a)_G \approx \left \{ \left [ -\left (
        \frac{497 \hspace{-0.5mm} \GeV}{M_G} \right )^2 -\left (
        \frac{527 \hspace{-0.5mm} \GeV}{M_G} \right )^4 -\left (
        \frac{508 \hspace{-0.5mm} \GeV}{M_G} \right )^6 -\left (
        \frac{668 \hspace{-0.5mm} \GeV}{M_G} \right )^8 \right ]
    c_{gG}^- \right.
  \nonumber \\
  & \hspace{1cm} \left. \phantom{xxx} + \left [ \left ( \frac{436
          \hspace{-0.5mm} \GeV}{M_G} \right )^4 +\left ( \frac{619
          \hspace{-0.5mm} \GeV}{M_G} \right )^6 -\left ( \frac{745
          \hspace{-0.5mm} \GeV}{M_G} \right )^8 +\left ( \frac{855
          \hspace{-0.5mm} \GeV}{M_G} \right )^{10} \right ] c_{GG}^-
  \right \} \hspace{0.5mm} {\rm pb} \,, \nonumber \\[2mm]
  & \left (\frac{d\sigma_a}{dM_{t \bar t}} \right)_G^{M_{t \bar t} <
    0.45{\rm TeV}} \approx \left \{ \left [ -\left ( \frac{840
          \hspace{-0.5mm} \GeV}{M_G} \right )^2 -\left ( \frac{601
          \hspace{-0.5mm} \GeV}{M_G} \right )^4 -\left ( \frac{338
          \hspace{-0.5mm} \GeV}{M_G} \right )^6 -\left ( \frac{565
          \hspace{-0.5mm} \GeV}{M_G} \right )^8 \right ] c_{gG}^-
  \right. \nonumber \\
  & \hspace{1cm} \left. \phantom{xxx} + \left [ \left ( \frac{477
          \hspace{-0.5mm} \GeV}{M_G} \right )^4 +\left ( \frac{589
          \hspace{-0.5mm} \GeV}{M_G} \right )^6 -\left ( \frac{590
          \hspace{-0.5mm} \GeV}{M_G} \right )^8 +\left ( \frac{612
          \hspace{-0.5mm} \GeV}{M_G} \right )^{10} \right ] c_{GG}^-
  \right \} \hspace{0.5mm} \frac{\rm fb}{\GeV} \,,  \nonumber \\[2mm]
  & \left (\frac{d\sigma_a}{dM_{t \bar t}} \right)_G^{M_{t \bar t} >
    0.45{\rm TeV}} \approx \left \{ \left [ -\left ( \frac{340
          \hspace{-0.5mm} \GeV}{M_G} \right )^2 -\left ( \frac{447
          \hspace{-0.5mm} \GeV}{M_G} \right )^4 -\left ( \frac{522
          \hspace{-0.5mm} \GeV}{M_G} \right )^6 -\left ( \frac{596
          \hspace{-0.5mm} \GeV}{M_G} \right )^8 \right ] c_{gG}^-
  \right. \nonumber \\
  & \hspace{1cm} \left. \phantom{xxx} + \left [ \left ( \frac{358
          \hspace{-0.5mm} \GeV}{M_G} \right )^4 +\left ( \frac{649
          \hspace{-0.5mm} \GeV}{M_G} \right )^6 -\left ( \frac{815
          \hspace{-0.5mm} \GeV}{M_G} \right )^8 +\left ( \frac{880
          \hspace{-0.5mm} \GeV}{M_G} \right )^{10} \right ] c_{GG}^-
  \right \} \hspace{0.5mm} \frac{\rm fb}{\GeV} \,.
\end{align} 
Notice that the coefficients in \eq{eq:gcombinations} labelled by $gG$
and $GG$ arise from the interference of the gluon with the axigluon
and the interference of the axigluon with itself. In order to obtain
the above formulas, the total decay width of the axigluon has been
fixed to $\Gamma_G/M_G = 10\%$, which is a typical value for massive
color-octet bosons with QCD-like couplings to quarks. In the specific
axigluon model taken as a benchmark in the main body of this work, one
has $\Gamma_G/M_G \approx 10\%$ in the regime of $M_G \approx 1 \TeV$,
$\theta \gtrsim 35^\circ$, and fourth-generation quark masses above
the Tevatron bounds \cite{Lister:2008is, Aaltonen:2011vr}. For smaller
values of the mixing angle $\theta$ the total decay width becomes
significantly larger, thereby effectively suppressing the size of the
axigluon contributions. In this case the formulas
\eq{eq:ttobservables} should not be applied, since they typically lead
to constraints in the $M_G\hspace{0.5mm}$--$\hspace{0.5mm}\theta$
plane that are too restrictive.  In our numerical analysis presented
in \Sec{sec:global}, we have therefore included the exact dependence
on $\Gamma_G$, employing the expressions for the partial widths given
in \eq{eq:GammaG}.

\section{\boldmath Matrix Elements for Dijet Production
    \unboldmath}
\label{app:dijet}

\renewcommand{\theequation}{D\arabic{equation}}
\setcounter{equation}{0}

This appendix contains the new-physics corrections to the dijet
tree-level matrix elements squared. Possible flavor-violating effects
are neglected throughout. In terms of the couplings \eq{eq:glgh} and
the partonic Mandelstam variables $\hat s$, $\hat t$, and $\hat u$, we
obtain the following results 
\bea \label{eq:dijetformula} 
\begin{split}
  \left ( \overline{\sum} \, \big | {\cal M} (q_i \bar q_i \to q_i
    \bar q_i) \big |^2 \right )_G & = \frac{4}{9} \, g_s^4 \, \bigg \{
  \, 2 \hspace{0.25mm} g_L^\ell g_R^\ell \bigg [ \frac{\hat t^2(\hat
    s-M_G^2)} {\hat s\, ((\hat s-M_G^2)^2+\Gamma_G^2 M_G^2)} + (\hat
  s\leftrightarrow \hat t)
  \bigg ]\\
  & \phantom{xx} + \, \left ( (g_L^\ell)^2+(g_R^\ell)^2 \right ) \bigg
  [ \frac{\hat u^2 (\hat s-M_G^2)}{(\hat s-M_G^2)^2+\Gamma_G^2 M_G^2}
  \left( \frac{1}{\hat s} - \frac{1}
    {3\,\hat t} \right) + (\hat s\leftrightarrow \hat t)\bigg ]\\
  &\phantom{xx} + \frac{1}{2} \, \frac{\left (
      (g_L^\ell)^4+(g_R^\ell)^4 \right ) \hat u^2}{(\hat
    s-M_G^2)^2+\Gamma_G^2 M_G^2} \left [ \, 1 + \frac{(\hat
      s-M_G^2)^2+\Gamma_G^2 M_G^2}{(\hat t-M_G^2)^2+\Gamma_G^2 M_G^2}
  \right. \\ & \hspace{4.75cm} \left. -\frac{2}{3} \, \frac{(\hat
      s-M_G^2)(\hat t-M_G^2)+\Gamma_G^2 M_G^2}{(\hat t-M_G^2)^2
      +\Gamma_G^2 M_G^2} \, \right ] \hspace{-1.5mm} \bigg \} ,\\[2mm]
  \left ( \overline{\sum} \, \big |{\cal M} (q_i \bar q_j \to q_i \bar
    q_j) \big |^2 \right )_G & = \frac{4}{9} \, g_s^4 \, \bigg \{
  \left [ \left ( (g_L^\ell)^2+(g_R^\ell)^2 \right ) \hat u^2 + 2
    \hspace{0.25mm} g_L^\ell g_R^\ell \, \hat s^2 \hspace{0.5mm}
  \right ]
  \frac{\hat t-M_G^2}{\hat t\, ((\hat t-M_G^2)^2+\Gamma_G^2 M_G^2)}\\
  & \phantom{xx} + \frac{1}{2} \left ( (g_L^\ell)^4+(g_R^\ell)^4
  \right ) \frac{\hat u^2}{(\hat t-M_G^2)^2+\Gamma_G^2 M_G^2} \, \bigg
  \}\,,
\end{split}
\eea 
where $g_s$ denotes the strong coupling constant. The result for $q_i
q_i \to q_i q_i$ ($q_i q_j \to q_i q_j$) follows from the expression
in the first (second) line by simply interchanging $\hat s$ with $\hat
u$, while the result for $q_i \bar q_i \to q_j \bar q_j$ is obtained
by interchanging $\hat s$ with $\hat t$ in the second line.  The
matrix elements in terms of $\chi = (1 + |\cos\hat{\theta}|)/(1 -
|\cos\hat{\theta}|)$ and $M_{jj}$ are easily obtained using $\hat s =
M_{jj}^2$, $\hat t = - (1-\cos\hat{\theta})\,\hat{s}/2$, and $\hat u =
- (1+\cos\hat{\theta})\,\hat{s}/2$. In order to obtain the matrix
elements after integrating out the axigluon, one simply neglects the
width $\Gamma_G$ and expands around the limit $\hat s, \hat t, \hat u
\ll M_G$.

\end{appendix}


\begin{thebibliography}{999}

\bibitem{Randall:1999ee}
  L.~Randall and R.~Sundrum,
  Phys.\ Rev.\ Lett.\  {\bf 83}, 3370 (1999)
  [arXiv:hep-ph/9905221].

\bibitem{Hill:2002ap}
  C.~T.~Hill and E.~H.~Simmons,
  Phys.\ Rept.\  {\bf 381}, 235 (2003)
  [Erratum-ibid.\  {\bf 390}, 553 (2004)] [arXiv:hep-ph/0203079].

\bibitem{Frampton:1987dn}
  P.~H.~Frampton and S.~L.~Glashow,
  Phys.\ Lett.\  B {\bf 190}, 157 (1987).

\bibitem{Frampton:1987ut}
  P.~H.~Frampton and S.~L.~Glashow,
  Phys.\ Rev.\ Lett.\  {\bf 58}, 2168 (1987).

\bibitem{Bagger:1987fz}
  J.~Bagger, C.~Schmidt and S.~King,
  Phys.\ Rev.\  D {\bf 37}, 1188 (1988).

\bibitem{Hill:1991at}
  C.~T.~Hill,
  Phys.\ Lett.\  B {\bf 266}, 419 (1991).

\bibitem{Hill:1994hp}
  C.~T.~Hill,
  Phys.\ Lett.\  B {\bf 345}, 483 (1995)
  [arXiv:hep-ph/9411426].

\bibitem{Hill:1993hs}
  C.~T.~Hill and S.~J.~Parke,
  Phys.\ Rev.\  D {\bf 49}, 4454 (1994)
  [arXiv:hep-ph/9312324].

\bibitem{Chivukula:1996yr}
  R.~S.~Chivukula, A.~G.~Cohen and E.~H.~Simmons,
  Phys.\ Lett.\  B {\bf 380}, 92 (1996)
  [arXiv:hep-ph/9603311].

\bibitem{Agashe:2003zs}
  K.~Agashe, A.~Delgado, M.~J.~May and R.~Sundrum,
  JHEP {\bf 0308}, 050 (2003)
  [arXiv:hep-ph/0308036].

\bibitem{Agashe:2006hk}
  K.~Agashe, A.~Belyaev, T.~Krupovnickas, G.~Perez and J.~Virzi,
  Phys.\ Rev.\  D {\bf 77}, 015003 (2008)
  [arXiv:hep-ph/0612015].

\bibitem{Lillie:2007yh}
  B.~Lillie, L.~Randall and L.~T.~Wang,
  JHEP {\bf 0709}, 074 (2007)
  [arXiv:hep-ph/0701166].

\bibitem{CDFnotetot} 
  E.~Thomson {\it et al.} [CDF~Collaboration], 
  Conference \ Note \ 9913, October 19, 2009,
  \href{http://www-cdf.fnal.gov/physics/new/top/2009/xsection/ttbar_combined_46invfb/}{\tt
    http://www-cdf.fnal.gov/physics/new/top/2009/xsection/ttbar\_combined\_46invfb/}

\bibitem{D0notetot}
  D{\O}~Collaboration, 
  Conference \ Note \ 5907-CONF, March 12, 2009, 	
  \href{http://www-d0.fnal.gov/Run2Physics/WWW/results/prelim/TOP/T79/}{\tt
   http://www-d0.fnal.gov/Run2Physics/WWW/results/prelim/TOP/T79/}

\bibitem{Aaltonen:2009iz}
  T.~Aaltonen {\it et al.}  [CDF~Collaboration],
  Phys.\ Rev.\ Lett.\  {\bf 102}, 222003 (2009)
  [arXiv:0903.2850 [hep-ex]].

\bibitem{Aaltonen:2011kc}
  T.~Aaltonen {\it et al.}  [CDF~Collaboration],
  arXiv:1101.0034 [hep-ex].

\bibitem{Schwarz:2006ud}
  T.~A.~Schwarz, FERMILAB-THESIS-2006-51, UMI-32-38081.

\bibitem{Abazov:2007qb}
  V.~M.~Abazov {\it et al.}  [{D\O}~Collaboration],
  Phys.\ Rev.\ Lett.\  {\bf 100}, 142002 (2008)
  [arXiv:0712.0851 [hep-ex]].

\bibitem{Aaltonen:2008hc}
  T.~Aaltonen {\it et al.}  [CDF~Collaboration],
  Phys.\ Rev.\ Lett.\  {\bf 101}, 202001 (2008)
  [arXiv:0806.2472 [hep-ex]].

\bibitem{D0brandnew} 
  {D\O}~Collaboration, Conference Note 6062-CONF, July 23, 2010, 
  \href{http://www-d0.fnal.gov/Run2Physics/WWW/results/prelim/TOP/T90/}{\tt
    http://www-d0.fnal.gov/Run2Physics/WWW/results/prelim/TOP/T90/}

\bibitem{CDFdileptonnote} 
  CDF~Collaboration, CDF Note 10398, March 10,
  2011, \href{http://www-cdf.fnal.gov/physics/new/top/2011/DilAfb/}{\tt 
    http://www-cdf.fnal.gov/physics/new/top/2011/DilAfb/}

\bibitem{LEPEWWG:2005ema}
  S. Schael {\it et al.} \ [ALEPH~Collaboration],
  Phys.\ Rept.\ {\bf 427}, 257 (2006)
  [arXiv:hep-ex/0509008].

\bibitem{Peskin:1991sw} 
  M.~E.~Peskin and T.~Takeuchi,
  Phys.\ Rev.\ D {\bf 46} (1992) 381.

\bibitem{Khachatryan:2010jd}
  V.~Khachatryan {\it et al.}  [CMS Collaboration],
  Phys.\ Rev.\ Lett.\  {\bf 105}, 211801 (2010)
  [arXiv:1010.0203 [hep-ex]].

\bibitem{Khachatryan:2011as}
  V.~Khachatryan {\it et al.}  [CMS Collaboration],
  Phys.\ Rev.\ Lett.\  {\bf 106}, 201804 (2011)
  [arXiv:1102.2020 [hep-ex]].

\bibitem{Collaboration:2011aj}
  G.~Aad {\it et al.}  [ATLAS Collaboration],
  New J.\ Phys.\  {\bf 13}, 053044 (2011)
  [arXiv:1103.3864 [hep-ex]]
  additional material available at  
  \href{https://atlas.web.cern.ch/Atlas/GROUPS/PHYSICS/PAPERS/exoticsdijets2010/}{\tt
    https://atlas.web.cern.ch/Atlas/GROUPS/PHYSICS/PAPERS/exoticsdijets2010/}

\bibitem{Frampton:2009rk}
  P.~H.~Frampton, J.~Shu and K.~Wang,
  Phys.\ Lett.\  B {\bf 683}, 294 (2010)
  [arXiv:0911.2955 [hep-ph]].

\bibitem{Cao:2010zb}
  Q.~H.~Cao, D.~McKeen, J.~L.~Rosner, G.~Shaughnessy and C.~E.~M.~Wagner,
  Phys.\ Rev.\  D {\bf 81}, 114004 (2010)
  [arXiv:1003.3461 [hep-ph]].

\bibitem{Chivukula:2010fk}
  R.~S.~Chivukula, E.~H.~Simmons and C.~P.~Yuan,
  Phys.\ Rev.\  D {\bf 82}, 094009 (2010)
  [arXiv:1007.0260 [hep-ph]].

\bibitem{Han:2010rf}
  T.~Han, I.~Lewis and Z.~Liu,
  JHEP {\bf 1012}, 085 (2010)
  [arXiv:1010.4309 [hep-ph]].

\bibitem{Bai:2011ed}
  Y.~Bai, J.~L.~Hewett, J.~Kaplan and T.~G.~Rizzo,
  JHEP {\bf 1103}, 003 (2011)
  [arXiv:1101.5203 [hep-ph]].

\bibitem{AguilarSaavedra:2011vw}
  J.~A.~Aguilar-Saavedra and M.~Perez-Victoria,
  JHEP {\bf 1105}, 034 (2011)
  [arXiv:1103.2765 [hep-ph]].

\bibitem{Gresham:2011pa}
  M.~I.~Gresham, I.~W.~Kim and K.~M.~Zurek,
  arXiv:1103.3501 [hep-ph].

\bibitem{Hewett:2011wz}
  J.~L.~Hewett, J.~Shelton, M.~Spannowsky, T.~M.~P.~Tait and M.~Takeuchi,
  arXiv:1103.4618 [hep-ph].

\bibitem{Barcelo:2011fw}
  R.~Barcelo, A.~Carmona, M.~Masip and J.~Santiago,
  arXiv:1105.3333 [hep-ph].

\bibitem{AguilarSaavedra:2011hz}
  J.~A.~Aguilar-Saavedra and M.~Perez-Victoria,
  arXiv:1105.4606 [hep-ph].

\bibitem{Antunano:2007da}
  O.~Antunano, J.~H.~K\"uhn, and G.~Rodrigo,
  Phys.\ Rev.\ D {\bf 77}, 014003 (2008),
  [arXiv:0709.1652 [hep-ph]].

\bibitem{Ferrario:2008wm}
  P.~Ferrario and G.~Rodrigo,
  Phys.\ Rev.\  D {\bf 78}, 094018 (2008)
  [arXiv:0809.3354 [hep-ph]].

\bibitem{Ferrario:2009bz}
  P.~Ferrario and G.~Rodrigo,
  Phys.\ Rev.\  D {\bf 80}, 051701 (2009)
  [arXiv:0906.5541 [hep-ph]].

\bibitem{Achard:2001qw}
  P.~Achard {\it et al.}  [L3~Collaboration],
  Phys.\ Lett.\  B {\bf 517}, 75 (2001)
  [arXiv:hep-ex/0107015].

\bibitem{Lister:2008is}
  A.~Lister  [CDF~Collaboration],
  arXiv:0810.3349 [hep-ex].

\bibitem{Aaltonen:2011vr}
  T.~Aaltonen {\it et al.}  [CDF Collaboration],
  Phys.\ Rev.\ Lett.\  {\bf 106}, 141803 (2011)
  [arXiv:1101.5728 [hep-ex]].

\bibitem{Collaboration:2011em}
  S.~Chatrchyan {\it et al.} [CMS~Collaboration],
  arXiv:1102.4746 [hep-ex].

\bibitem{Buchalla:1995dp}
  G.~Buchalla, G.~Burdman, C.~T.~Hill and D.~Kominis,
  Phys.\ Rev.\  D {\bf 53}, 5185 (1996)
  [arXiv:hep-ph/9510376].

\bibitem{Burdman:2000in}
  G.~Burdman, K.~D.~Lane and T.~Rador,
  Phys.\ Lett.\  B {\bf 514}, 41 (2001)
  [arXiv:hep-ph/0012073].

\bibitem{Martin:2004ec}
  A.~Martin and K.~Lane,
  Phys.\ Rev.\  D {\bf 71}, 015011 (2005)
  [arXiv:hep-ph/0404107].

\bibitem{Buras:2001ra}
  A.~J.~Buras, S.~J\"ager and J.~Urban,
  Nucl.\ Phys.\  B {\bf 605}, 600 (2001)
  [arXiv:hep-ph/0102316].

\bibitem{Laiho:2009eu}
  J.~Laiho, E.~Lunghi and R.~S.~Van de Water,
  Phys.\ Rev.\  D {\bf 81}, 034503 (2010)
  [arXiv:0910.2928 [hep-ph]].

\bibitem{Lubicz:2008am}
  V.~Lubicz and C.~Tarantino,
  Nuovo Cim.\  B {\bf 123}, 674 (2008)
  [arXiv:0807.4605 [hep-lat]].

\bibitem{Buras:2010pza}
  A.~J.~Buras, D.~Guadagnoli and G.~Isidori,
  Phys.\ Lett.\  B {\bf 688}, 309 (2010)
  [arXiv:1002.3612 [hep-ph]].

\bibitem{Charles:2004jd}
  J.~Charles {\it et al.}  [CKMfitter Group],
  Eur.\ Phys.\ J.\ C {\bf 41}, 1 (2005) [arXiv:hep-ph/0406184],
  updated results available at
  \href{http://ckmfitter.in2p3.fr}{\tt http://ckmfitter.in2p3.fr}

\bibitem{Brod:2010mj}
  J.~Brod and M.~Gorbahn,
  Phys.\ Rev.\  D {\bf 82}, 094026 (2010)
  [arXiv:1007.0684 [hep-ph]].

\bibitem{Nakamura:2010zzi}
  K.~Nakamura {\it et al.}  [Particle Data Group],
  J.\ Phys.\ G {\bf 37}, 075021 (2010), updated results available at
  \href{http://pdglive.lbl.gov}{\tt http://pdglive.lbl.gov}

\bibitem{TheHeavyFlavorAveragingGroup:2010qj}
  D.~Asner {\it et al.} [The Heavy Flavor Averaging Group],  
  arXiv:1010.1589 [hep-ex], updated results available at 
  \href{http://www.slac.stanford.edu/xorg/hfag}{\tt http://www.slac.stanford.edu/xorg/hfag}

\bibitem{Hill:1994di}
  C.~T.~Hill and X.~m.~Zhang,
  Phys.\ Rev.\  D {\bf 51}, 3563 (1995)
  [arXiv:hep-ph/9409315].

\bibitem{Abazov:2011ws}
  V.~Abazov  {\it et al.} [D{\O}~Collaboration],
  arXiv:1104.4590 [hep-ex].

\bibitem{Baur:2004uw}
  U.~Baur, A.~Juste, L.~H.~Orr and D.~Rainwater,
  Phys.\ Rev.\  D {\bf 71}, 054013 (2005)
  [arXiv:hep-ph/0412021].

\bibitem{Baur:2005wi}
  U.~Baur, A.~Juste, D.~Rainwater and L.~H.~Orr,
  Phys.\ Rev.\  D {\bf 73}, 034016 (2006)
  [arXiv:hep-ph/0512262].

\bibitem{Berger:2009hi}
  E.~L.~Berger, Q.~H.~Cao and I.~Low,
  Phys.\ Rev.\  D {\bf 80}, 074020 (2009)
  [arXiv:0907.2191 [hep-ph]].

\bibitem{Field:1997gz}
  J.~H.~Field,
  Mod.\ Phys.\ Lett.\ A {\bf 13}, 1937 (1998)
  [arXiv:hep-ph/9801355].

\bibitem{Arbuzov:2005ma}
  A.~B.~Arbuzov {\it et al.},
  Comput.\ Phys.\ Commun.\ {\bf 174}, 728 (2006)
  [arXiv:hep-ph/0507146].

\bibitem{Group:1900yx}
  The Tevatron Electroweak Working Group [CDF and D{\O}~Collaboration],
  arXiv:1007.3178 [hep-ex].

\bibitem{Burdman:1999us}
  G.~Burdman, R.~S.~Chivukula and N.~J.~Evans,
  Phys.\ Rev.\  D {\bf 61}, 035009 (2000)
  [arXiv:hep-ph/9906292].

\bibitem{He:2001tp}
  H.~J.~He, N.~Polonsky and S.~f.~Su,
  Phys.\ Rev.\  D {\bf 64}, 053004 (2001)
  [arXiv:hep-ph/0102144].

\bibitem{Bridgeman:2008zz}
  A.~Bridgeman, FERMILAB-THESIS-2008-50.

\bibitem{Aaltonen:2008dn}
  T.~Aaltonen {\it et al.}  [CDF~Collaboration],
  Phys.\ Rev.\  D {\bf 79}, 112002 (2009)
  [arXiv:0812.4036 [hep-ex]].

\bibitem{Collaboration:2010bc}
  G.~Aad {\it et al.}  [ATLAS~Collaboration],
  Phys.\ Rev.\ Lett.\  {\bf 105}, 161801 (2010)
  [arXiv:1008.2461 [hep-ex]].

\bibitem{Abazov:2009mh}
  V.~M.~Abazov {\it et al.}  [D{\O}~Collaboration],
  Phys.\ Rev.\ Lett.\  {\bf 103}, 191803 (2009)
  [arXiv:0906.4819 [hep-ex]].

\bibitem{Collaboration:2010eza}
  G.~Aad {\it et al.}  [ATLAS Collaboration],
  Phys.\ Lett.\  B {\bf 694}, 327 (2011)
  [arXiv:1009.5069 [hep-ex]].

\bibitem{Randall:2001gb}
  L.~Randall and M.~D.~Schwartz,
  JHEP {\bf 0111}, 003 (2001)
  [arXiv:hep-th/0108114].

\bibitem{Bauer:2010iq}
  M.~Bauer, F.~Goertz, U.~Haisch, T.~Pfoh and S.~Westhoff,
  JHEP {\bf 1011}, 039 (2010)
  [arXiv:1008.0742 [hep-ph]].

\bibitem{MCFM}
  J.~Campbell and R.~K.~Ellis, 
  \href{http://mcfm.fnal.gov}{\tt http://mcfm.fnal.gov}

\bibitem{Nason:1987xz}
  P.~Nason, S.~Dawson and R.~K.~Ellis,
  Nucl.\ Phys.\  B {\bf 303}, 607 (1988).

\bibitem{Martin:2009iq}
  A.~D.~Martin, W.~J.~Stirling, R.~S.~Thorne and G.~Watt,
  Eur.\ Phys.\ J.\  C {\bf 63}, 189 (2009)
  [arXiv:0901.0002 [hep-ph]].

\bibitem{Ellis:1991qj}
  R.~K.~Ellis, W.~J.~Stirling and B.~R.~Webber,
  Camb.\ Monogr.\ Part.\ Phys.\ Nucl.\ Phys.\ Cosmol.\  {\bf 8}, 1 (1996).

\bibitem{Eichten:1983hw}
  E.~Eichten, K.~D.~Lane and M.~E.~Peskin,
  Phys.\ Rev.\ Lett.\  {\bf 50}, 811 (1983).

\bibitem{Lane:1996gr}
  K.~D.~Lane,
  arXiv:hep-ph/9605257.

\bibitem{Gao:2011ha}
  J.~Gao, C.~S.~Li, J.~Wang, H.~X.~Zhu and C.~P.~Yuan,
  Phys.\ Rev.\ Lett.\  {\bf 106}, 142001 (2011)
  [arXiv:1101.4611 [hep-ph]].

\bibitem{Albajar:1988rs}
  C.~Albajar {\it et al.}  [UA1 Collaboration],
  Phys.\ Lett.\  B {\bf 209}, 127 (1988).

\bibitem{Abe:1993it}
  F.~Abe {\it et al.}  [CDF Collaboration],
  Phys.\ Rev.\ Lett.\  {\bf 71}, 2542 (1993).

\bibitem{Carone:1994aa}
  C.~D.~Carone and H.~Murayama,
  Phys.\ Rev.\ Lett.\  {\bf 74}, 3122 (1995)
  [arXiv:hep-ph/9411256].

\bibitem{Holdom:1995id}
  B.~Holdom,
  Phys.\ Lett.\  B {\bf 351}, 279 (1995)
  [arXiv:hep-ph/9502273].

\bibitem{Davydychev:1992mt}
  A.~I.~Davydychev and J.~B.~Tausk,
  Nucl.\ Phys.\  B {\bf 397}, 123 (1993).

\end{thebibliography}
\end{document}